%% file: Main.tex
\input{LatexHeader}

\begin{document}

\include{Title}

\include{Introduction}

\include{Electron}

\include{MuonTau}

\include{neutrino}

\include{Conclusion}

\input{output.bbl}

\end{document}

%% file: LatexHeader.tex
\documentclass[onecolumn,sort&compress,numbers]{els-mrw} 

\usepackage{amsmath,amssymb,amsfonts,amsthm,makeidx,graphicx}
\usepackage{txfonts}
\usepackage{helvet}
\usepackage{indentfirst}
\usepackage{subcaption}
\usepackage[compat=1.1.0]{tikz-feynman}
\usepackage{contour}
\usetikzlibrary{arrows,shapes,positioning}
\usetikzlibrary{decorations.markings}
\usetikzlibrary{decorations.pathmorphing}
\usetikzlibrary{decorations.pathreplacing}
\usetikzlibrary{patterns}
\usetikzlibrary{plotmarks}
\usetikzlibrary{shadows}
\tikzfeynmanset{/tikzfeynman/warn luatex = false}
\usepackage{silence}
\usepackage{breqn}
\usepackage{siunitx}

%% file: Title.tex

\chapter{Measured Lepton Magnetic Moments}\label{LeptonChapter}


\author[1]{Gerald Gabrielse}%
\author[2,3]{Graziano Venanzoni}%

\address[1]{\orgname{Center for Fundamental Physics at Northwestern University}, \orgdiv{Department of Physics and Astronomy}, \orgaddress{Technology Center F131, 2145 Sheridan Avenue, Evanston, IL 60208}}
\address[2]{\orgname{University of Liverpool}, \orgaddress{ Liverpool L69 3BX, United Kingdom}}
\address[3]{\orgname{Istituto Nazionale di Fisica Nucleare (INFN), Sezione di Pisa}, \orgaddress{Largo Bruno Pontecorvo 3, 56127 Pisa, Italy}}
\articletag{Lepton Magnetic Moments}

\maketitle

\begin{abstract}[Abstract]  

The electron and muon magnetic moments have played, and continue to play, important roles in testing the fundamental mathematical description of physical reality called the Standard Model of particle physics (SM).  The electron magnetic moment is the most precisely measured property of an elementary particle and the most precise SM prediction, setting up the most precise confrontation ever between experiment and theory. It enables the most precise test of quantum field theory, and of the fundamental CPT symmetry invariance of the SM with leptons.  The stable electron is studied with quantum methods while the electron remains for months in its quantum ground and first excited states.  The muon magnetic moment is one of the most precisely measured property of an unstable elementary particle.  Although less precise measured than the electron, it provides greater sensitivity to physics beyond the Standard Model -- a powerful tool for testing the existence of new particles and forces. Because muons decay quickly, they must be studied as they orbit at nearly the speed of light in a large storage ring.  The extremely high precision of the electron and muon magnetic moment measurements  has driven major advances in theoretical physics, inspiring new techniques in quantum field theory, precision calculations, and lattice gauge theory.  
Only experimental limits currently exist on the size of the magnetic moments of the tau and neutrino leptons.
\end{abstract}

\begin{keywords}
Lepton magnetic moments\sep Precision tests of the Standard Model \sep 
Quantum technology \sep Quantum electrodynamics \sep Physics beyond the Standard Model 
\end{keywords}

\begin{figure}[h]
\centering
\framebox(7cm,4cm){
\includegraphics[width=7cm]{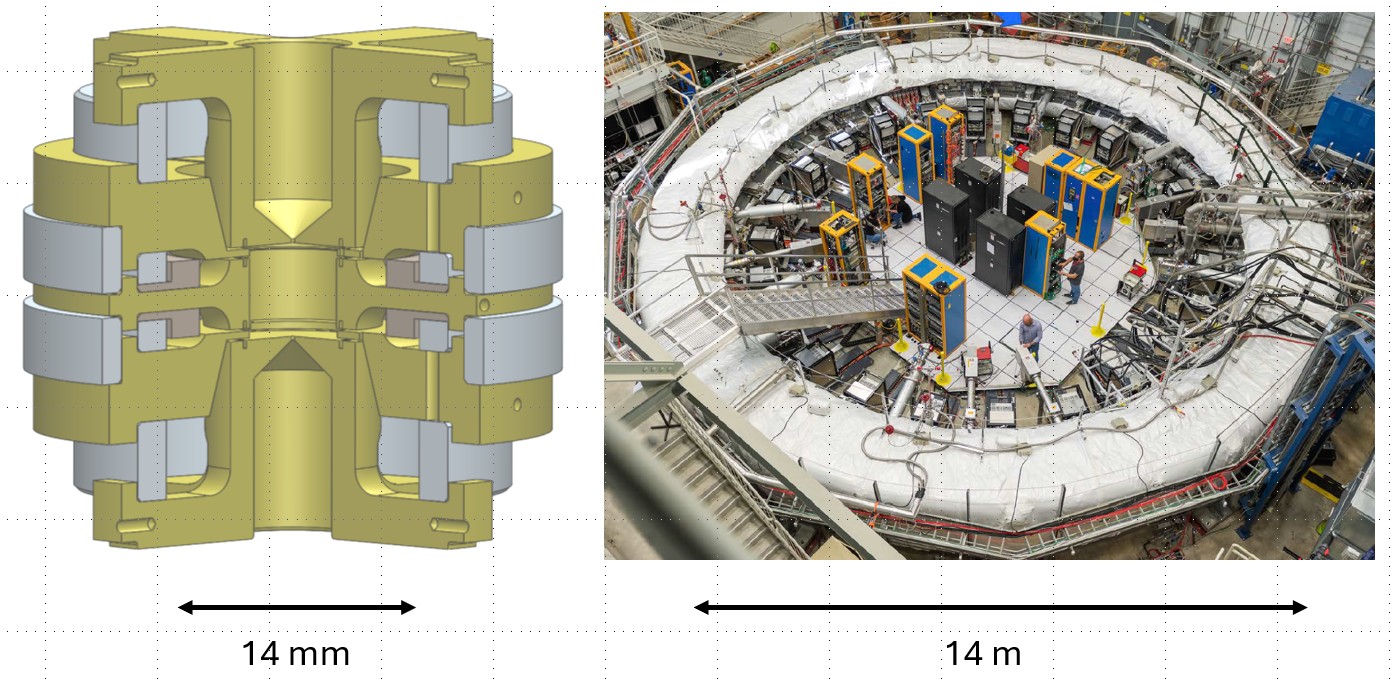}
}
\caption{The electron and muon magnetic moments are measured within structures that differ in size by a factor of $10^3$, at energies and storage times that differ by a $10^{13}$ and $10^9$.   Left: a roughly 14-mm cylindrical Penning trap cavity confines  a single electron (or positron) for months, with excitations from its cyclotron ground state to its 0.6 meV excited state used to measure its magnetic moment.  Right: a 14-m diameter storage ring within which the magnetic moments of approximately 5000 muons with a momentum of 3.1 GeV/c are stored for up to 700 $\mu s$ every 90 ms on average. }
\label{fig:titlepage}
\end{figure}


\begin{glossary}[Nomenclature]
	\begin{tabular}{@{}lp{34pc}@{}}
        $a$ & Magnetic moment anomaly (also called anomalous magnetic moment); a probe of quantum fluctuations as predicted by quantum field theory and of possible BSM physics\\
        BNL & Brookhaven National Laboratory, Upton, New York, United States \\
        BSM & Beyond the Standard Model, {\it e.g.} new particles, forces, or interactions missing in the Standard Model \\
        CERN & European Organization for Nuclear Research, the world’s largest particle physics laboratory, located near Geneva, on the border between Switzerland and France \\
        FNAL & Fermi National Accelerator Laboratory (commonly known as \emph{Fermilab}), a major particle physics laboratory near Chicago, United States \\
        $g$ & Gyromagnetic factor (also known as the g-value or g-ratio) \\      
        $g/2$ & Dimensionless magnitude of a magnetic moment scaled by its natural unit \\
        $g -2$ & Difference of the gyromagnetic factor $g$ from the Dirac prediction ($g=2$) for leptons \\
        $e\,\hbar/2m_{\mu}$ & Natural scale of the magnetic moment for the muon \\
        LHC & Large Hadron Collider, the world’s largest and most powerful particle collider, located at CERN near Geneva \\
        $\mu$ & Magnetic moment \\
        $\mu_B = e\,\hbar/2m_e$ & Natural unit for the magnetic moment of the electron (or positron), called the Bohr magneton \\
        NMR & Nuclear Magnetic Resonance technique; used to precisely measure the magnetic field in experiments determining the muon magnetic moment \\
        SM & Standard Model of particle physics \\
        $T_a = \frac{2\pi}{\omega_a}$ & Anomalous precession period (also referred to as the $g\!-\!2$ precession period) \\
        $\omega_a$ & Angular anomaly frequency, or anomalous spin precession frequency \\
        $\omega_c$ & Cyclotron angular frequency \\
        $\omega_p$ & Spin precession frequency (also Larmor precession frequency) of protons which are contained in NMR probes to precisely determine the magnetic field in the Muon $g-2$ experiment\\       
        $\omega_s$ & Angular spin precession frequency \\
	\end{tabular}
\end{glossary}

\section*{Objectives}
\begin{itemize}
\item The electron and the negative muon are lepton particles believed to differ only in that the muon mass is 207 times greater, allowing the muon to live for only about two millionths of a second while the electron is stable.
\item The stable electron is studied while it is at rest in its lowest quantum 
states in a Penning trap that is about 14 mm in diameter.
\item The muon is studied before it decays while moving at almost the speed of light in an apparatus a thousand times larger.  
\item The magnetic moments of an electron, positron and muon provide extremely important tests of the Standard Model of particle physics
\item The electron magnetic moment, the most precisely measured property of an elementary particle, tests the most precise prediction of the Standard Model
\item The muon magnetic moment provides one of the most sensitive tests for new particles that is ``beyond''  the Standard Model.   
\end{itemize}

%% file: Introduction.tex
\section{Introduction and overview}\label{sec:Introduction}

\subsection{Standard Model of particle physics}

The Standard Model of Particle Physics (SM) is the fundamental mathematical description of physical reality.  The SM is a collection of elementary particles (with no internal structure), interaction forces (electromagnetic, strong and weak), and symmetries (including charge-conjugation C, parity P and time-reversal T).  These are embedded within the mathematics of quantum field theory.  The SM predicts that physical reality is not invariant under C, P and T transformations separately, but is invariant under simultaneous CPT transformations.

The SM is the great triumph, and also the great frustration, of modern physics 
\cite{Gabrielse2013}.  It is able to make amazingly precise predictions that are confirmed by measurements of unprecedented precision -- most notably for the lepton moments that are the focus here.  At the same time, the great frustration of the SM is that it is not able to account for how basic features of the universe could emerge in a big-bang cosmology.  The predictions and measurements of the size of the tiny magnet that the electron and muon particles possess have been central in demonstrating the triumph of the SM.  They are also essential in the modern quest to search for new physics that may be missing from the SM, often called beyond the standard model (BSM) physics.

\subsection{The SM leptons and anti-leptons}

The lepton and anti-lepton particles of the SM experience the electromagnetic and the weak interactions, but not the strong interaction. The three charged leptons of the SM are the electron ($e^-$), muon ($\mu^-$) and tau (or tauon, $\tau^-$).  Each has an antimatter counterpart, sometimes called an anti-lepton. The anti-electron is called the positron ($e^+$).  The anti-muon is generally called the positive muon $\mu^+$, and the anti-tau is generally called the positive tau ($\tau^+$).  All SM electrons are indistinguishable from each other.  The same is true for all of the leptons and anti-leptons.  

A collision between an electron and a positron annihilates both particles, turning their mass energy into two or more photons, and other particles and anti-particles if the available energy is sufficient. More generally, each lepton is assigned a lepton number of $+1$ and each anti-lepton is assigned a lepton number of $-1$.  The total lepton number is the sum of the lepton numbers for all of the leptons and anti-leptons present in a collection of particles.  For any interaction in the SM, the lepton number before and after the particles interact is conserved (i.e.\ remains the same). 

The three charged leptons have charge $-e$, where $e$ is the quantum of electric charge.  The anti-leptons have the opposite sign charge, $e$.  The leptons and anti-leptons are all spin one-half fermions, with quantized angular moment of $\pm \hbar/2$ along any measurement axis. These charged leptons differ greatly in their masses.  The mass of the electron and the positron, $m_e$, is the smallest of the lepton masses.  The $\mu^-$ and $\mu^+$ have a much larger mass, $m_\mu \approx 207 \, m_e$.  The $\tau^-$ and $\tau^+$ mass is even larger, $m_\tau \approx 3500 \, m_e$.

The lepton mass hierarchy leads to very different lifetimes for lepton decay via the weak interaction.  A lepton that decays must produce a lepton to conserve lepton number, and the new lepton must have less mass than the first to conserve energy. Electrons and positrons, with no less massive lepton to decay to,  are thus stable. The $\mu^\pm$ can decay to the less massive $e^\pm$, and they live only $2.2 \, \mu \rm{s}$ in their rest frames on average.   The $\tau^\pm$ live only $3 \times 10^{-13}$s in their rest frames.

\newcommand{\muvec}{\boldsymbol{\mu}}
\newcommand{\Svec}{\mathbf{S}}

\subsection{Lepton and anti-lepton magnetic moments\label{sec:LeptonMoments}}

The leptons and anti-leptons in free space each act as a tiny magnet even though they have no known internal structure. The Dirac equation that is part of the SM predicts the basic size of this magnet.  The SM also predicts a part per thousand correction that occurs because the lepton interacts with ``empty space'', the latter actually being filled with particles and antiparticles that the uncertainty principle permits to exist for extremely short times.  A particle with a  charge $q$ and mass $m$, in a circular orbit with angular momentum $\hbar$, has a magnetic moment $q \,\hbar/2 m$.  A natural scaling for a magnetic moment is thus
\begin{equation}
\mu_B \equiv \frac{e \, \hbar}{2m},
\end{equation}
where $e$ is the fundamental unit of charge.  For an electron or positron, $\mu_B$ with $m=m_e$ is called the Bohr magneton.  This overview uses a generalization whereby  $\mu_B=e\,\hbar/2m_\mu$ is the natural scale for muons, etc.  The electron magnetic moment is thus naturally the largest of the lepton moments, with the muon and tau moments smaller by a factor of $m_\mu/m_e \approx 207$  and of $m_\tau/m_e \approx 3500$.  

Because these leptons (with charge $\pm e$, mass $m$ and spin $\hbar/2$) have no internal structure, each of their magnetic moment vectors $\muvec^\pm$ is proportional to their spin vector, $\Svec$ so that  
\begin{equation}
\muvec^\pm = 
\pm \frac{g}{2} 
\, \mu_B \, \frac{\Svec}{\hbar/2}.
\label{eq:LeptonMagneticMoment}
\end{equation}
Thus $g/2$ is the dimensionless magnitude of a particle's magnetic moment scaled to $\mu_B$, 
\begin{equation}
\frac{g}{2} =  \frac{|\muvec|}{\mu_B} = 1 + a,
\end{equation}
with $g$ often called the ``g-value''.  The once unexpected ``anomaly''  $a \approx 10^{-3}$ is a very small correction for the leptons because of their lack of internal structure.  (Protons and neutrons with quark ingredients have large g-values.)  The most precise lepton measurements measure $a$ to determine $g/2$ more accurately, and are sometime referred to as $g-2$ measurements because $a=(g-2)/2$.

All sectors of the SM (electromagnetic, strong and weak) provide contributions to the SM prediction of
the dimensionless magnetic moments $g/2=1+a$,    
\begin{equation}
\frac{g}{2}= 1 + a^{\rm{QED}} + a^{\rm{hadronic}} + a^{\rm{weak}}. 
\label{eq:SMTheory}
\end{equation}
Physics that is beyond the SM could conceivably add an  $a^{\rm{new}}$ to this sum.  The Dirac prediction \cite{DiracTheoryOriginal} is the first term on the right. The calculation of the smaller contributions to the SM prediction by many theorists is an ongoing theory tale, appropriately told in its own chapter.  As this work is continuing, some surprises may yet emerge.

  Quantum field theory, called quantum electrodynamics (QED), provides $a^{\rm{QED}} \approx 0.001$ as a series in powers of the fine structure constant $\alpha \equiv e^2/4\pi\epsilon_0\,\hbar c$, 
\begin{equation}
a^{\rm{QED}} = C_2\left(\frac{\alpha}{\pi}\right) +C_4\left(\frac{\alpha}{\pi}\right)^2
+C_6\left(\frac{\alpha}{\pi}\right)^3
+C_8\left(\frac{\alpha}{\pi}\right)^4
+C_{10}\left(\frac{\alpha}{\pi}\right)^5 + ...\, .
\label{eqQED}
\end{equation}
The constants $C_2=1/2$ \cite{Schwinger},   $C_4$ \cite{Petermann,Sommerfield:1958},  
$C_6$ \cite{Kinoshita:1995,LaportaRemiddi:1996} and $C_8$ \cite{QED_C8_Lapo} are calculated exactly using QED field theory, but require measured lepton mass ratios as input \cite{KURZ2014_HeavyLeptons}.  The exact calculation of the leading, mass-independent contributions to the 6th order $C_6$  and the 8th order $C_8$ (which includes 891 Feynman diagrams) are remarkable recent accomplishments, as is the numerical evaluation of the 10th order $C_{10}$ (which includes 12,672 Feynman diagrams) \cite{TenthOrderQED2012,Aoyama:2012wk,Laporta:1994md,Baikov:2013ula,QED_C10_nio,Volkov:2019phy,Volkov:2024yzc,Aoyama:2024aly}.  
(A Feynman integral is a particular kind of calculus integral of multidimensional rational functions that may have as many as 20 dimensions).

Hadronic contributions and weak interaction contributions are all smaller. It is customary to write the  hadronic contribution  as the sum of 4 terms~\cite{Jegerlehner:2009ry,Jegerlehner:2017gek,Aoyama:2020ynm,Aliberti:2025beg},
\begin{equation}
 a^{\rm{hadronic}} = a^{\rm{HVP-LO}} + a^{\rm{HVP-NLO}} + a^{\rm{HVP-NNLO}} + a^{\rm{HLBL}},   
\end{equation}
(e.g.\ in Figure \ref{amu22}). These contributions are  currently the biggest calculational challenge to the SM prediction of the lepton magnetic moments \cite{Aliberti:2025beg}. ``HVP'' stand for ``hadronic vacuum polarization'' and ``HLBL'' refers to what are called ``hadronic light-by-light'' Feynman diagrams.  ``LO'' is ``leading order'', ``NLO'' is ``next-to-leading order'', and ``NNLO'' means ``next-to-next-leading order''. Higher order terms are neglected at the current levels of precision.
The hadronic contribution is a much bigger challenge for the muon than for the electron because it is larger for the muon than the electron by about a factor of the square of the masses of these particles, $(m_\mu/m_e)^2 \approx 4 \times 10^4$.

The weak interaction is often written as the sum of two terms \cite{WeakCondtribution1,WeakCondtribution2,WeakCondtribution3,WeakCondtribution4,Jegerlehner:2009ry,Gnendiger:2013pva,Aoyama:2020ynm,Aliberti:2025beg},
\begin{equation}
  a^{\rm{weak}}  = a^{\rm{EW1}} + a^{\rm{EW2}},
\end{equation} 
  where ``EW1'' refers to the leading order one-loop Feynman diagrams, and ``EW2'' refers to higher order corrections.
Any $a^{\rm{new}}$ from beyond-the-standard model physics is also expected to be larger in the muon system by the same factor of $(m_\mu/m_e)^2 \approx 4 \times 10^4$.

The QED field theory calculations are now calculated and checked at the level of precision of current and proposed electron and muon magnetic moment measurements.  The mass independent parts of $C_6$ and $C_8$ are both exactly calculated and checked with numerical calculations.  QED field theory is tested through the 8th order by both electron and muon measurements since both $C_6$ and $C_8$ are needed to make a SM prediction that is as precise as the electron and the muon measurements. QED field theory is tested through 10th order by the electron measurement insofar as $C_{10}$ is crucial for attaining a precision in the electron prediction that equals that of the electron measurement.  The calculation precision, in fact, would suffice for the ten times improved precision that is the goal of a new Northwestern electron measurement.   

The SM prediction of the dimensionless magnetic moment, $g/2$, is a function of $\alpha$.  The biggest challenge to making a SM prediction of the dimensionless electron magnetic moment at the electron measurement precision is not theoretical.  It is to obtain an accurate measured value of the fine structure constant $\alpha$ that must be  put into Equation \ref{eq:SMTheory} to make the SM prediction. Two determinations of $\alpha$ report measurements with the needed precision \cite{MullerAlpha2018,RbAlpha2020}, but unfortunately these disagree by more than 5 standard deviations.  To test the SM prediction at the current precision of the electron measurement requires that this discrepancy be resolved.  If a new electron measurement underway reduces the uncertainty by a factor of 10, fully testing the SM prediction requires that the uncertainty in $\alpha$ be reduced by this factor as well.  The measured value of $\alpha$  is not a problem for the SM prediction of the muon magnetic moment because the muon measurement precision is a thousand times lower than for the electron.

The uncertainty in the hadronic contribution $a^{\rm{hadronic}}$ for the electron is below the current measurement uncertainty for $\mu_e/\mu_B$ \cite{Aliberti:2025beg}.  It now seems that a SM prediction sufficient for a new electron measurement with a ten times higher precision will require a reduction in the  uncertainty in this contribution.   The realization that a  new electron measurement is underway will hopefully stimulate theoretical progress, as it has in the past.  For the muon, the calculation of 
$a^{\rm{hadronic}}$, particularly  $a^{\rm{HVP-LO}}$, is and has been the biggest challenge to the SM prediction of the muon moment. Different evaluations of $a_\mu^{\rm{HVP-LO}}$ (the ``subscript'' $\mu$ indicates the muon) based on electron-positron data at low-energy colliders, and computer-based simulations of strong interactions using lattice Quantum Chromodynamics (QCD)  currently disagree despite an enormous effort going into its evaluation, for reasons not yet understood. An exciting new development is that QCD lattice gauge calculations are approaching the precision needed to evaluate this term, and may be resolving a longstanding discrepancy between measured muon magnetic moment and the SM prediction~\cite{Jegerlehner:2017gek,Aoyama:2020ynm}.  The latest calculations are summarized in \cite{Aliberti:2025beg} and in the muon section that follows.  

Finally, the intrinsic CPT symmetry invariance of the SM can be stringently tested by comparing lepton and anti-lepton magnetic moments.  A consequence of CPT invariance is that the magnetic moments of a particle and antiparticle are identical in magnitude but opposite in sign, which means that their $g/2$ would be the same.  The unprecedented high precision with which the electron and positron magnetic moments can be measured makes comparing these magnetic moments to be the most precise test of lepton CPT invariance by far.

\subsection{Different methods, precisions and purposes}

It is often said that the muon is essentially a heavy electron, identical to the electron except for a larger mass. Practical considerations, however, require that electrons and muons be studied very differently (Figure~\ref{fig:titlepage}).

\begin{itemize}

\item Electrons are readily available in every laboratory, while muons must be created in high energy collisions at a large particle physics facility.

\item A single electron is suspended nearly  at rest within a particle trap for months at a time, within a volume less than 1 mm$^3$, is cooled to the quantum ground state of its cyclotron motion.  Many quantum methods can then be used, including quantum jump spectroscopy to a first excited state that is only 0.6 meV higher in energy. 
Muons at 3.1 GeV, an energy $5 \times 10^{12}$ greater,  travel at essentially the speed of light in a 14 m diameter storage ring until they decay in only 64 $\mu$s on average in the lab frame.

\item An electron with the energy of its first excited cyclotron state is highly nonrelativistic in that its  relativistic factor $\gamma$ differs from 1 by only a part in $10^9$.  Muons at 3.1 GeV are highly relativistic with $\gamma \approx 29$.  Special relativity is actually critical for both electron and muon measurements but in completely different ways.

\end{itemize}

\noindent
Not surprisingly, 
the electron magnetic moment is measured much more precisely than that of the muon, currently by a factor of a thousand, because measurements can take as much time as needed, the particle environment can be much better controlled over a distance scale  that is smaller by more than a factor of
$10^4$, and because quantum methods can be used.

The measured electron and muon magnetic moments both provide important tests of the SM, despite the large precision difference, because they test the SM differently.  The measured electron moment, because of its extraordinary $1$ part in  $10^{13}$ precision,  tests the field theory structure of the SM well into the 10th order in the electromagnetic coupling constant.  It also tests the new analytic and numerical evaluation methods for 13,000 Feynman diagrams that were required to reach the measurement precision.  The remarkable agreement between measurement and prediction is important for establishing the validity of the SM, much as was the discovery of the Higgs particle.  Quantum electrodynamics is now established as the prototype of a successful quantum field theory as a result in a way that its inventors never imagined would be possible, in the words of Freeman Dyson, one of the inventors of this field theory.  The accurate prediction of the measured electron magnetic moment is arguably the great triumph of the Standard Model.

The muon magnetic moment plays a different important role.  It much more sensitively tests the SM prediction for the hadronic contribution,  $a^{\rm{hadronic}}$, and for the weak contribution, $a^{\rm{weak}}$. It is also expected to be more sensitive than the electron moment to new physics in the form of particles omitted from the SM, for example, by a factor of approximately $(m_\mu/m_e)^2 \approx 40,000$. This leaves the muon measurements 40 times more sensitive to such new physics that is the electron moment, even when the precision lower by a factor of 1000 is taken into account. Great interest was generated by the possibility that a 3 standard deviation discrepancy that persisted for many years between the SM prediction and the measured magnetic moment of the muon might be caused by new physics. This hope is fading a bit as QCD lattice gauge calculations (stimulated by the discrepancy of theory and experiment) are approaching a useful precision, and seem to be shifting $a^{\rm{had}}$ to make the predicted and measured magnetic moments agree much better. The muon magnetic moment may yet turn into a great triumph of the Standard Model, rather than an indication of its breakdown.

It is worth noting that the measured electron magnetic moment is used to help determine the measured value of the muon magnetic moment in several ways.  A muon search for new physics requires that the Dirac and QED contributions to the SM prediction be accurately calculated, and the electron measurement tests this.  The agreement between the predicted and measured electron magnetic moment is also an important consistency check on the measured fine structure constant that must be used to make a SM prediction of the muon magnetic moment. Equation \ref{eq:MuonaDetermineation} will also show that the measured 
electron $g_e/2$ is one of the direct inputs used to determine the muon magnetic moment.  

The electron community generally quotes $g/2$ while the muon community prefers $a$.  A given fractional accuracy on $a$ corresponds to a fractional accuracy on $g/2$ that is smaller by about a factor of 1000.

%% file: Electron.tex
\newcommand{\nuap}{\nu_a^\prime}
\newcommand{\nucp}{\nu_c^\prime}
\newcommand{\nusbar}{\bar{\nu}_s}
\newcommand{\nuab}{\bar{\nu}_a}
\newcommand{\nucb}{\bar{\nu}_c}
\newcommand{\nuzb}{\bar{\nu}_z}
\newcommand{\numb}{\bar{\nu}_m}
\newcommand{\fcb}{\bar{f}_c}

\newcommand{\wc}{\omega_c}
\newcommand{\ws}{\omega_s}
\newcommand{\wa}{\omega_a}
\newcommand{\wap}{\omega_a^\prime}
\newcommand{\wcp}{\Omega_c^\prime}
\newcommand{\wsb}{\bar{\omega}_s}
\newcommand{\wab}{\bar{\omega}_a}
\newcommand{\wcb}{\bar{\omega}_c}
\newcommand{\wzb}{\bar{\omega}_z}
\newcommand{\wmb}{\bar{\omega}_m}

\newcommand{\zhat}{\bf \hat{z}}

\section{Electron and positron magnetic moments}\label{sec:Electron}

The one-century history of the electron magnetic moment began in 1922 with the deflection of silver atoms in a magnetic field gradient observed by Stern and Gerlach \cite{SternGerlach}.  The deflection provided the first indication that an electron can behave as a tiny magnet.
Section \ref{sec:EarlyYears} summarizes the many measurements of the electron magnetic moment made between 1947 and 1987.  The most precise of these made use of classical electron cyclotron orbits  to produce a value for the electron magnetic moment in Bohr magnetons that stood for nearly two decades \cite{DehmeltMagneticMoment}. 
Section \ref{sec:QuantumMeasurements} summarizes the era of quantum measurements of 
the electron magnetic moment that began in 1987.  
A century after Stern-Gerlach, Gabrielse and his students reported in 2023 the use of a quantum cyclotron \cite{QuantumCyclotron} to measure the  
electron magnetic moment to a precision of $1$ part in $10^{13}$ \cite{NorthwesternMagneticMoment2023}.  This is the most precisely  measured  property of any elementary particle and also the most precise prediction of the Standard Model of particle physics, setting up the most precise confrontation of theory and experiment ever made.  In the future, new quantum methods now being pursued have the goal to improve the measurement precision by an additional factor of ten. 

\subsection{Early years (1947 - 1987) \label{sec:EarlyYears}}

As befits the property of an elementary particle that is the most precisely measured and the most precisely predicted, the electron magnetic moment has been measured many times. In the 40 years between 1947 and 1987, there were 24 measurements of increasing precision \cite{HistoryOfElectrong} using three different experimental approaches.  

{\bf Atomic spectroscopy:} Between 1947 - 1954, the size of the electron's magnetic moment was deduced from the splitting of atomic spectral lines in a magnetic field. The smallest fractional uncertainty achieved was $4 \times 10^{-5}$. 
What was actually being measured, of course, was the electron magnetic moment as it was modified by its interaction with the surrounding bound electrons.  

{\bf Free precession:}  Between 1954 and 1972, Crane and then Rich at the University of Michigan led a series of measurements using free electrons.  Spin-polarized electrons made circular cyclotron orbits as they traveled along helical orbits in the magnetic field of a solenoid. The electron spins precessed as they made circular cyclotron orbits.  The electron magnetic moment was deduced  by comparing the measured spin polarization after a measured number of cyclotron orbits.  The fractional uncertainty eventually achieved was $3.5 \times 10^{-9}$.  Compared to the atomic spectroscopy measurements, this was a precision that was better by a factor of $10^4$.

{\bf Trapped electrons:} In parallel with the free precession measurements,  electrons suspended in Penning traps were explored between 1958 and 1972 as a possible alternative.  Graff at the  University of Mainz, and Dehmelt at the University of Washington (UW), all developed methods to measure the electron magnetic moment using a large number of electrons stored in a Penning trap
\cite{GraffMeasuringCyclotronAndSpinResonancesForElectrons1968,DehmeltWalls1968,GraffAnomalyResonanceForElectrons1969,WallsSteinAnomalyForElectrons1973}. The electron magnetic moment measurements, with a fractional precision as good as $2.4 \times 10^{-6}$, did not achieve nearly the precision of the free precession measurements.

{\bf One-trapped electron:} The demonstration of the direct detection of a single trapped electron at the University of Washington in 1973 was a critical step forward \cite{FirstSingleElectron1973}.  
The magnetic field was along the symmetry axis of a Penning trap with trap electrodes shaped along the hyperbolic contours of an electrostatic quadrupole potential \cite{OrthogonalCompensate,Gabrielse84h}.  A classical cyclotron orbit was driven as in the  precession measurements, but this time from thermal  equilibrium near 4.2 K.  With enough averaging time, the stable electron spin state could be determined.  The last classical cyclotron measurement in 1987 determined the electron magnetic moment to a fractional precision of $4.3 \times 10^{-12}$.  This improvement on the free precession measurements by a factor of about 800 stood for nearly two decades.  
Later (e.g.\ in Figure \ref{fig:QuantumMeasurements}) we will see that 
these classical cyclotron measurements disagreed with the much more precise quantum cylotron measurements to be discussed below by about 1.7 standard deviations.

\subsection{Modern era: one-particle quantum cyclotron:  1987 - 2025\label{sec:QuantumMeasurements}}

The era of quantum-mechanical measurements of the electron and positron magnetic moments began in 1987 when the Gabrielse research group at Harvard University set out to realize a one-electron quantum cyclotron.  This is one electron suspended in a magnetic field whose quantum cyclotron states can be unambiguously resolved (see Section~\ref{sec:QuantumCyclotron}).  The quantum cyclotron, which they first realized in 1999 \cite{QuantumCyclotron}, made it possible to  use quantum jump spectroscopy, quantum non-demolition detection, and other quantum methods to attain a much higher precision than was possible using classical cyclotron motion.  

The electron magnetic moment, the most precisely measured property of an elementary particle, is measured within a tiny quantum cyclotron and not using a large particle physics facility.  Sections \ref{sec:QuantumCyclotron}-\ref{sec:QuantumJumpSpectroscopy} discuss the quantum methods and the extreme environments.  The most precise confrontation of any theory and experiment
is the comparison of this measurement with the most precise prediction of the SM (Section \ref{sec:Confrontation}).  The  most sensitive test of lepton CPT invariance takes place when an electron and positron are compared within the same tiny volume in a university laboratory.

\subsubsection{One-particle quantum cyclotron}\label{sec:QuantumCyclotron}

\begin{figure}[htbp!]
\centering
\includegraphics[width=0.25\columnwidth]{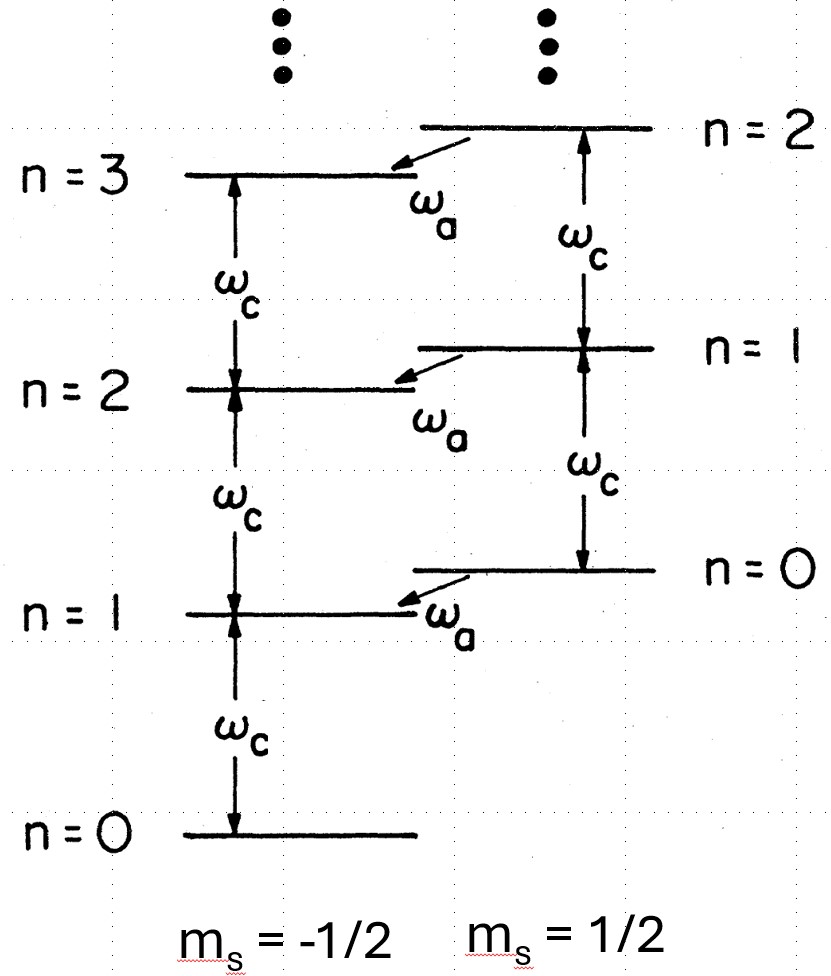}
\caption{Lowest cyclotron and spin energy levels of the quantum cyclotron.  }
\label{fig:EnergyLevelsFreeSpace}
\end{figure}

A one-electron quantum cyclotron \cite{QuantumCyclotron} is a single electron suspended in a magnetic field $\mathbf{B}=B\hat{z}$ whose individual cyclotron and spin energy levels, 
\begin{equation}
E = h \wc (n + \tfrac{1}{2}) + h \ws m_s,
\end{equation}
can be clearly resolved.  As illustrated in Figure \ref{fig:EnergyLevelsFreeSpace}, these are the superimposed energy levels of a harmonic oscillator and a two-state, spin 1/2 system.  
The first term gives the cyclotron energy levels for a cyclotron frequency, $\wc=eB/m$.  The cyclotron quantum number is  $n=0,1,...$.  The second term is the spin energy $-\boldsymbol{\mu} \cdot \boldsymbol{B}$ for an electron spin magnetic moment (Equation \ref{eq:LeptonMagneticMoment}), 
\begin{equation}
\boldsymbol{\mu}=- \frac{g}{2} \, \mu_B \, \frac{\mathbf{S}}{\hbar/2}. 
\end{equation}
that is proportional to its spin $\mathbf{S}$ and normalized to its spin eigenvalue, $\hbar/2$.  The resulting spin precession frequency is $\ws = (g/2)\wc$ and the spin quantum number is $m_s=\pm 1/2$.   

For the quantum cyclotron, the cyclotron frequency $\wc$ and the spin frequency $\ws$ are the transition frequencies between the cyclotron energy levels, and the spin energy levels, and the anomaly frequency is the difference $\wa \equiv \ws-\wc$.  In terms of ratios of these frequencies, the magnetic moment of the negative electron is
\begin{equation}
-\frac{\mu}{\mu_B} = \frac{g}{2} = \frac{\ws}{\wc} =
1 + \frac{\wa}{\wc}, 
\label{eq:FreeSpacegOver2}
\end{equation}
and the magnetic moment ``anomaly'' is $a = \omega_a/\omega_c.$
The negative sign is not present for a positron so $g$ is positive for both the electron and positron.   Measuring the ratio $\omega_a/\omega_c$ rather than $\ws/\wc$, often called a $g-2$ measurement, gives an uncertainty in the measured $g/2$ which is smaller by a factor of 1000.   

One important advantage for an electron measurement compared to a muon measurement is that the electron cyclotron frequency can be measured directly using the same electron, at the same location in the apparatus, at almost the same time. The electron serves as its own magnetometer insofar as $B$ cancels out of the frequency ratio, $\omega_a/\omega_c$, and no other measurements are required to determine $a$ and $g$.  For muon measurements, 
the frequency ratio  $\omega_a/\omega_p$ is measured instead, where $\omega_p$ is the frequency of protons in 
NMR probes averaged over the muon orbits in a large storage ring.  Other measurements are then used to convert $\omega_p$ to $\omega_c$ to deduce $a$.

\subsubsection{Extremely cold temperature eliminates blackbody photons}

\begin{figure}[htbp!]
\centering
\includegraphics[width=0.85\columnwidth]{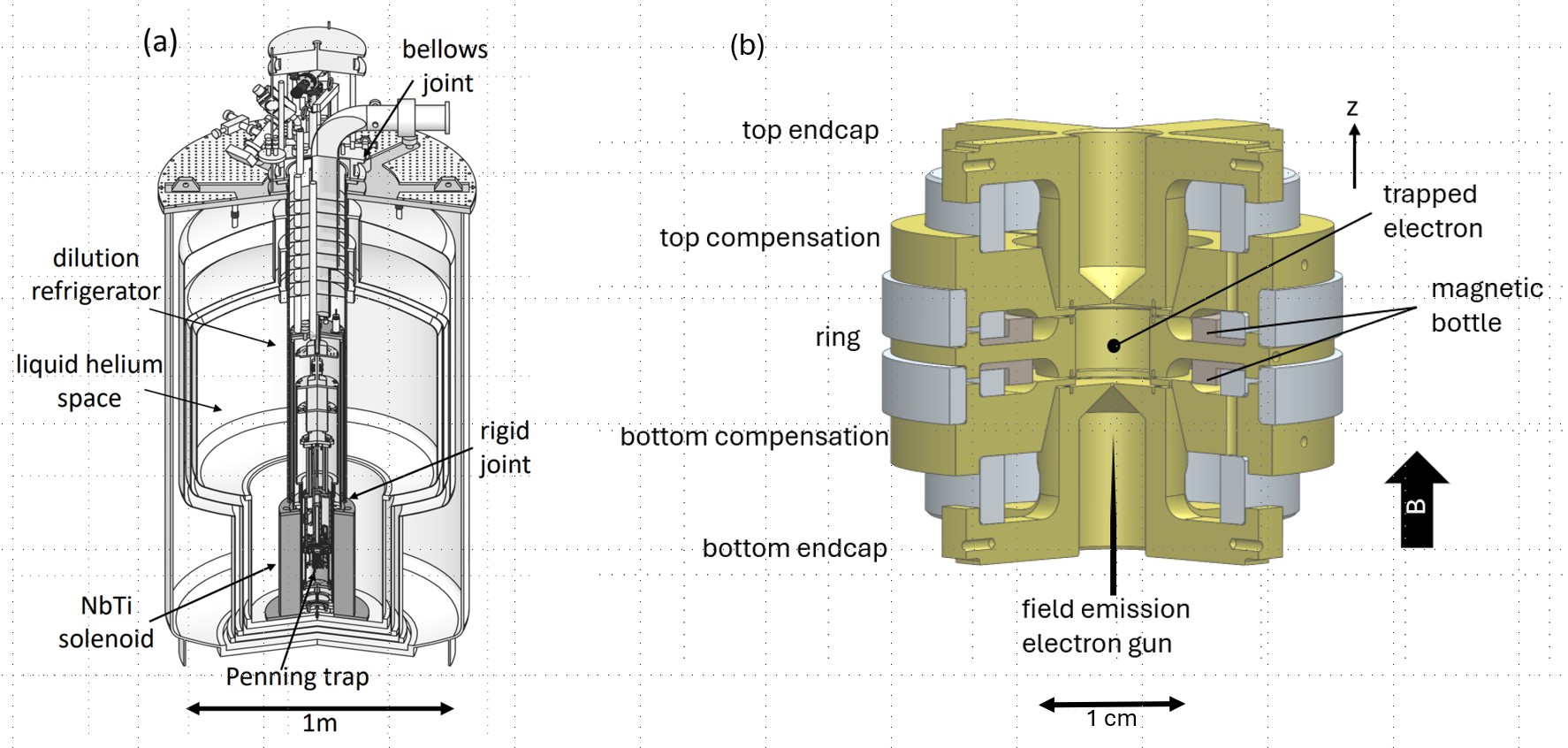}
\caption{(a) Cryogenic system supports a 50 mK electron trap upon a 4.2 K solenoid to provide a very stable magnetic field.
  (b) Silver electrodes of a cylindrical Penning trap cavity.  From \cite{NorthwesternMagneticMoment2023}. }
\label{fig:FridgeTrap}
\end{figure}

Before the quantum era, electron magnetic moment measurements were carried out at temperatures of 4.2 K and above. Realizing a quantum cyclotron required temperatures below 0.1 K. The low temperature is needed to prepare the electron in the quantum ground state of its cyclotron motion, and eliminate the blackbody photons that could excite the electron out of its ground state. Figure \ref{fig:FridgeTrap}a represents the dewar and dilution refrigerator used to provide the low temperature for the most precise electron magnetic moment measurement \cite{NorthwesternMagneticMoment2023}.  Deep inside the dewar is the cylindrical Penning trap cavity represented in Figure \ref{fig:FridgeTrap}b.  A single electron (or positron) can be suspended at the center of this trap cavity for as long as desired.       

Thermal equilibrium is used to prepare the quantum cyclotron in its quantum ground state.  For cyclotron motion in equilibrium with its trap cavity environment at temperature $T_c$, the probability to be in the first cyclotron excited state over the probability to be in the cyclotron ground state is the ratio of Boltzmann factors, $exp(- \hbar \wc/kT_c)$.  A temperature $T_c \ll \hbar \wc /k = 7.1\,$K is thus required if thermal equilibrium is to prepare only the ground state of the quantum cyclotron rather than an incoherent superposition of many cyclotron states.  The $7.1$ K is for a typical magnetic field of $B=5.3$ T.

\begin{figure}[htbp!]
\centering
\includegraphics[width=0.5\columnwidth]{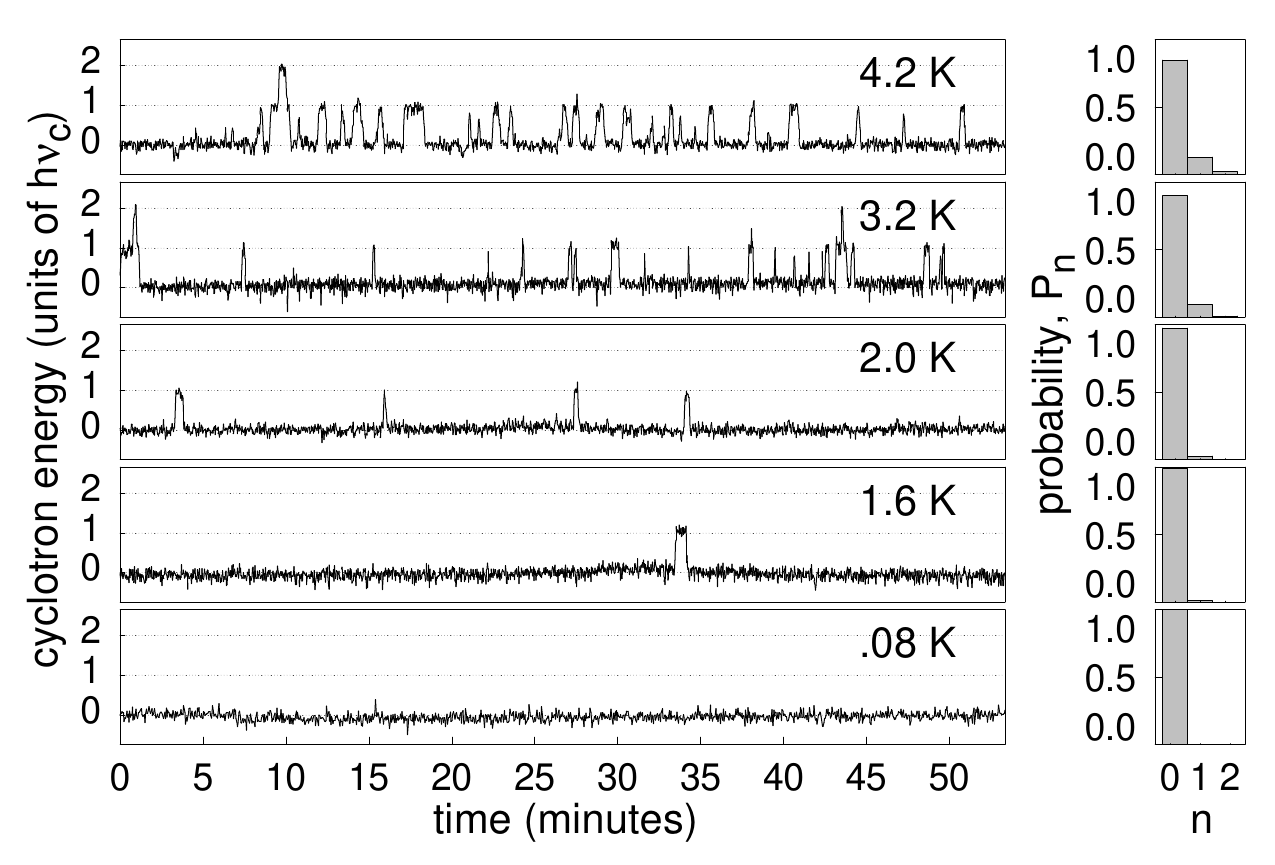}
\caption{The blackbody photons that stimulate cyclotron excitations for higher temperatures go away at lower trap cavity temperatures.  From \cite{QuantumCyclotron}.}
\label{fig:QuantumJumps}
\end{figure}

Figure \ref{fig:QuantumJumps} shows the measured cyclotron energy for about one hour of time for a one-electron quantum cyclotron enclosed within a trap cavity that is at five temperatures between 4.2 K and 80 mK \cite{QuantumCyclotron}.   Blackbody photons from the trap cavity excite quantum jumps upward  from the cyclotron ground state ($n=0$) to the first excited state ($n=1$), with several examples of $n=2$ excitations.  The downward quantum jumps are either the emission of a photon stimulated by a blackbody photon, or spontaneous emission of synchrotron radiation.  At 1.6 K, there are only enough blackbody photons to provide only one excitation during this hour.  The decay that follows  is very likely a spontaneous emission since so few blackbody photons are still present.  For temperatures between 80 mK and 8 mK, no quantum jumps at all are observed for many days.  The quantum cyclotron is prepared and remains in its quantum ground state until nearly resonant photons are sent into the trap from the outside.    

Thermal equilibrium between the trap cavity at temperature and the cyclotron motion temperature $T_c$ is clearly observed, but only when the state of the quantum cyclotron is observed over many hours.  The measured distribution of the time spent in each of the observed quantum states is plotted to the right in Figure \ref{fig:QuantumJumps}.  At the higher temperatures, a Boltzmann distribution is observed with a temperature that agrees with the measured temperature of the trap cavity.  At temperatures far below 7.1 K, ``thermal equilibrium'' consists of the quantum cyclotron occupying only its ground state with $n=0$ and ``cyclotron temperature'' ceases to be a useful description.  

At the lowest temperatures, with no excitations at all are observed over many days, the quantum cyclotron is ready for quantum jump spectroscopy.  Nearly resonant photons are sent into the trap cavity and the quantum jump rate is measured as a function of the frequency of these photons.  The quantum cyclotron has also been used as a background-free detector for ``dark photons'' dynamically mixed with ordinary photons of the same frequency to set a new dark photon limit \cite{DarkPhotonFromQuantumCyclotron}.

\subsubsection{Self-shielded superconducting 
solenoid}

The perfect cancellation of the magnetic field from the measured frequency ratio  $\wa/\wc$ that determines $\mu/\mu_B = -g/2$ for the electron takes place only if the frequencies $\wa$ and $\wc$ can both be measured before the magnetic field drifts.  Because of the unusually high measurement precision, magnetic field stability is still a significant challenge. 

The first of two novel technical solutions is a "self-shielding" solenoid that was invented to cancel fluctuations in the ambient laboratory field that penetrates to the electron by factors of 200 and more \cite{SelfShieldingSolenoid}.  For this solenoid design geometry (now implemented in most MRI imaging systems to allow their operation near elevators and other magnetic disturbances), a fluctuation in the ambient laboratory magnetic field induces current in the superconducting windings of the solenoid that cancels the field fluctuation at the location of the electron.  

The second solution is an unusual dilution refrigerator design \cite{Atoms2019TowardImprovedMeasurement} that allows the electron trap to be cooled to 10 to 50 mK in temperature while it is hanging below the dilution refrigerator.  This allows the trap that confines the electron to rest mechanically upon the much warmer 4.2 K solenoid form, to eliminate relative motion of the electron trap through residual gradients in the magnetic field produced by the solenoid.  A field drift of $2\times 10^{-9}$ per day was demonstrated \cite{Helium3NMR2019}.

\subsubsection{Inhibited spontaneous emission}

The inhibition of the spontaneous emission of synchrotron radiation from the cyclotron excited state is essential to realizing a quantum cyclotron.  To observe quantum jumps between the lowest quantum states requires that the electron lifetime in its first excited cyclotron state is shorter than the averaging time needed to measure the state of the quantum cyclotron.  For a 5.3 T magnetic field in free space, for example, the first excited cyclotron state decays to the ground state in only 0.1 seconds on average via the spontaneous emission of synchrotron radiation.

The cylindrical Penning trap (Fig.~\ref{fig:FridgeTrap}b), invented for this purpose, inhibits this spontaneous emission \cite{CylindricalPenningTrap,CylindricalPenningTrapDemonstrated} by observed factors of 200 to 500.  An electron at the center of the trap is surrounded by a microwave cavity that is cylindrical with flat ends.  The resonant electromagnetic modes within the cavity modify the density of states for synchrotron emission.  When the electron cyclotron frequency is between the frequencies of the nearest resonant modes of the cavity that couple to the cyclotron motion of a centered electron, the spontaneous emission is inhibited.  

\begin{figure}[htbp!]
\centering
\includegraphics[width=0.4\columnwidth]{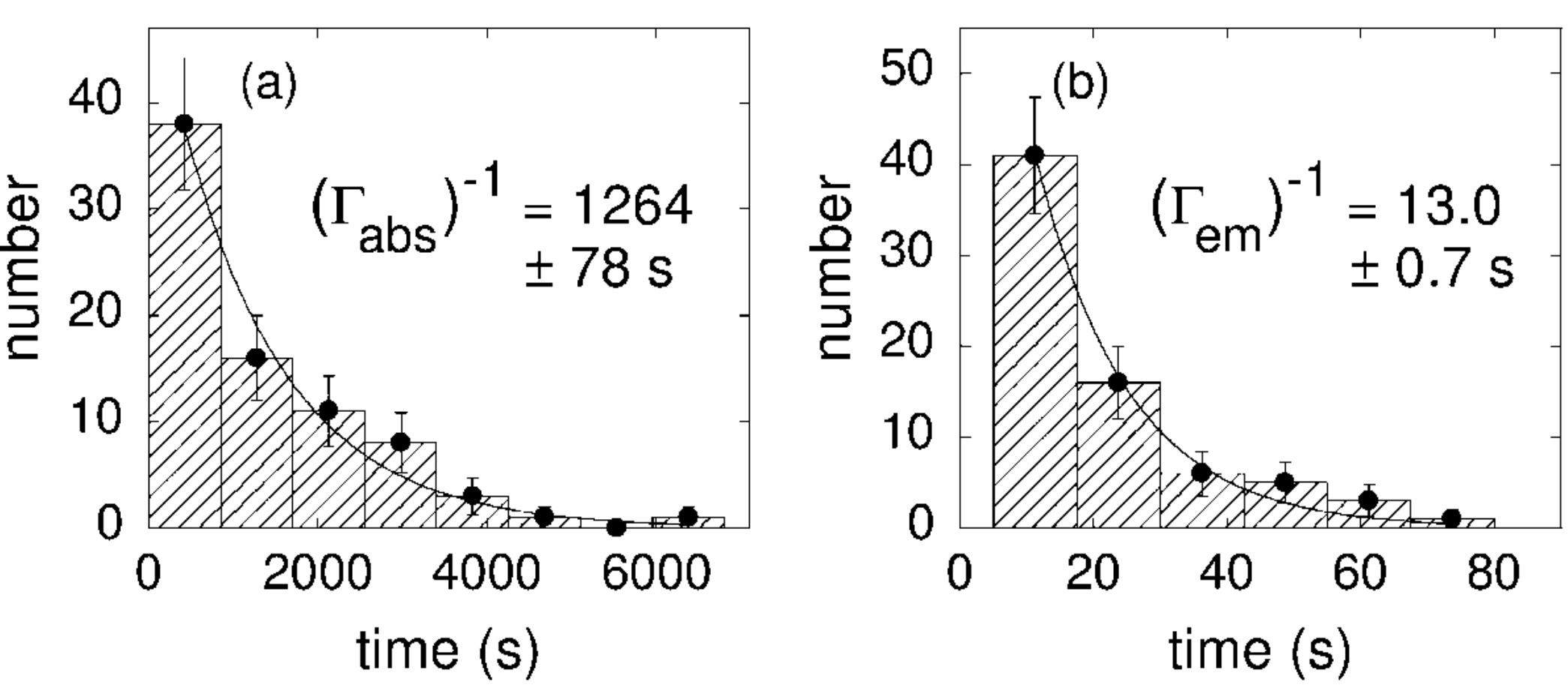}
\caption{(a) Histogram of the measured time it takes for a blackbody photon to be absorbed by the quantum cyclotron in its ground state.  (b) Histogram of the measured time it takes for a quantum cyclotron in its first excited state to decay to the ground state.   From \cite{QuantumCyclotron}.}
\label{fig:InhibitedEmission}
\end{figure}

\begin{figure}[htbp!]
\centering
\includegraphics[width=0.4\columnwidth]{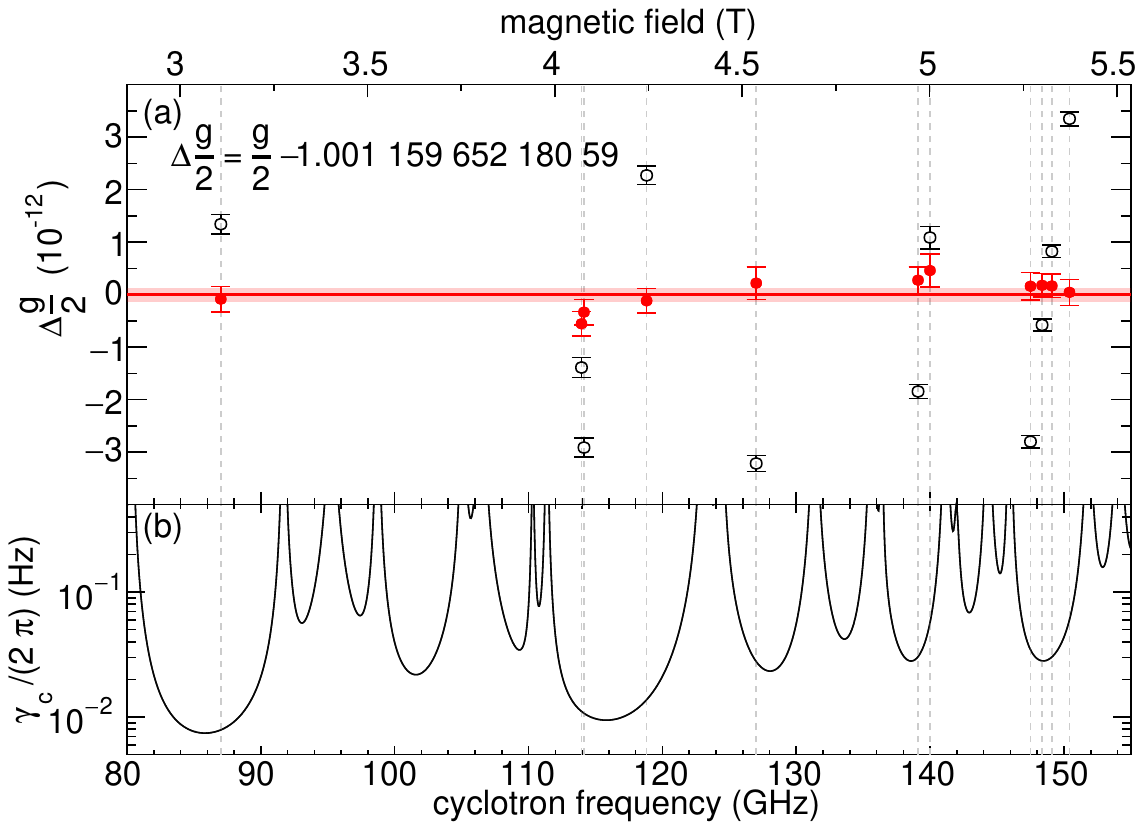}
\caption{(a) Shift in the measured electron g/2 before
(white) and after (red) cavity-shift correction. (b) Measurements
take place in valleys of the cyclotron damping rate $\gamma_c$ where
spontaneous emission is inhibited. From \cite{NorthwesternMagneticMoment2023}.}
\label{fig:CavityShifts}
\end{figure}

Figure \ref{fig:InhibitedEmission} is a direct measurement of the inhibited spontaneous emission \cite{QuantumCyclotron}.  It is based upon many hours of observed quantum jumps, one hour of which is shown in Figure \ref{fig:QuantumJumps} for a trap cavity at 1.6 K. Figure \ref{fig:InhibitedEmission}a is a histogram of the time it takes before an electron in the ground state of the quantum cyclotron is excited to its first excited state by absorbing a blackbody photon.  At 1.6 K there are so few blackbody photons in the cavity that it takes about 1300 s on average for an excitation to take place.  Figure \ref{fig:InhibitedEmission}b is a histogram of the time it takes the first excited state of the quantum cyclotron to decay to the ground state.  At 1.6 K there are very few blackbody photons available to stimulate emission. The measured  average of 13 s is thus essentially the lifetime for the spontaneous emission of synchrotron radiation from the excited state.  Without the cavity present, this state would radiate in about 0.1 s, so this example demonstrates the inhibition of spontaneous emission by a factor of 130.  This inhibition provides the averaging time needed to detect the quantum cyclotron state. Measurements of the electron magnetic moment are thus carried out in valleys of the cyclotron damping rate $\gamma_c$  where spontaneous emission is inhibited (Figure \ref{fig:CavityShifts}b).

One more advantage arises from the inhibition of the spontaneous emission of cyclotron radiation, though this has not yet been used to full advantage.  For a state with a mean lifetime $\tau$ and an energy $\hbar \omega$ above a stable ground state, transition frequency measurements will have a spread $\Delta \omega \sim 1/\tau$. A longer lifetime produces a narrower resonance linewidth.  For a frequency $\omega/2\pi = 140$ GHz, for example, a state that lives a free-space lifetime of 0.1 s has a fraction frequency linewidth $\Delta \omega / \omega \sim 1/\omega \tau \sim 7 \times 10^{-11}$.  For the 13 s lifetime demonstrated in Figure \ref{fig:InhibitedEmission}b, this fractional width decreases to $6 \times 10^{-13}$.  The measured linewidths to be illustrated (e.g.\ in Figure \ref{fig:Lineshapes})  are broader than this so far due to the magnetic field gradient introduced for QND detection (to be discussed in Section \ref{sec:QND}). However, this will change in a new generation of measurements now underway that eliminates this gradient.

\subsubsection{Cavity 
shifts\label{sec:CavityShifts}}

The inhibition of spontaneous emission of cyclotron radiation made possible the one-particle quantum cyclotron, and will allow narrower resonance lineshapes to be observed in the future.  However, this inhibition does not come for free.  The cyclotron oscillator is coupled to the oscillators that are the resonant modes of the trap cavity that cause the inhibition.  It is well known that coupled oscillators pull (i.e.\ shift) each other's resonant frequencies.  This unwanted cavity shift of a measured cyclotron frequency must thus be corrected, and the correction adds uncertainty to the determination of the electron magnetic moment.       

The cavity shifts are calculated for a perfect cylindrical cavity \cite{RenormalizedModesPRL,RenormalizedModesPRA} with cyclotron damping added \cite{HarvardMagneticMoment2008,HarvardMagneticMoment2011}.  A renormalized calculation was required to avoid the infinite result that occurs if the cavity shifts from all of the individual cavity modes are simply added together. A procedure was also devised to include the differing quality factors of the resonant modes of the trap cavity that coupled to an electron centered in the trap \cite{HarvardMagneticMoment2008,HarvardMagneticMoment2011}.  For the last measurement, the dimensionless magnetic moment $g/2$ was measured at 11 different values of the magnetic field, for cyclotron frequencies ranging from 85 to 150 GHz.  The cavity shifted values (open circles in Figure \ref{fig:CavityShifts}a) all line up after the cavity shift correction is applied  (red points in Figure \ref{fig:CavityShifts}a).

\subsubsection{Energy eigenstates and the Brown-Gabrielse invariance theorem}

A spatially uniform magnetic field $B\hat{z}$ is directed along the symmetry axis of the cylindrical cavity, perpendicular to the flat ends.  The conducting walls of the microwave cavity are slit to form electrodes that are visible in Figure~\ref{fig:FridgeTrap}b. When properly designed and biased \cite{CylindricalPenningTrap}, the resulting electric field and the magnetic field form a Penning trap that can suspend an electron  at the center of the Penning trap cavity to inhibit spontaneous emission.  The cylindrical Penning trap cavity was a substantial departure from traps with hyperbolic electrodes used before the quantum cyclotron \cite{DehmeltMagneticMoment}, which were not well-characterized microwave cavities.

The electron will oscillate along the magnetic field direction, $\hat{z}$, at the axial frequency $\wzb/2\pi\approx 114$ MHz in a recent example \cite{NorthwesternMagneticMoment2023}.  To make this a harmonic oscillation that can be cleanly observed requires that the relative geometry of the trap electrodes be carefully designed  to produce a high quality electrostatic quadrupole potential, $V \propto z^2 - x^2/2  - y^2/2$.  An  ``orthogonalized'' electrode geometry made it possible to tune the purity of the electrostatic quantum potential without changing its strength \cite{CylindricalPenningTrap}, and to  cleanly observe a single electron \cite{CylindricalPenningTrapDemonstrated}.   In experiments so far, the ``axial'' oscillation is a classical oscillation.  Measured shifts in this frequency reveal spin flips and cyclotron excitations of the electron, as will be discussed. 

The fields of the Penning trap produce a third motion of a trapped electron, in addition to its cyclotron and axial motions.  For $B=5.3$ T,  this magnetron motion is a circular magnetron motion at $\wmb/2\pi = 43$ kHz.  This motion is cooled by axial sideband cooling  \cite{VanDyckMagnetronCoolingLimit,Review} to minimize its effect during measurements. 

The  trapped electron is an artificial atom in which the electron is bound to the exterior ``nucleus'' that the Penning trap provides.  The binding is so weak that the measured  oscillation frequencies of the bound electron are only slightly shifted from the oscillation frequencies of a free electron quantum cyclotron described above.  The spin precession frequency, $\ws/2\pi \approx 149$ GHz  for $B=5.3$ T, is not shifted by the trap.   The cyclotron frequency is the same as the spin frequency to about 1 part per thousand.  The trap shifts the frequency from $\wc$ to a $\wcb$, by about 49 kHz for a 5.3 T field.  The trap-modified anomaly frequency is then $\wab \equiv \ws - \wcb \approx 173$ MHz for the same example field.  

The magnetic moment of an electron or positron is determined from measurements of the anomaly and cyclotron frequencies, $\wa=\ws - \wc$ and $\wc$ via Equation \ref{eq:FreeSpacegOver2}.  
The Penning trap shifts $\wc$ but not $\ws$.  In addition, the magnetic field is always slightly misaligned with the symmetry axis of the electrodes.  The trap shift, the shift from the mentioned imperfection, and others, would make it impossible to use the quantum cyclotron for precise magnetic moment measurements. The measured particle oscillation frequencies $\wcb$, $\wzb$ and $\wmb$, are all shifted by amounts that are difficult to measure and impossible to calculate.  The  
Brown-Gabrielse Invariance Theorem \cite{InvarianceTheorem} 
\begin{equation}
\wc = \sqrt{\wcb^2 + \wzb^2 +\wmb^2},   
\end{equation} fortunately provides a way to directly measure the angular cyclotron frequency $\wc$ needed to determine the magnetic moment in Bohr magnetons directly from the shifted frequencies that are measured.  
Another consequence of this theorem is that the   most precise mass spectroscopy is now carried out in a Penning trap \cite{InvarianceTheoremApplications2,TrueCyclotronFrequency}.

\subsubsection{Relativistic one-electron quantum cyclotron \label{sec:RelativitisticQuantumCyclotron}}

\begin{figure}[htbp!]
\centering
\includegraphics[width=0.25\columnwidth]{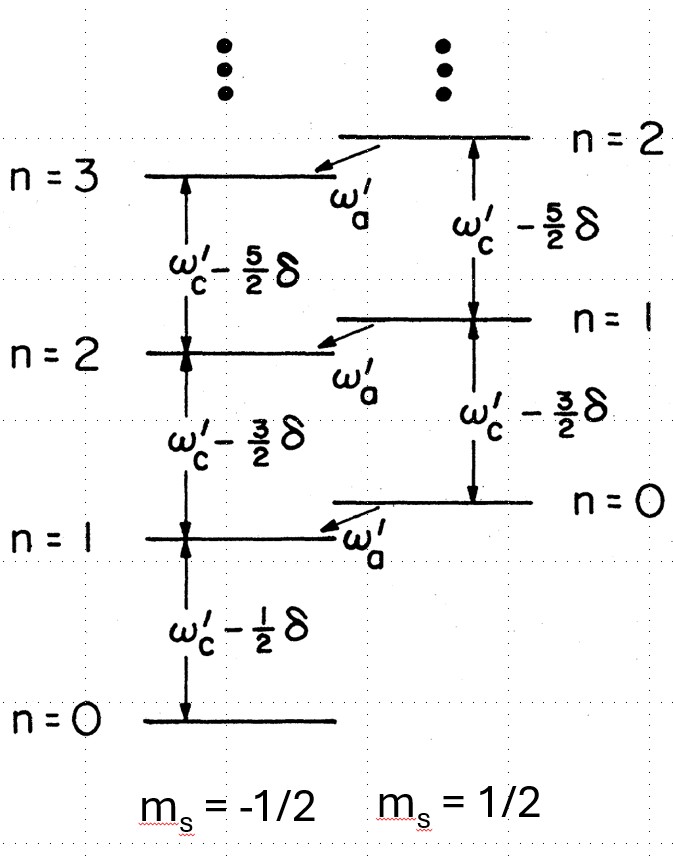}
\caption{Lowest cyclotron and spin energy levels of the relativistic quantum cyclotron.  }
\label{fig:EnergyLevels}
\end{figure}

Special relativity plays a crucial role for both electron and muon magnetic moment measurements, but in dramatically different ways.  The quantum cyclotron energy levels for the electron in a magnetic field are relativistically shifted by only about 1 part in $10^9$.  However, this tiny shift is what makes is possible to use only the lowest energy levels of the quantum cyclotron to make magnetic moment measurements. Equally spaced cyclotron energy levels cannot be individually addressed because a resonant drive will excite all cyclotron states.  
Special relativity turns the equally spaced levels 
into pairs of individually addressable qubits. A weak drive applied to the ground state will thus excite only the first excited state, for example.  
When classical cyclotron orbits were excited in the pre-quantum era, then the cyclotron velocity would need to be measured to determine the relativistic shift of the cyclotron frequency.  
When only the lowest quantum energy levels are used, no velocity needs to be measured to determine the relativistic shift. Instead, the shift is a fixed and measurable property of each energy level, as illustrated in Figure \ref{fig:EnergyLevels}.  

In a bit more detail, special relativity shifts the cyclotron and spin energy levels shown in Figure \ref{fig:EnergyLevelsFreeSpace} (with quantum numbers $m_s$ and $n$)  by \cite{Review,Gabrielse85e_RelativisticMassBistableHysteresis}
\begin{equation}
   \Delta E = -\frac{1}{2}\hbar \delta (n+1+m_s)^2, 
\end{equation} 
where   $ \delta/\omega_c = \hbar \omega_c/mc^2$, 
$mc^2$ is the rest energy of the electron, and typically $\delta/\omega_c  \approx 10^{-9}$.
Simply by measuring the cyclotron frequency, this shift $\delta$ is determined much more precisely than is needed, so the shift adds no uncertainty to a magnetic moment measurement.  Figure \ref{fig:EnergyLevels} represents the shifted energy levels and the transition frequencies between them. 
The cyclotron transition frequency between state $(n+1,m_s)$ and $(n,m_s)$  is relativistically shifted by $-\delta (n+1+m_s)$, as is the spin transition frequency between states  $(n,m_s=1/2)$ and $(n,m_s=-1/2 )$.  The lowest cyclotron state for $m_s=-1/2$ is stable.  The lowest energy cyclotron state for $m_s=1/2$ lives so long that it is effectively stable \cite{Review}.  For the measurements, both of these are effectively stable ground states. The tiny relativistic shifts of the lowest excited states are nonetheless much larger than the damping widths of these states.  The lowest cyclotron states of the quantum cyclotron are essentially resolved, and will be more so when the magnetic bottle gradient is removed for experiments currently underway.     

The anomaly frequency is a relativistic invariant that does not depend upon the cyclotron quantum number $n$.  This is because the relativistic shifts are the same for the two states involved in an anomaly transition, $(n, m_s=1/2)$ and $(n+1,m_s=-1/2)$.  The anomaly frequency is a relativistic invariant even at the extremely large classical cyclotron orbits used for the muon measurements, to be discussed, for which the average cyclotron quantum number is much too large to be useful.

\subsubsection{Quantum non-demolition (QND) detection\label{sec:QND}}

A quantum non-demolition (QND) measurement of a quantum state does not change the quantum state when the measurement is repeated.  The Hamiltonian has a term describing the system, another describing the detector, and a third term that couples them.  For a QND measurement, the coupling term commutes with the system Hamiltonian.

The QND detector that determines the energy of the cyclotron and spin states without modifying them is the monitored frequency $\wzb$ of the classical axial oscillation along the magnetic field direction.   The frequency shift,   $\Delta \wzb = \delta_z (n + m_x)$, is proportional to the cyclotron and spin quantum numbers.  The detected shift for a one-quantum cyclotron excitation has typically been about 1 Hz out of 100 MHz.  The QND coupling  is the result of a magnetic bottle gradient added by encircling the trap with a small nickel ring that saturates when $B\hat{z}$ is applied.  The energy on the vertical axis in Figure \ref{fig:QuantumJumps} is deduced from such an observed frequency shift.   

\subsubsection{Quantum jump spectroscopy\label{sec:QuantumJumpSpectroscopy}}

Quantum jump spectroscopy is counting the number of quantum jumps between two quantum states as a function of the frequency of the photons sent to the quantum system to excite it. The maximum number of quantum jumps roughly takes place when the photon frequency and the transition frequency are equal, and generally there is a lineshape model that describes how to extract the transition frequency from the quantum  jump lineshape.

\begin{figure}[htbp!]
\centering
\includegraphics[width=0.5\columnwidth]{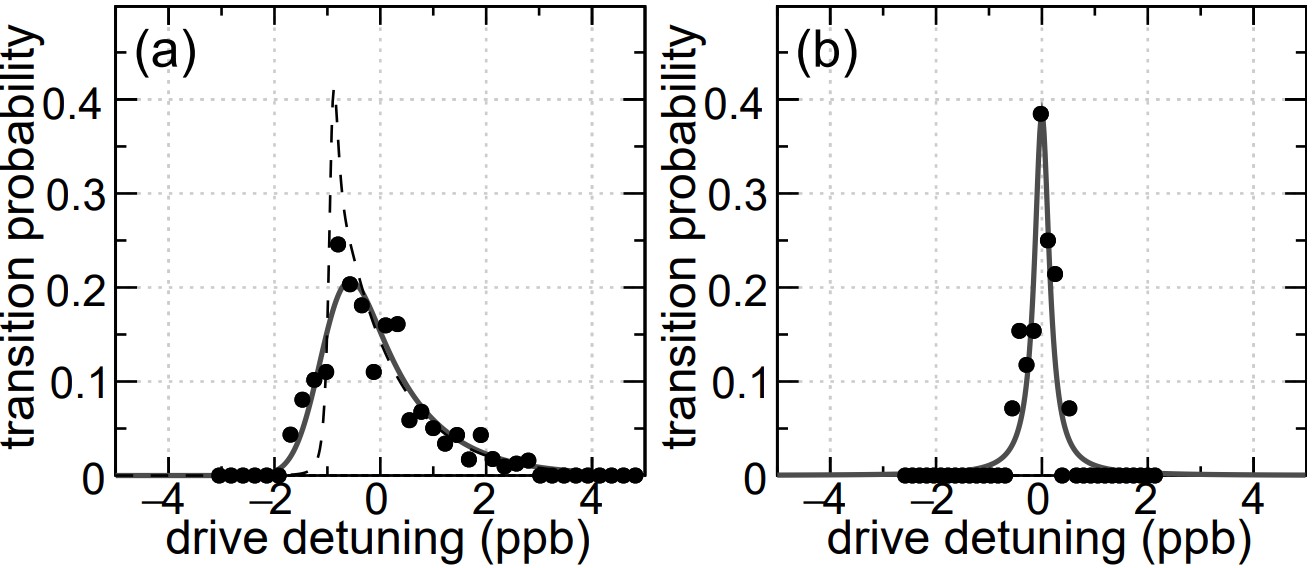}
\caption{Quantum jump spectroscopy lineshapes that determine the cyclotron frequency $\wcb$ (a) and the anomaly frequency $\wab$ (b).  The measured points are compared to what is predicted (dashed) and fit (solid) as a function of the fractional detuning of the drive frequency from resonance for the most precise measurement of the electron magnetic moment \cite{NorthwesternMagneticMoment2023}.}
\label{fig:Lineshapes}
\end{figure}

Figure \ref{fig:Lineshapes} shows the quantum jump lineshapes for the most precise measurement of the electron magnetic moment \cite{NorthwesternMagneticMoment2023}.  The observed probabilities of excitation from the ground to the excited state are plotted as a function of the fractional detuning of the drive frequency from resonance for the most precise measurement of the electron magnetic moment \cite{NorthwesternMagneticMoment2023}. The measured and predicted lineshapes do not agree so well for the cyclotron lineshape.  This increased the uncertainty that comes from extracting $\wcb$ from this lineshape.

\subsection{Quantum measurements of unprecedented precision}

\begin{figure}[htbp!]
\centering
\includegraphics[width=0.4\columnwidth]{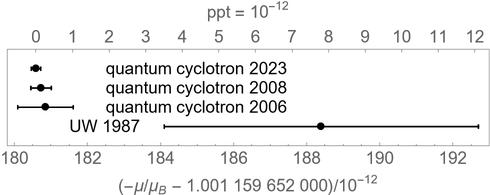}
   \caption{Three quantum cyclotron measurements \cite{HarvardMagneticMoment2006,HarvardMagneticMoment2008,NorthwesternMagneticMoment2023}  of the electron $-\mu/\mu_B = g_e/2$ greatly improved upon the precision attained with classical cyclotron motion (UW1987 \cite{DehmeltMagneticMoment}). A ppt (part per trillion) corresponds to $10^{-12}$.
     }  
   \label{fig:QuantumMeasurements}
\end{figure}

The substantial increase in precision of the electron $g_e^-/2 = -\mu_e/\mu_B$ that was enabled by the quantum cyclotron is illustrated by the  three quantum cyclotron measurements of the electron magnetic moment  represented in Figure \ref{fig:QuantumMeasurements}. 
The three quantum measurements agree well.  They are displayed on a scale that is broad enough to compare the best precision realized using  classical cyclotron motion (UW1987).

The first quantum cyclotron measurement of the electron magnetic moment was reported in 2006 \cite{HarvardMagneticMoment2006}, followed by a second in 2008 \cite{HarvardMagneticMoment2008}. A third quantum cyclotron measurement was reported in 2023 \cite{NorthwesternMagneticMoment2023}, after Gabrielse relocated to Northwestern University. 
The three quantum measurements are compared in Figure \ref{fig:QuantumMeasurements}. The methods and measurements were developed one PhD thesis at a time, in a sequence of 14 PhD theses, over 35 years.  A fourth measurement, underway as this review is written, aims to be an order of magnitude more precise.

The largest uncertainty in the most precise measurement of the electron magnetic moment in Bohr magnetons comes from the inconsistency of the observed quantum jump lineshape for the cyclotron resonance, and the predicted lineshape. The source of the broadened lineshape seems to be associated with the  sensitivity of the superconducting solenoid to very small mechanical vibrations.  Improved vibration isolation for the 2023 measurement narrowed the anomaly resonance line to what was expected from theory.  However, it did not proportionately narrow the cyclotron resonance line, even though both $\wcb$  and $\wab$ have the same fractional dependence upon the size of magnetic field fluctuations.  The reason may well be that the fluctuations of the magnetic field are averaged very differently during the measurements because the two frequencies are so different.  This is under study.

A second source of systematic uncertainty for the most precise measurement comes from the interaction of the cyclotron motion and the resonant electromagnetic modes of the cylindrical trap cavity. This interaction is critical for realizing the quantum cyclotron because it inhibits spontaneous emission (as discussed above), thereby providing the averaging time needed to resolve a one-quantum spin flip or cyclotron transition.  However, the interaction of these cavity mode oscillators also pulls the frequency of the cyclotron oscillator.  The calculated cavity-shift correction that must be applied must be properly renormalized to avoid a sum that goes to infinity. A procedure to correct a renormalized calculation that can be done using experimental input has been devised and applied successfully.  A more general way to carry out this calculation was just reported \cite{Day:2025waz}.

\subsection{Most precisely measured property of an elementary particle confronts the most precise SM prediction\label{sec:Confrontation}}

\begin{figure}[htbp!]
\centering\includegraphics[width=0.375\columnwidth]{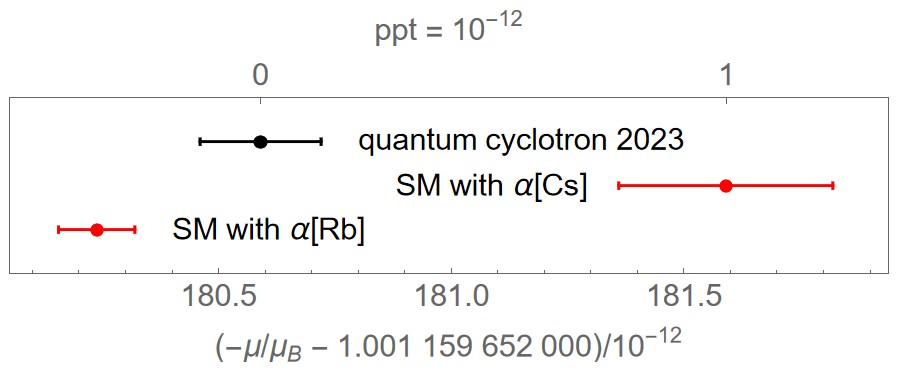}
   \caption{The most precise confrontation of measurement and theory is a comparison of the measured electron magnetic moment, $-\mu/\mu_B = g/2$, and the SM prediction of this moment.   
    SM predictions are functions of discrepant $\alpha$ measurements \cite{MullerAlpha2018,RbAlpha2020}. A ppt is $10^{-12}$.
   }  
   \label{fig:CompareToSM}
\end{figure}

The measurements are compared with two SM predictions that disagree with each other (Figure~\ref{fig:CompareToSM}).  The problem here is not the SM. Instead, the different SM predictions arise because two discrepant measured values of the fine structure constant $\alpha$ \cite{MullerAlpha2018,RbAlpha2020} are used as the required input to predict $\mu_e/\mu_B = - g_e/2$ in  Eq.~\eqref{eq:SMTheory}.   Both the quantum cyclotron measurement of the electron magnetic moment and the SM calculation are precise enough to test the SM 10 times more precisely, once the discrepancy in measured $\alpha$ values  is resolved.  Until the discrepancy is resolved, the best that can be said is that the predicted and measured $\mu/\mu_B$ agree to about $\delta (g_e/2) = 0.7\times10^{-12}$, half of the  $\alpha$ discrepancy.    

\begin{figure}[htbp!]
\centering
\includegraphics[width=0.45\columnwidth]{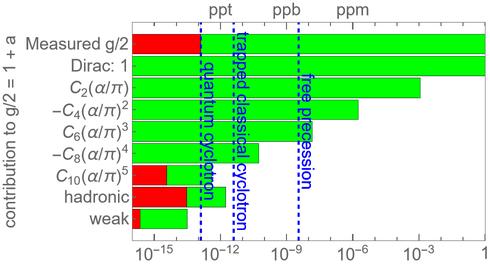}
\caption{The quantum cyclotron measurement of the electron $g_e/2=-\mu/\mu_B$ compared to the size of the various SM theory contributions. Measurement and prediction uncertainties are red.
The blue dashes compare the  measurement precision attained using the quantum cyclotron, to what was attained using classical cyclotron motion in a trap and 
using free precession. 
QED field theory is tested to an unprecedented tenth order.}
\label{fig:SMContributions}
\end{figure}

Nonetheless, the agreement between the SM prediction and the quantum cyclotron measurements is arguably the great triumph of the SM \cite{GreatistTriumph}, and the most precise confirmation ever made of the SM and its field theory framework.  In a congratulatory letter to Gabrielse, quoted in Physics Today \cite{DysonLetter}, Freeman Dyson expressed his profound surprise that this framework was able to make a much more precise prediction than he and the other inventors of QED field theory ever imagined. Figure \ref{fig:SMContributions} compares the measured value of the magnetic moment and its experimental uncertainty to the Dirac prediction, the QED field theory contributions through 10th order, along with both the hadronic and weak interaction contributions.

\subsection{Most precise lepton test of the fundamental CPT invariance of the Standard Model}

The field theory of the Standard Model has an unavoidable CPT symmetry.  C stands for a  ``charge conjugation'' transformation that changes particles into their antiparticles.  C would thus change an electron into a positron, and a positron into an electron.  P stands for a parity transformation which is defined to be spatial inversion through a point.  T stands for time reversal transformation that reverses the direction of all motions.  According to the SM, the weak force keeps all of these  transformations from leaving physical reality  unchanged.  Remarkably, however, the SM predicts that all of physical reality is left unchanged after all three transformations are applied to it.  The SM is in this sense invariant under CPT transformations and this is sometimes referred to as the CPT theorem.  

The most precise test of CPT invariance that has been carried out with a lepton system is the comparison of the electron and positron magnetic moments.  The prediction of CPT invariance is that the electron and positron would have exactly the same value of $g_e/2$. 
The most precisely measured values for the electron, the positron, and their difference are given by 
\begin{align}
g_e^-/2 &= 1.001 ~ 159 ~ 652 ~ 180~ 59~~(13) 
&[000.13~\rm{ppt}]\label{eq:geOver2}\\ 
g_e^+/2 &=1.001 ~ 159 ~ 652 ~ 187~ 90~(430) 
&[004.30~\rm{ppt}]\\
g_e^+/2 - g_e^-/2 &= 0.000 ~ 000 ~ 000 ~ 007~ 31~(430) 
&[004.30~\rm{ppt}]
\end{align}
The uncertainties in the last digits are in parenthesis.
These two values agree within two standard deviations, limited by the much larger positron uncertainty, which should soon be dramatically reduced by new measurements underway.

The electron magnetic moment was measured in 2023 \cite{NorthwesternMagneticMoment2023} using the one-electron quantum cyclotron, as summarized above. The  positron magnetic moment measurement, alas, has not been measured during the quantum era.  Last measured in 1987  \cite{DehmeltMagneticMoment}, using the classical cyclotron orbits, it is less precise by a factor of 30.  A quantum cyclotron measurement at the current level of precision would thus increase the precision of the lepton CPT test by a factor of 30.  If the new methods outlined in the next section succeed in improving the measurement precision for an electron and positron by a factor of ten, then the lepton CPT test would be 300 times more sensitive than the current test.

\subsection{Bright future with special relativity and quantum-limited detection}

A new measurement in an entirely new apparatus at Northwestern University seeks to measure the electron magnetic moment 10 times more precisely, and to compare the electron and positron moments 30 times more precisely. Given that measurements of the electron magnetic moment have been vigorously pursued for a century, it seems remarkable that such a large sensitivity improvement seems possible.

The first enabling new idea is to realize a quantum cyclotron that employs special relativity as a QND coupling to a detector rather than the magnetic gradient employed so far.  If one-quantum cyclotron and spin transitions can be so detected, the measured linewidths of the measured anomaly and cyclotron resonances will be greatly reduced and the systematic errors from magnetic gradient broadening will be largely eliminated. The second enabling new idea is to use a quantum-limited detector rather than the cryogenic HEMT detector whose backaction heats the trapped electron.  A 200 MHz SQUID detector is now operating, despite the challenge that its superconducting elements will not operate in the high magnetic field required for a quantum cyclotron. This should reduce the electron detector temperature by a factor of 25. The third enabling new idea is to use detector backaction circumvention \cite{Fan2020BackActionPRL} to make it possible to detect the quantum and spin state of the system only when the axial detection motion is in its quantum ground state.  

Technical improvements include a smaller Penning trap cavity for improved detection efficiency, and a more harmonic cylindrical trap design that promises to increase detector sensitivity.  Improved vibration isolation should reduce shifts that arise from vibrations of the superconducting solenoid that produces the large magnetic field.  A more general renormalized calculation of trap cavity shifts should reduce the systematic uncertainty in making the cavity-shift correction. The new apparatus is capable of both positron and electron measurements.  

Success in this ambitious program immediately would allow a lepton test of CPT invariance that is more sensitive by a factor of 30.  Testing the SM to the full potential of such a measurement would require improved QED field theory calculations at the tenth order.  The existing discrepancy between measured $\alpha$ values must be fixed, of course.  More precise measurements of the $\alpha$ will also be required based upon more precise measurements of atomic masses and atom recoils.  The hope is that much more precise measurements of the electron and positron magnetic moments will stimulate the theoretical and experimental progress, as it has in the past.

%% file: MuonTau.tex
\section{Muon magnetic moment}\label{sec:Muon}

\subsection{Muon decay}
\label{muondecay}

The concept of measuring the magnetic moment of the muon
was first suggested by Lee and Yang as a consequence of their hypothesis on parity violation in weak interactions \cite{Lee:1956qn}. This parity violation makes muons particularly valuable for experimental studies: it is relatively straightforward to produce beams of polarized muons because muons generated from pion decays are naturally polarized along their motion. Additionally, the spin direction of a muon can be determined by analyzing the angular distribution of the electrons produced in its decay. In essence, nature provides muons with an inherent 
``polarizer'' and ``polarimeter'' at no extra cost.  This can be easily understood by taking the two energy extremes for the emitted electron in the three-body weak decay of the muon $\mu^{-} \rightarrow e^{-}+ \nu_{\mu}+ \bar \nu_e$  
at rest.

\begin{figure}[!htb]
    \centering
    \setlength{\tabcolsep}{0pt} 
    \begin{tabular}{cc}
        \begin{subfigure}[t]{0.5\textwidth}
            \centering
            \begin{tikzpicture}
            \def\blobbymuon{(0,0) circle (0.5)};
            \draw[magenta, pattern=crosshatch, pattern color=magenta]\blobbymuon;
            \node[above] at (0.2,0.6) {\Large $\mu^{-}$};
            \draw[->, thick, blue] (-0.5,-1.)--(0.5,-1.);
            \node[below] at (0.,-1.1) {\large $\vec{s}_{\mu^-}$};
            \draw[->, thick, red] (1.,1.+0.2)--(3.,1.+0.2);
\node[above] at (3.5,0.95+0.15) {\large $\vec{p}_{\bar{\nu}_{e}}$};
\draw[->, thick, blue] (1.5,0.5+0.2)--(2.5,0.5+0.2);
\node[below] at (3.5,0.95+0.15) {\large $\vec{s}_{\bar{\nu}_{e}}$};
\draw[->, thick, red] (1.,-1.-0.2)--(3.,-1.-0.2);
\node[above] at (3.5,-0.95-0.3) {\large $\vec{p}_{\nu_{\mu}}$};
\draw[->, thick, blue] (2.5,-1.5-0.2)--(1.5,-1.5-0.2);
\node[below] at (3.5,-0.95-0.3) {\large $\vec{s}_{\nu_{\mu}}$};

\draw[->, thick, red] (-1.,0.)--(-3.,0.);
\node[above] at (-3.5,0.-0.15) {\large $\vec{p}_{e^-}$};
\draw[->, thick, blue] (-2.5,-0.5)--(-1.5,-0.5);
\node[below] at (-3.5,0.-0.15) {\large $\vec{s}_{e^-}$};
            \end{tikzpicture}
            \caption{}
        \end{subfigure}
        &
        \begin{subfigure}[t]{0.5\textwidth}
            \centering
            \begin{tikzpicture}
            \def\blobbymuon{(0,0) circle (0.5)};
            \draw[magenta, pattern=crosshatch, pattern color=magenta]\blobbymuon;
            \node[above] at (0.2,0.6) {\Large $\mu^{-}$};
            \draw[->, thick, blue] (-0.5,-1.)--(0.5,-1.);
            \node[below] at (0.,-1.1) {\large $\vec{s}_{\mu^-}$};
\draw[->, thick, red] (-1.,0)--(-3.,0);
\node[above] at (-3.5,-0.1) {\large $\vec{p}_{\nu_{\mu}}$};
\draw[->, thick, blue] (-2.5,-.5)--(-1.5,-0.5);
\node[below] at (-3.5,-0.1) {\large $\vec{s}_{\nu_{\mu}}$};

\draw[->, thick, red] (1.,-1.-.2)--(2.,-1-0.2);
\node[above] at (2.5,0.-0.95-.5) {\large $\vec{p}_{e^-}$};
\draw[->, thick, blue] (1.5+0.5,-1.5-0.2)--(1.5-0.5,-1.5-0.2);
\node[below] at (2.5,0.-0.95-.5) {\large $\vec{s}_{e^-}$};

\draw[->, thick, red] (1.,0.)--(3.,0.);
\node[above] at (3.5,0.-0.15) {\large $\vec{p}_{\bar{\nu}_{e}}$};
\draw[->, thick, blue] (1.5,-0.5)--(2.5,-0.5);
\node[below] at (3.5,0.-0.15) {\large $\vec{s}_{\bar{\nu}_{e}}$};
            \end{tikzpicture}
            \caption{}
        \end{subfigure}
    \end{tabular}
    \caption{Muon decay into an electron and two neutrinos, with particle spin directions, $\vec{s}$,  and momentum directions  $\vec{p}$. (a) Negative muon decaying into a high energy electron. (b) Negative muon decaying into a low energy electron.
}
    \label{fig:mu-decay}
\end{figure}
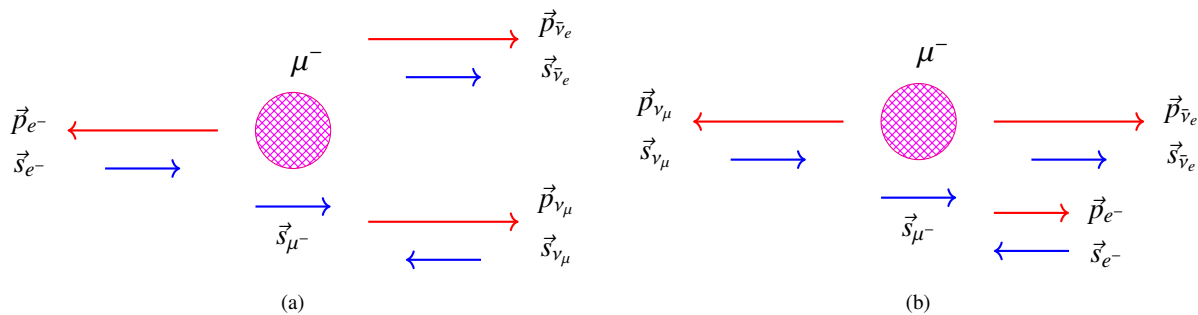

When a muon decays into
an electron with nearly maximum energy ($E_{max}= m_\mu/2 \approx 53$ MeV), the neutrino and anti-neutrino must be emitted in the direction opposite to that of the electron to conserve momentum. Since the neutrino is left-handed and the anti-neutrino is right-handed, the $\nu\bar\nu$ pair carries zero total angular momentum. Due to the weak interaction, the high-energy electron is preferentially emitted left-handed, meaning its momentum is predominantly directed opposite to the muon spin\footnote{In the case of $\mu^+$ decay, the highest-energy positrons are instead emitted parallel to the muon spin.} (Figure~\ref{fig:mu-decay}a). When the electron energy is close to zero ($E_{min}\sim m_e$), the neutrino and anti-neutrino are emitted back-to-back with a total spin of one. In this situation, angular momentum conservation requires that the electron be emitted along the muon spin direction\footnote{For $\mu^+$ decay, low-energy positrons are instead emitted anti-parallel to the muon spin.}   
 (Figure~\ref{fig:mu-decay}b).

Between these limits, the behavior of the distribution of the positrons emitted in the decay of a polarized muon\footnote{In the following, whereas not explicitly mentioned, we will use the term ``muon'' to refer to the anti-muon $\mu^+$, as it is mostly used in the experiment, and ``decay'' positrons to indicate the positrons emitted in the decay of $\mu^+$.} at rest is encoded in two functions. The first is a decay probability that describes how frequently positrons of a given energy are produced.  
 The second, the so-called ``asymmetry'',  
quantifies how strongly their emission direction is aligned or anti-aligned with the muon’s spin. Both the decay probability and asymmetry increase with energy, reaching their peak values at the kinematic limit, where the correlation between spin and momentum becomes strongest. 

The first measurements of the muon magnetic moment exploited this angular correlation between the muon spin and the direction of the decay positron. When a polarized muon beam is stopped in a target, the rate of high-energy positrons exhibits a time-dependent modulation. This modulation arises because the muon spin precesses in the presence of a magnetic field due to its magnetic moment. The modulation frequency corresponds to the Larmor precession of a magnetic moment \(\mu = g e \,\hbar/ 4m_\mu\), allowing for the extraction of the g-factor.

In early experiments, the measured precession frequency was consistent with \(g = 2\), as predicted for a point-like spin-\(1/2\) particle in Dirac theory (see Section~\ref{g-section}). Small deviations from this value were later understood as arising from quantum loop corrections. 
To measure these tiny effects with high precision, dedicated storage ring techniques in which muons circulate at relativistic speeds in highly uniform magnetic field were developed, leading to successive generations of muon \(g\!-\!2\) experiments such as CERN-III, E821 at BNL, and E989 at Fermilab (see Sections~\ref{sec:bfield}--\ref{sec:e989}).
These experiments measured the muon anomaly \(a_\mu = (g_\mu - 2)/2\) (sometimes also called the ``muon anomalous magnetic moment''\footnote{The muon anomaly \(a_\mu = (g_\mu - 2)/2\) is a dimensionless quantity 
while the anomalous magnetic moment is the dimensionful quantity \(\mu_{\mu, \mathrm{anomalous}} = a_\mu \, \frac{e\hbar}{2m_\mu}\).})
with increasing precision over time.

In a storage ring, for relativistic muons, the energy of the decay positrons depends not only on its original emission direction but also on the angle between the muon spin and its momentum. When the muon spin points in the same direction as its momentum, decay positrons are more likely to be emitted forward and thus receive a boost in energy. Conversely, when the spin and momentum are anti-aligned, positrons tend to emerge with lower energies.

\begin{figure}[!htb]
    \centering
    \begin{subfigure}[t]{0.45\textwidth}
        \centering
        \includegraphics[width=\textwidth,height=3.7cm,trim=0 50 0 20,clip]{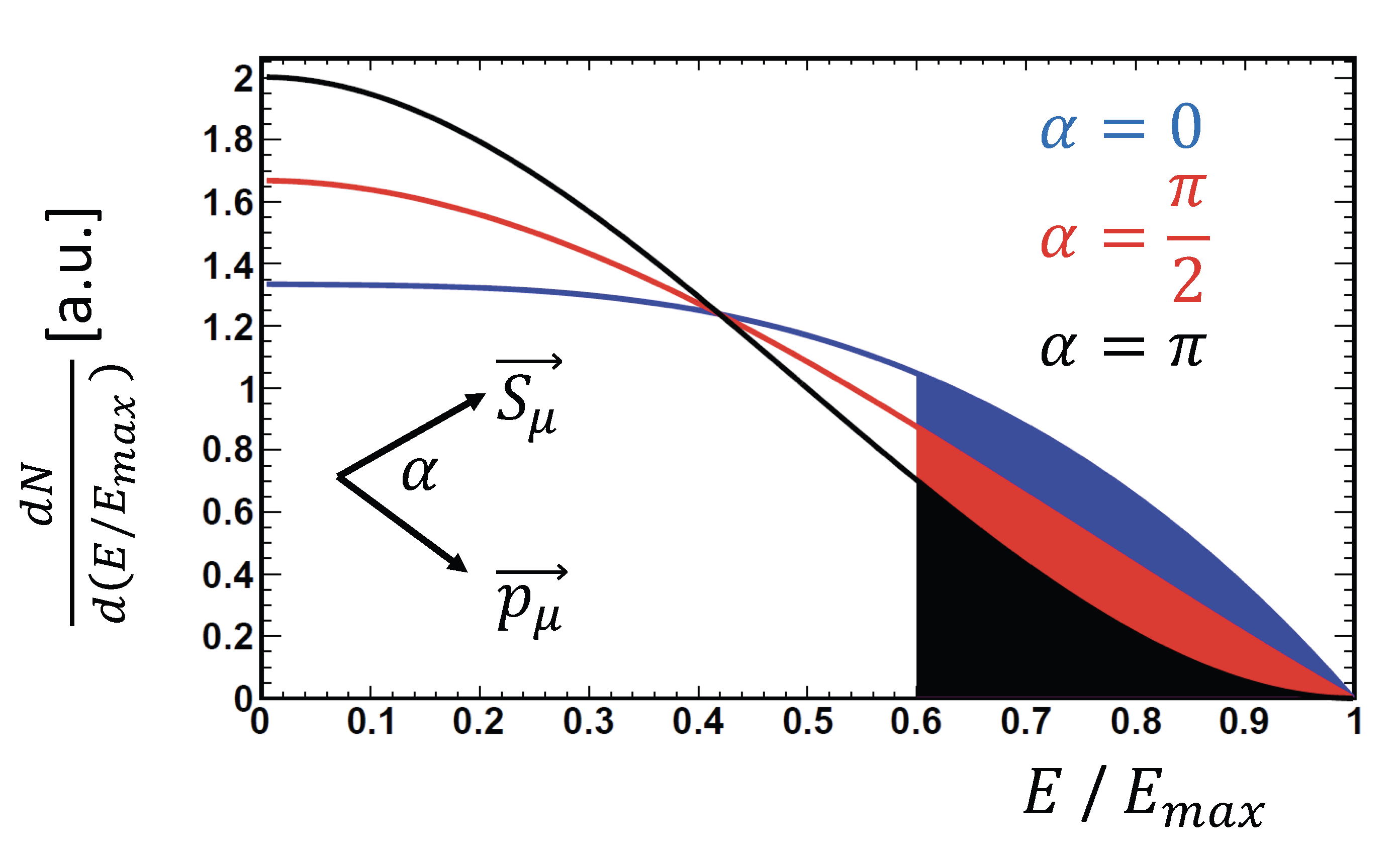}
        \caption{}
        \label{fig:three_folded_wiggles_a}
    \end{subfigure}
    \hfill
    \begin{subfigure}[t]{0.47\textwidth}
        \centering
        \includegraphics[width=\textwidth,height=4cm]{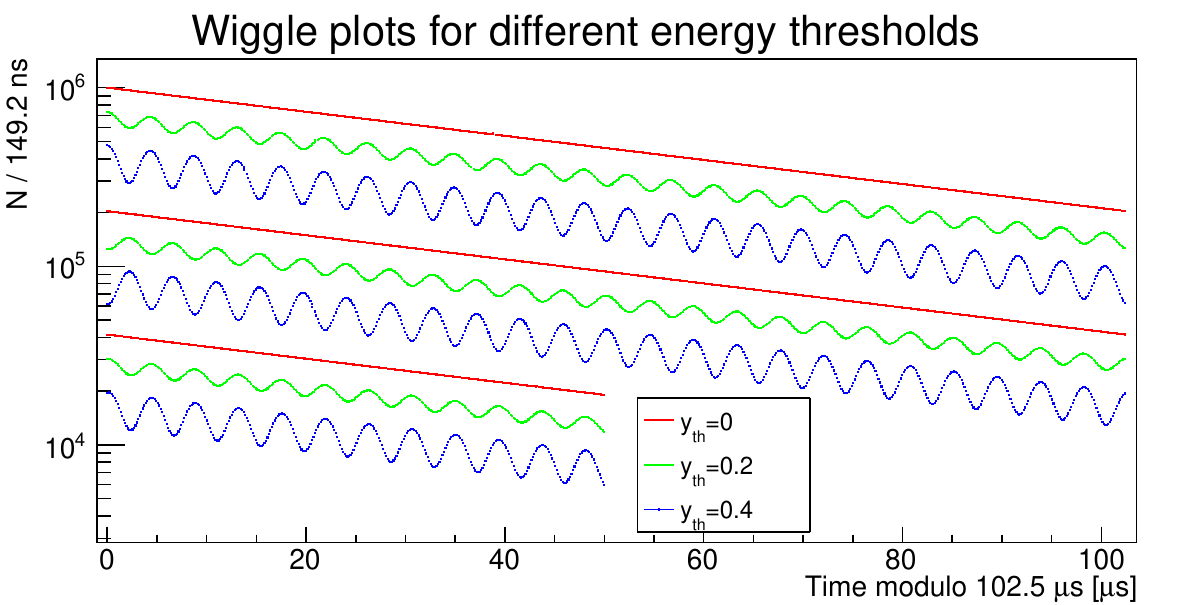}
        \caption{}
        \label{fig:three_folded_wiggles_b}
    \end{subfigure}
    \caption{(a) Muon decay energy distribution in the lab frame for three different values of the angle \( \alpha \) between the muon spin and its momentum. When spin and momentum are aligned, the energy distribution shifts toward higher energies. The shaded regions above the threshold \( E / E_{\text{max}} \approx 0.6 \) illustrate how the number of positrons depends on \( \alpha \), and thus oscillates at \( \omega_a \). From~\cite{Fienberg:2019nwt}. (b) Number of positrons (the ``wiggle plot'') as a function of time during which muons circulate in the storage ring before decaying (``storage time''), for three energy thresholds \( y_{th} = E / E_{\text{max}} \). The cyclotron period, {\it i.e.} the time required for one full revolution of a muon in the ring, for a typical energy of \SI{3.1}{GeV}, is 149.2 ns.}
    \label{fig:three_folded_wiggles}
\end{figure}

As a result, the energy spectrum of positrons in the lab frame depends on the angle between the muon spin and its momentum $\alpha$, as shown in Figure~\ref{fig:three_folded_wiggles}a. Since muons are circulating in a (uniform) magnetic field, their spin precesses relative to their momentum and the angle $\alpha$ evolves in time as  $\alpha = \omega_a t + \phi_a$ where $\omega_a$ is known as the angular 
``anomalous precession frequency'' (or anomaly frequency)  and $\phi_a$ is the initial spin-momentum phase at $t = 0$. 
Since the energy of each positron depends on $\alpha$ the positron energy spectrum is modulated in time.
By selecting high-energy positrons—those most strongly correlated with the spin direction—and tracking how their detection rate varies over time, the experiment extracts $\omega_a$, which directly encodes the anomalous magnetic moment of the muon, as shown in the following.
This precession leads to a time-dependent modulation in the positron count rate  $N(t)$ above a certain energy threshold $
y_{th}=
E/E_{max}$, commonly referred to as a ``wiggle plot'':
\begin{equation}
    N(t)=N_0(y_{th})e^{-t/\gamma\tau_\mu}\left[1+A(y_{th})\cos(\omega_at+\phi_a)\right].
    \label{eqw1}
\end{equation}
The lifetime of the muon at rest is $\tau_\mu=2.2\,\mu s$,  and $\gamma$ accounts for the relativistic time dilation ($\gamma\tau_\mu \approx 64.4\,\mu$s is the typical value in a storage ring, corresponding to a muon momentum of 3.1 GeV/\si{c}). 
Higher thresholds increase the oscillation amplitude but reduce event counts as shown in Fig.~\ref{fig:three_folded_wiggles}b. The statistical precision of the measurement of the anomalous  
precession frequency is   
\begin{equation}
    \frac{\delta\omega_a}{\omega_a}=\frac{\sqrt{2}}{\omega_a\gamma\tau_\mu\mathcal{P}\sqrt{NA^2(y_{th})}},
\end{equation}
where $\mathcal{P}$ is the average muon polarization and $N$ the total number of positrons with energy above threshold. Since $N\propto N_0$ to optimize the statistical precision of $ \omega_a $, the figure of merit $ N_0 A_0^2 $ must be maximized, suggesting an optimal threshold near $ y_{th} \approx 0.6 $, or about $1.86$ GeV at a storage ring. In experimental conditions, detector resolution shifts this optimal threshold slightly lower to around $1.7$ GeV.

\subsection{Early muon $g$ measurements}
\label{g-section}
\begin{figure}[!htb]
    \centering
    \begin{subfigure}[t]{0.35\textwidth}
        \centering
        \includegraphics[width=\textwidth]{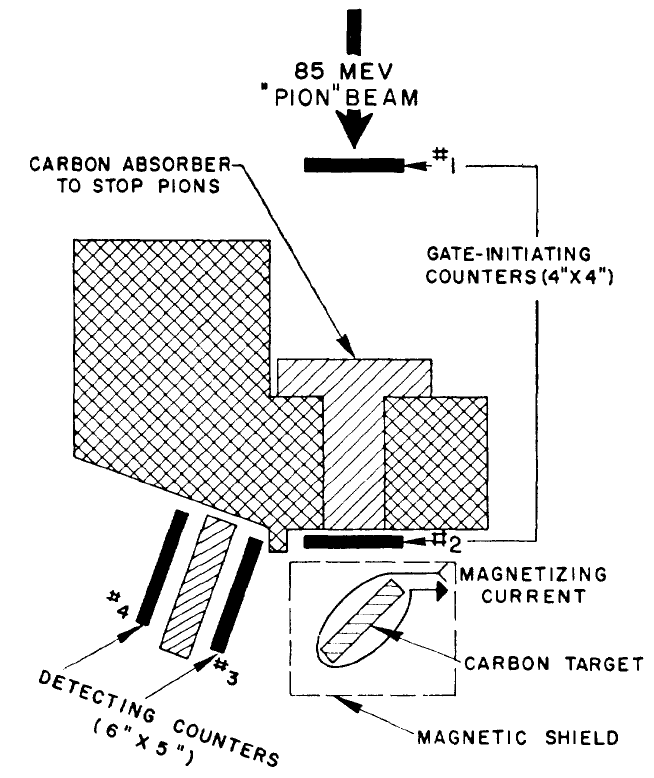}
        \caption{}
        \label{fig:history_lederman_garwin_weinrich_a}
    \end{subfigure}
    \begin{subfigure}[t]{0.35\textwidth}
        \centering
        \includegraphics[width=\textwidth]{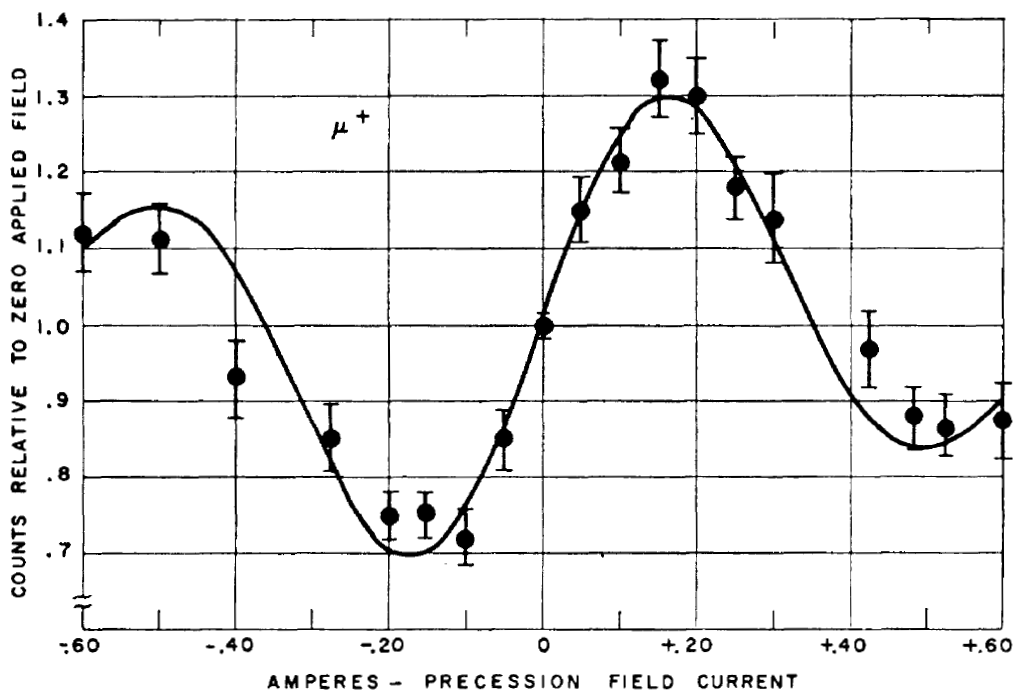}
        \caption{}
        \label{fig:history_lederman_garwin_weinrich_b}
    \end{subfigure}
    \caption{(a) Experimental setup at Nevis. (b) Number of positron counts as a function of the applied field. From~\cite{Garwin:1957hc}.}
    \label{fig:history_lederman_garwin_weinrich}
\end{figure}

The first attempt to measure the muon gyromagnetic ratio, \( g_{\mu} \), was made in 1957 at the Nevis Laboratories cyclotron in Irvington, New York~\cite{Garwin:1957hc}.
Parity violation made  $85\,\si{MeV}/c$ pions decay into the polarized muons upon striking  a carbon target.  A scintillator telescope tracked positrons emitted when the muons decayed after their spins precessed (rotated) in an external magnetic field (Figure~\ref{fig:history_lederman_garwin_weinrich}). The variation in positron emission with increasing magnetic field strength revealed \( g_{\mu} \approx 2.00 \pm 0.10 \).  This was not precise enough to reveal $a_\mu \approx 10^{-3}$. 
However, it suggested that the muon is fundamental and  point-like  (since, for composite particles, \( g\) can significantly deviate from 2).  The measurement also reinforced the evidence for P and C violation in the decays of pions and muons.  In 1957, Cassels et al.~\cite{Cassels:1957} achieved 1\% precision using a Liverpool cyclotron and a more advanced detection system.

 Three years later, in 1960, Garwin et al.~\cite{Garwin:1960zz} provided the first evidence of a small deviation from $g_\mu=2$, measuring $g_{\mu} = 2(1.00122\pm0.00008)$. 
With an uncertainty of approximately 7\%, this measurement confirmed the theoretical prediction of the first two terms on the right-hand side of Equation~\ref{eqQED}, specifically that \( C_2 = \frac{1}{2} \).
Equivalently, this corresponds to $a_{\mu} = \alpha/{2\pi} \sim 0.001$.  
 As noted in Section \ref{sec:LeptonMoments},  \( a_{\mu} \) is nonzero because of the interactions between the muon and the quantum vacuum as described by quantum field theory.

\subsection{Cyclotron and spin precession of a muon in a magnetic Field}
\label{sec:bfield}

The precision of muon magnetic moment measurements increased when the anomaly $a_\mu$ was measured directly \cite{Louisell,Schupp:1961zz,Cassels:1957} instead of $g_\mu$.  A non-relativistic muon moving in a circular orbit perpendicular to a constant and spatially uniform magnetic field, with an angular frequency
\begin{equation}
\omega_c= \frac{eB}{m_\mu}.
\end{equation}
The muon spin simultaneously precesses with an angular spin frequency 
\begin{equation}
\omega_s = 
\frac{g_\mu}{2} \omega_c = (1 + a_\mu) \, \omega_c.
\label{omegac}
\end{equation}
For $g_\mu=2$ (i.e.\ $a_\mu=0$), the angle between the muon's momentum and spin would remain constant throughout its motion.
The SM prediction that $a_\mu\approx 10^{-3}$ for charged leptons, means that the spin precession at $\omega_s$ is slightly bigger than the cyclotron rotation at $\omega_c$,  
The difference, the anomalous precession frequency,
\begin{equation}
\omega_a \equiv \omega_s - \omega_c = a_\mu \omega_c,
\end{equation}
is about 1000 times smaller than either $\omega_s$ or $\omega_c$.  
Measuring $\omega_a$ is the foundation for all measurements of the anomalous magnetic moment of the muon. Because $\omega_a$ is so much smaller than $\omega_c$, the muons must live and be stored for many cyclotron revolutions in a storage ring.  

The muon's short lifetime of only $2.2\,\mu s$  is mitigated by using relativistic muons, whose lifetime is extended by the Lorentz factor $\gamma$ due to time dilation.  
For relativistic muons, the cyclotron frequency  becomes:
\begin{equation}
\omega_C = \frac{\omega_c}{\gamma}.
\end{equation}
The spin precession frequency 
in the muon's rest frame is still described by Equation \ref{omegac}, but now we must account for the fact that the rest frame of the muon, 
in the laboratory frame rotates  at the much smaller Thomas precession~\cite{Thomas:1926dy,Thomas:1927yu} angular frequency,
\begin{equation}
\omega_T = \left( 1 - \frac{1}{\gamma} \right) \omega_c,
\end{equation}
The spin precession angular frequency observed in the lab frame is thus
\begin{equation}
\omega_S = \omega_s - \omega_T 
= \left( a_\mu + \frac{1}{\gamma} \right) \omega_c.
\end{equation}
The anomalous precession frequency in the lab frame, the angular velocity with which the spin rotates relative to momentum,  is then
\begin{equation}
\omega_a = \omega_S - \omega_C = a_\mu \omega_c = a_\mu \frac{eB}{m_\mu }.\label{eq:MuonAnomalyFrequency}
\end{equation}
Thus, we arrive at the remarkable result that the frequency $\omega_a$ is not affected by Lorentz transformations.  This relativistic invariance is discussed in the quantum limit for the electron in Section \ref{sec:RelativitisticQuantumCyclotron}.

To determine $a_\mu$ using Equation \ref{eq:MuonAnomalyFrequency}, 
the anomalous precession frequency in the laboratory reference frame, $\omega_a$, must be measured.  Previous sections have discussed how parity violation provides the polarized muons that are required, and following sections give more details. For the electron measurements, the cyclotron frequency $\omega_c$ can be measured directly to determine $a_e=\omega_a/\omega_c$. A big difference for the muon measurements is that an accurate direct measurement of $\omega_c$ is not possible.  
Instead, the average magnetic field \(B\) experienced by the muons as they orbit in the large storage ring is used, measured precisely using both fixed and mobile nuclear magnetic resonance (NMR) probes.
The NMR technique relies on the principle that the magnetic moments of protons in a material precess at a frequency proportional to the strength of the magnetic field. By placing NMR probes filled with a hydrogen-rich substance (typically water or petroleum jelly) at various locations in the storage ring, the precession frequency of the protons is measured. This frequency, known as the Larmor frequency, is directly related to the local magnetic field. A trolley system carrying multiple NMR probes moves through the storage ring to map the magnetic field profile, allowing to determine the average magnetic field experienced by the stored muons with great accuracy, see Sections~\ref{sec:e821} and~\ref{sec:e989}.

\subsection{The CERN experiments}
\begin{figure}[htbp]
    \centering
    \includegraphics[width=0.5\textwidth]{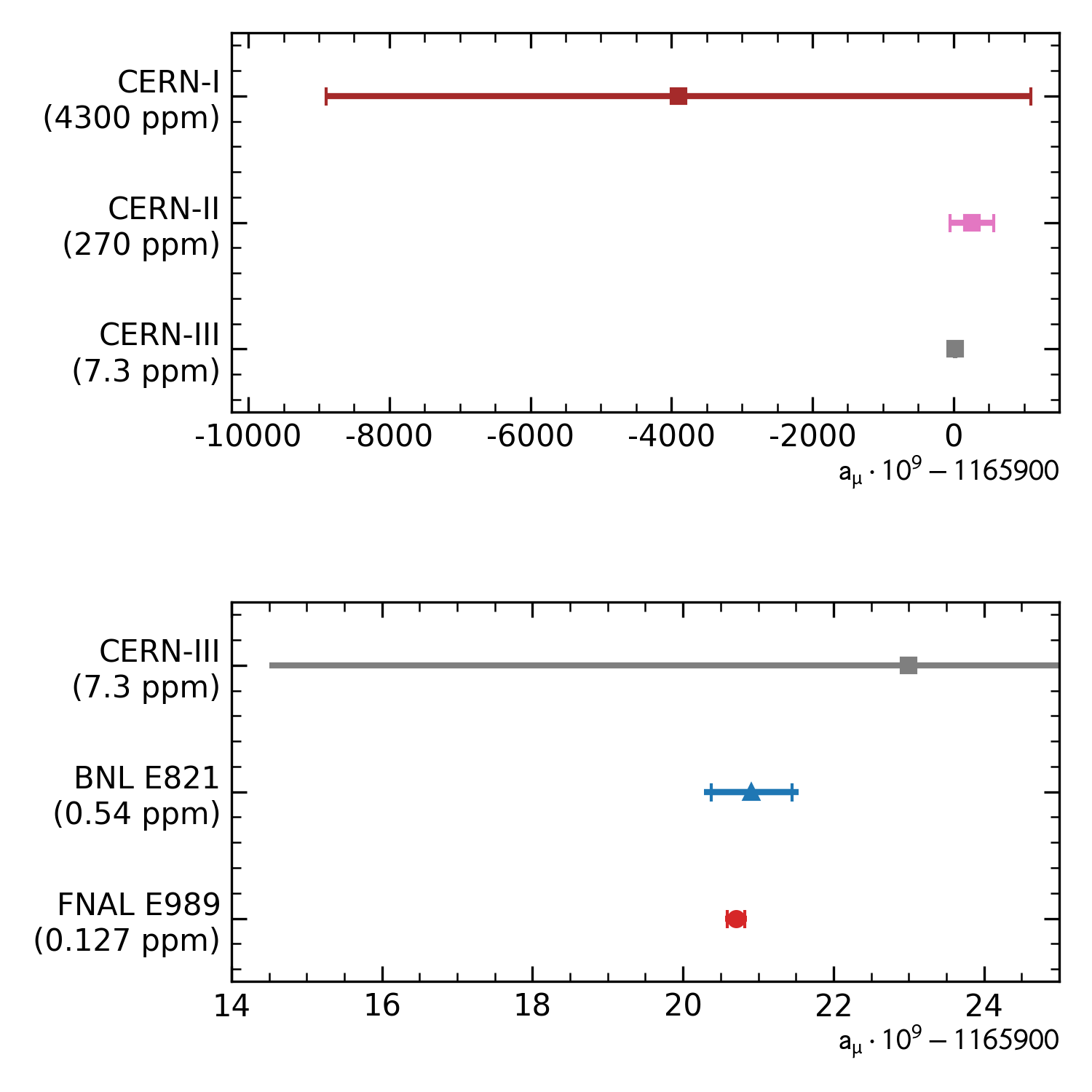}
 \caption{
Comparison of the muon anomalous magnetic moment measurements from different experiments.
The top panel shows results from the three CERN experiments (CERN-III, CERN-II, CERN-I), with their corresponding errors of 7.3~ppm, 270~ppm, and 4300~ppm, respectively.
The bottom panel displays the high-precision experiments: CERN-III (7.3~ppm), BNL E821 (0.540~ppm), and FNAL E989 (0.127~ppm).
The horizontal axis represents the measured value of \(a_\mu \times 10^9\) after subtracting an offset of 1165900.
}
    \label{fig:muon_g_minus_2_comparison}
\end{figure}

In the $1960-1970$s, a series of measurements of the muon magnetic anomaly $a_{\mu}$, with increasing precision, were carried out at CERN, with the aim of testing the validity of the QED prediction~\cite{Farley:1979yb,Farley:2004hp} (Figure~\ref{fig:muon_g_minus_2_comparison}). These experiments followed a common approach: a polarized muon beam, produced via pion decay, was subjected to a magnetic dipole field, and the time distribution of decay positrons was analyzed to extract $a_{\mu}$. In the following, we summarize the key advances of each experiment.

\subsubsection{CERN-I}

\begin{figure}[!htb]
    \centering
    \begin{subfigure}[t]{0.48\textwidth}
        \centering
        \includegraphics[width=\textwidth]{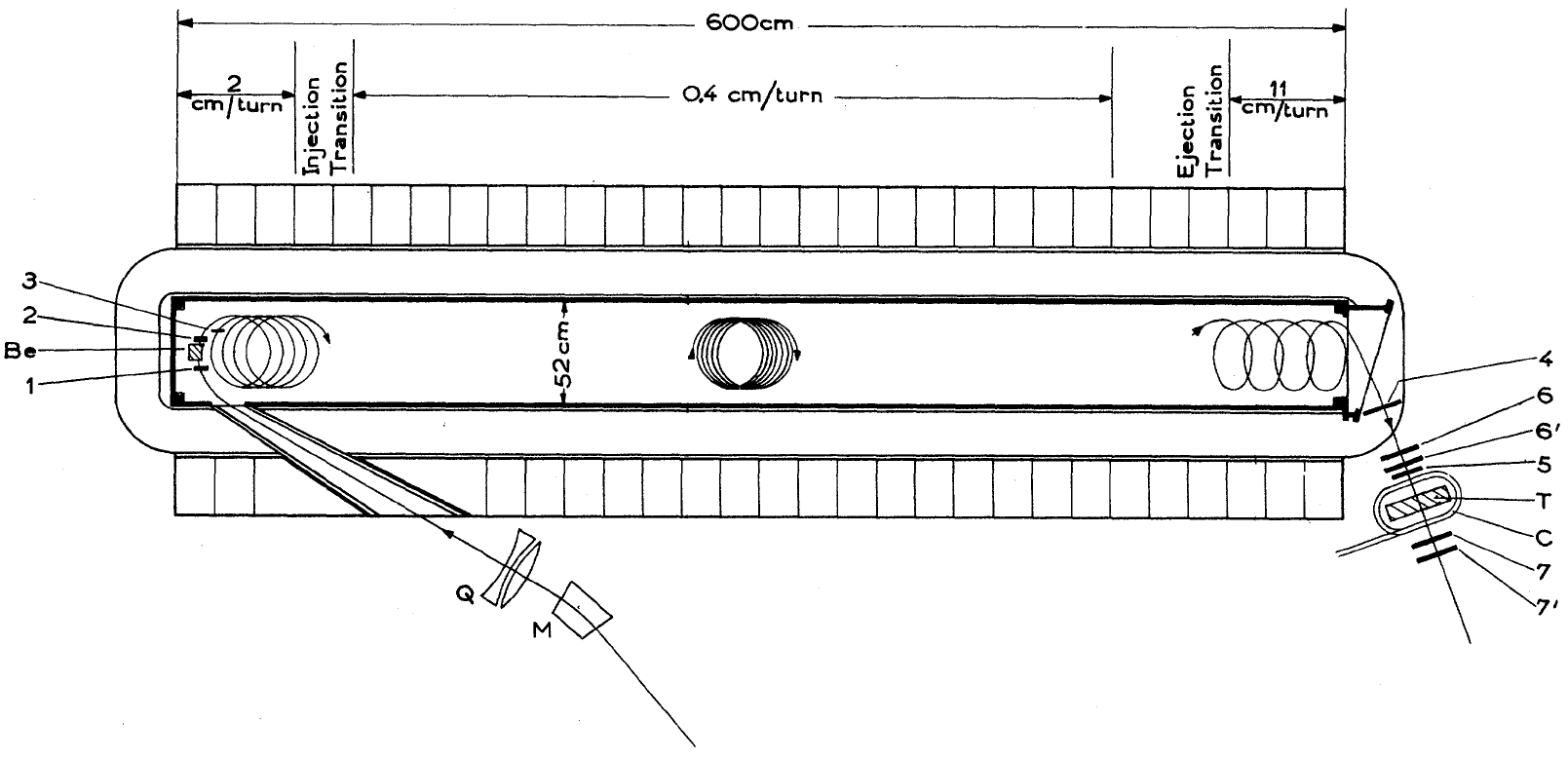}
        \caption{}
        \label{fig:cern1_left}
    \end{subfigure}
    \begin{subfigure}[t]{0.48\textwidth}
        \centering
        \includegraphics[width=\textwidth]{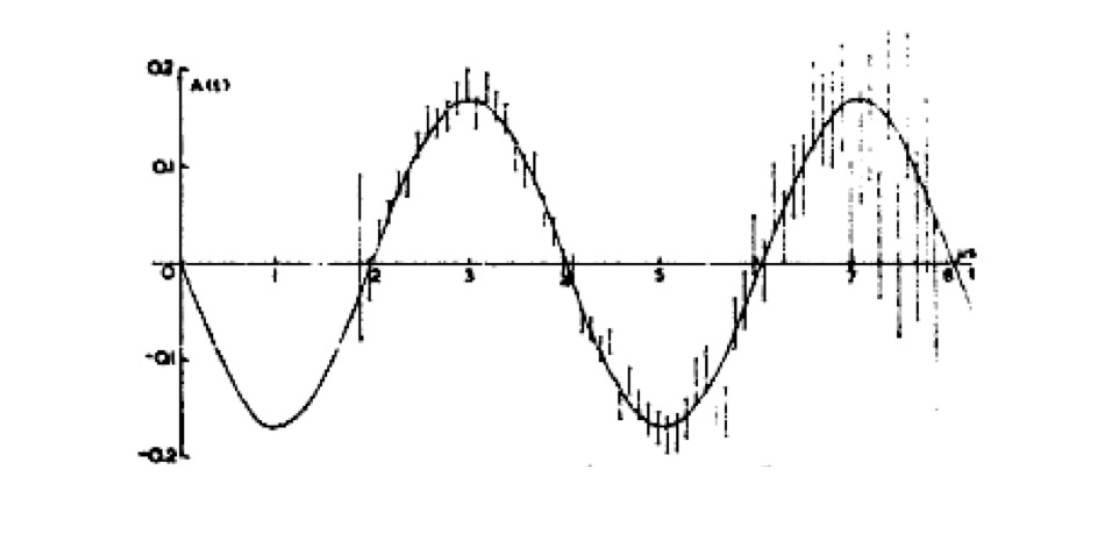}
        \caption{}
        \label{fig:cern1_right}
    \end{subfigure}
    \caption{(a) Overview of the CERN-I experimental setup: a $\mu^{+}$ beam was injected in the dipole magnet and the coincidence of $1$, $2$ and $3$ counters defined the injection signal; ejected muons stopped in target T, where the stop signal was a coincidence of counters $4$, $5$, $6$, $6^\prime$ and a veto on counter $7$; after the stop signal, the coincidence of $6-6^\prime$ or $7-7^\prime$ defined forward and backward emitted positrons. (b) Measured distributions as a function of storage time of the forward-backward positron asymmetry, with the best fit according to Equation \ref{cern1eq}. From~\cite{Charpak:1961mz,Charpak:1962zz}.}
    \label{fig:cern1}
\end{figure}

The first $g-2$ experiment (CERN-I~\cite{Charpak:1961mz,Charpak:1962zz}) employed a dipole magnet 6 meters long, 52 cm wide, with a 14 cm gap (Figure~\ref{fig:cern1}a). A $\mu^+$ beam, produced from in-flight decays of $\pi^+$, entered the magnet through a field-free channel. They were captured in the $1.6$ T magnetic field by abruptly reducing their momentum from $150\,$MeV/c to $90\,$MeV/c
in a beryllium absorber. A magnetic field gradient guided the circular orbits (radius $\sim 19\,$cm) from one end of the magnet to the other while also focusing the muons in the horizontal plane. Muons typically remained in the magnet for $2$ to 8 $\mu$s, completing several revolutions before stopping in a polarization analyzer.

The muon spin direction was flipped by $\pm\,\ang{90}$ in successive runs, by means of a pulsed vertical magnetic field produced in the first microsecond after the muon arrival on target: this way, forward ($c^+$) and backward ($c^-$) emitted positrons (respect to the direction of the spin) were counted by the same pair of telescope counters (indicated with $6-6^\prime$ and $7-7^\prime$ in Figure~\ref{fig:cern1}a), instead of using two different telescopes with different efficiencies. 
The measured forward-backward positron asymmetry, 
\begin{equation}
    A(t)=\frac{c^+(t)-c^-(t)}{c^+(t) + c^-(t)} = A_0 \sin(\omega_a t + \phi).
    \label{cern1eq}
\end{equation}
was fit to extract the anomalous precession frequency, $\omega_a$ that is needed to use Equation \ref{eq:MuonAnomalyFrequency} to determine the muon anomaly.  
The average magnetic field that is also needed was measured to a  precision of approximately 
$5 \times 10^{-4}$~\cite{Charpak:1961mz,Charpak:1962zz}. 

Only a couple of anomaly precision ``wiggles'' were observed (Figure~\ref{fig:cern1}b) before the muon decayed, but CERN-I nonetheless determined \( a_{\mu} \) with an uncertainty  of \( 5 \times 10^{-6} \) (4300~\si{ppm}).  This is comparable in magnitude to the 4th order QED contribution \( C_4 \,\left(
{\alpha}/{\pi}\right)^2
\sim 4.1\times 10^{-6} \) in Equation \ref{eqQED} and Figure~\ref{amu22}.

\subsubsection{CERN-II}  

\begin{figure}[h]
    \centering
    \begin{subfigure}[t]{0.38\textwidth}
        \centering
        \includegraphics[width=\textwidth]{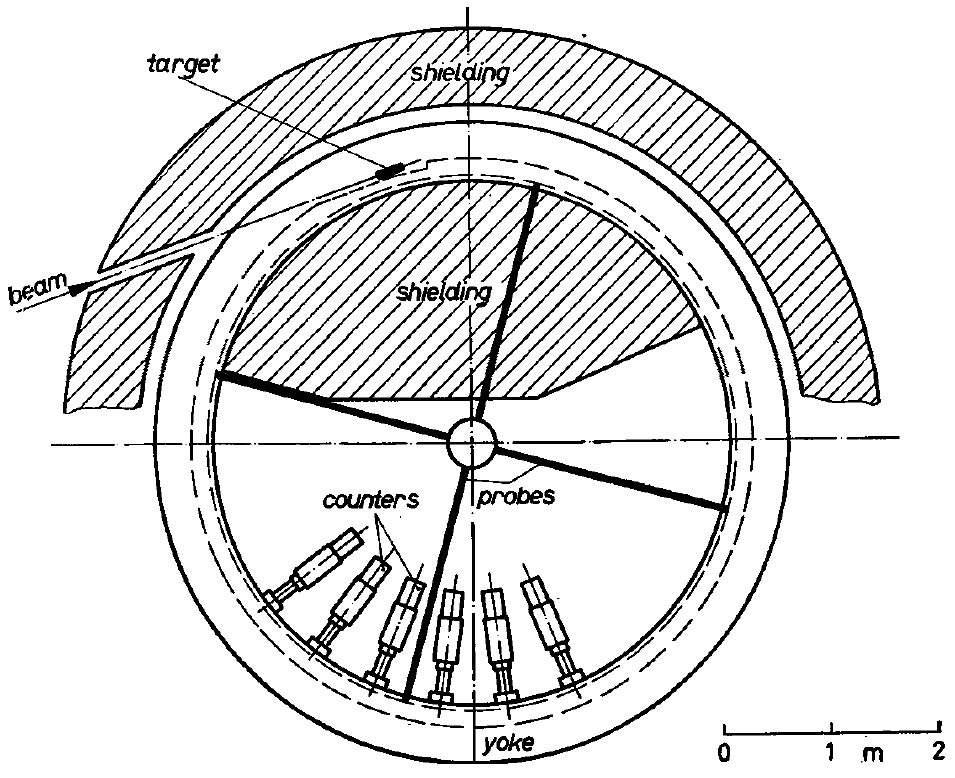}
        \caption{}
        \label{fig:cern2_left}
    \end{subfigure}
    \hspace{0.02\textwidth} 
    \begin{subfigure}[t]{0.3\textwidth}
        \centering
        \includegraphics[width=\textwidth]{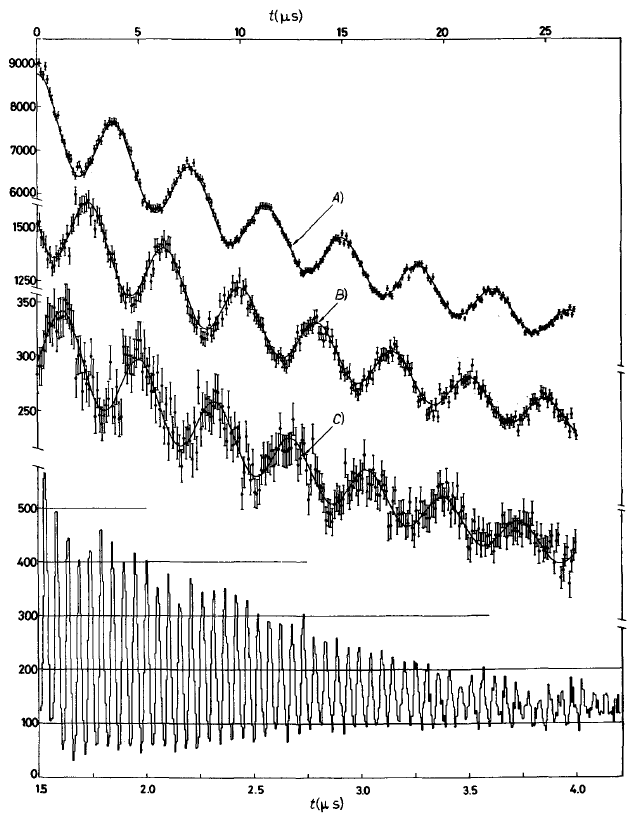}
        \caption{}
        \label{fig:cern2_right}
    \end{subfigure}
    \caption{(a) $5$-m diameter ring magnet used in the CERN-II experiment: protons entered the yoke and hit a target inside the magnetic field. (b) Decay positron distribution as a function of the storage time, showing the \( g{-}2 \) modulation with a period of approximately 3.7~\(\mu\)s. 
Upper time scale: curve A) from 20~\(\mu\)s to 45~\(\mu\)s, curve B) from 65~\(\mu\)s to 90~\(\mu\)s, and curve C) from 105~\(\mu\)s to 130~\(\mu\)s. 
Lower time scale: the muon rotation frequency at early times (from 1.5 to 4.5~\(\mu\)s). It shows a 19~MHz modulation caused by the rotation of the muon bunch around the storage ring. 
As the bunch spreads out, this modulation gradually disappears. This effect is used to determine the radial distribution of the muon orbits. From~\cite{Bailey:1968rxd,Bailey:1972eu}.}
    \label{fig:cern2}
\end{figure}

The CERN-II experiment \cite{Farley66,Bailey:1968rxd,Bailey:1972eu} used higher momentum $\mu^+$ at 1.27 GeV/c to extend the muon lifetime in the lab reference frame by a factor of  $\gamma = 12$ over its rest-frame lifetime.  About 200 muons at a time \cite{Farley:2004hp}, stored for the first time in a storage ring, came from the decay of pions initially trapped in the ring.  The 5-m diameter ring was made of 40, 1.7-T bending magnets. Six lead-scintillator counters along a quarter of the inner radius of the ring (Figure~\ref{fig:cern2}a) detected the number of decay positrons as a function of time to produce a ``wiggle plot'' similar to Equation \ref{eqw1}, but with some technical corrections to account for a high hadronic background from a pion production target inside the storage ring. 
The number of observed $\omega_a$ oscillations increased from a few to 50 (Figure~\ref{fig:cern2}b).  The anomalous precession frequency $\omega_a$ was extracted, and the magnetic field was measured periodically with 4 NMR probes, to determine $a_\mu$, again using Equation \ref{eq:MuonAnomalyFrequency}. The measurement uncertainty of $31 \times 10^{-8} $ (\SI{270}{ppm}) is comparable in size to the 6th order QED contribution $ C_6 \left( {\alpha}/{\pi} \right)^3 \sim 30 \times 10^{-8} $
in Equation \ref{eqQED} and in Figure~\ref{amu22}.

\newpage

\subsubsection{CERN-III}
\label{sec:cern3}

A key innovation for CERN-III~\cite{cernIII,CERNMuonStorageRing:1975gyf,CERNMuonStorageRing:1977bbe,CERN-Mainz-Daresbury:1978ccd}, the final measurement in the CERN series, was a storage ring with an electrostatic quadrupole electric field $\vec{E}$ added to the magnetic field $\vec{B}$ to better control the muon trajectories.  The significant cost for a muon with velocity $\vec{\beta}$ is that the observed anomalous precession frequency of Eq.~\ref{eq:MuonAnomalyFrequency} acquires two additional terms,
\begin{equation}
\vec{\omega}_a = \vec{\omega}_S - \vec{\omega}_C = \frac{e}{m_\mu } \left[a_{\mu} \vec{B} - a_{\mu} \frac{\gamma}{\gamma+1} (\vec{\beta} \cdot \vec{B}) \vec{\beta} - \left(a_{\mu} - \frac{1}{\gamma^2-1}\right) \frac{\vec{\beta} \times \vec{E}}{c} \right].
\label{eq:VectorAnomalyFrequency}
\end{equation}
The frequency vectors indicate the direction of rotation via the right hand rule.
The additional term going as $\vec{\beta}\cdot\vec{B}$ vanishes insofar as $\vec{\beta}$ and $\vec{B}$ are perpendicular.  
The second additional term, going as $\vec{\beta} \times \vec{E}$, vanishes with the astute choice  
\begin{equation}
\gamma = \sqrt{1 + \frac{1}{a_{\mu}}} \approx 29.3.
\end{equation}
At the corresponding ``magic momentum'' of \SI{3.098}{GeV/c},    the anomaly frequency becomes identical to Equation \ref{eq:MuonAnomalyFrequency},
\begin{equation}
\omega_a =  a_\mu \omega_c = a_\mu \frac{eB}{m_\mu },\label{eq:MuonAnomalyFrequency2}
\end{equation}
the form that pertained before an electric field was added to a magnetic field.  
A more detailed analysis that includes corrections  for muon orbits not perfectly perpendicular to the magnetic field, and for muon momenta that  deviate slightly from the magic value, requires the vector Equation \ref{eq:VectorAnomalyFrequency}.

\begin{figure}[!htb]
    \centering
    \begin{subfigure}[t]{0.48\textwidth}
        \centering
        \includegraphics[width=\textwidth]{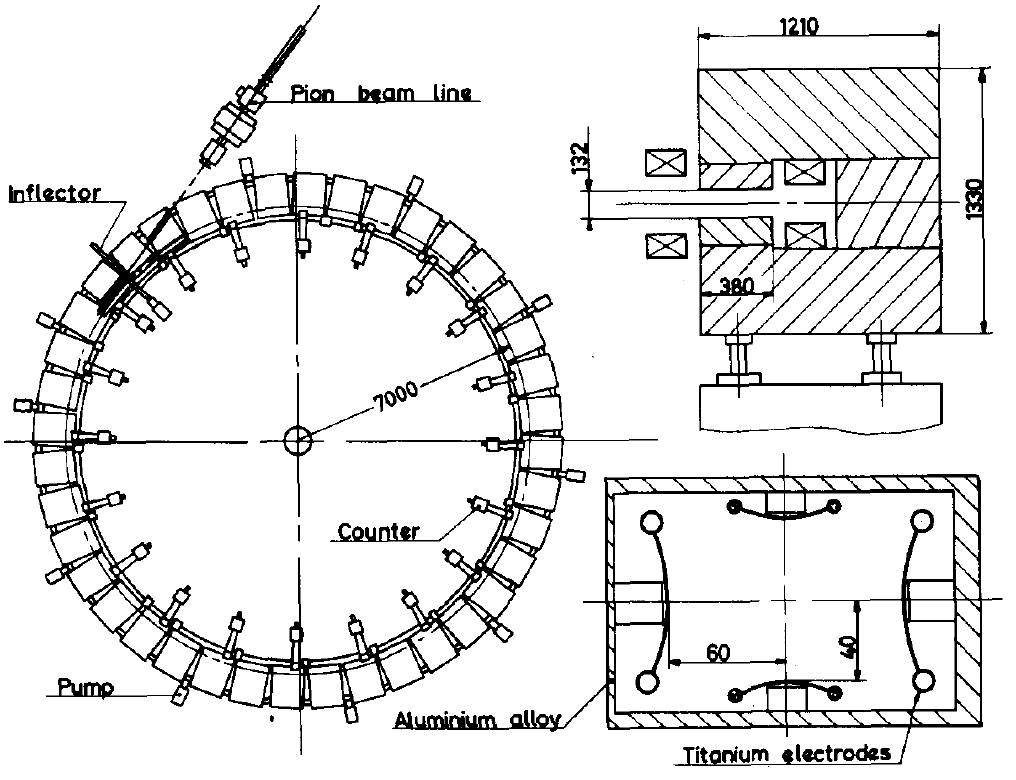}
        \caption{}
        \label{fig:CERN-III_left}
    \end{subfigure}
    \hfill
    \begin{subfigure}[t]{0.48\textwidth}
        \centering
        \includegraphics[width=\textwidth]{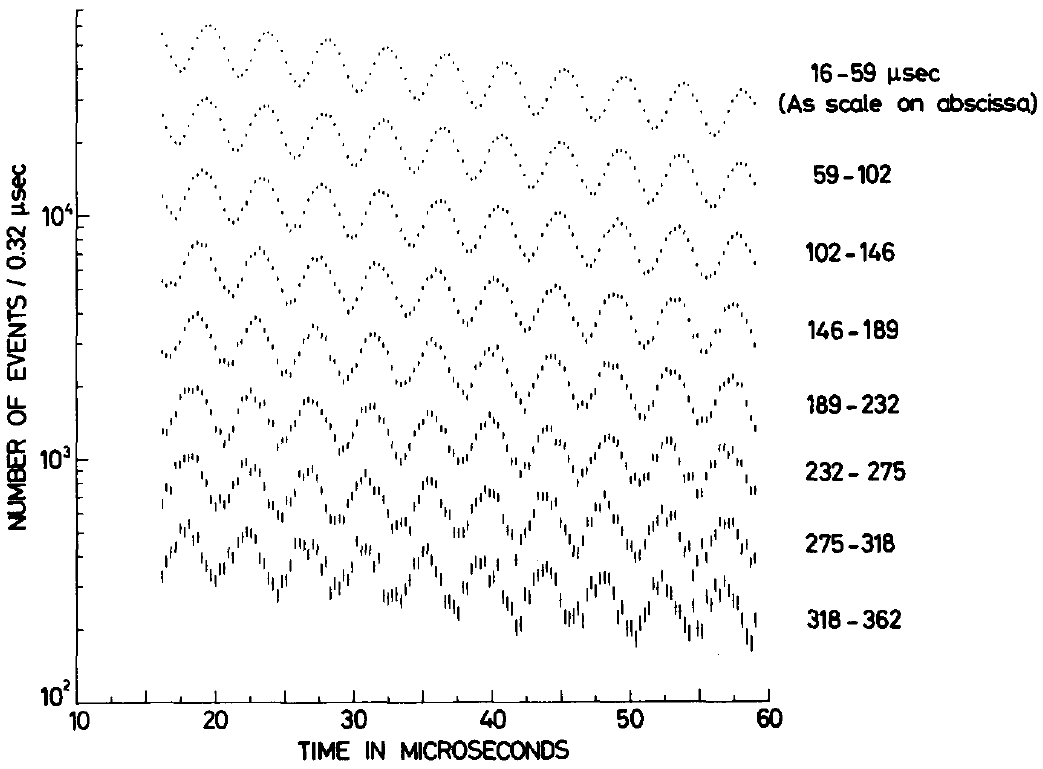}
        \caption{}
        \label{fig:CERN-III_right}
    \end{subfigure}
    \caption{(a) Overview of the CERN-III $14$-m diameter $40$-magnet ring. On the top right the C-shaped yoke is shown, the open side faced the centre of the ring; on the bottom right the cross section of the vacuum chamber and the electric quadrupoles are shown. (b) Time distribution of electrons from muon decays, from a portion of CERN-III data (the first $80$ $g-2$ cycles are shown). From~\cite{CERNMuonStorageRing:1975gyf,CERN-Mainz-Daresbury:1978ccd}.}
    \label{fig:CERN-III}
\end{figure}

The 14-m diameter storage ring (Fig. \ref{fig:CERN-III}a) used 40 dipole magnets in direct contact to form a highly homogeneous and stable 1.47-T magnetic field.   
The beam target was placed outside the ring, reducing the background from the hadronic flash. Pions were transported to the ring through a beamline to improve the initial muon polarization.
An inflector canceled the storage field at injection to minimize beam deflection. Thirty-two electric quadrupoles focused the muon beam.  The injected pions were momentum-selected before entering the ring to increase both the pion flux and the longitudinal muon polarization. 
Detectors, made of lead-scintillator sandwiches, were positioned inside the ring, where the C-shaped magnets were open.
Positrons from muon decays were detected using lead-scintillator counters placed inside the ring. These  have a lower momentum than the stored muons, and thus follow a smaller curvature radius so they emerge towards the inner part of the ring. 

Out of $10^6$ injected pions, approximately 200 muons were successfully stored and about 35  decay electrons were detected per injection. 
At the magic momentum, the muon lifetime  is \SI{64.4}{\micro\second} in the lab frame, and anomaly oscillations are observed for about 655 $\mu s$ (Fig. \ref{fig:CERN-III}b) to greatly improve the precision with which $\omega_a$ is extracted from measured wiggle plot  like that in Figure~\ref{fig:CERN-III}b. 
The magnetic field experienced by the muons was measured at \SI{1.5}{ppm} (see also Table~\ref{tab:experiment_summary}) using proton NMR so that $a_\mu$ could be determined by Equation \ref{eq:MuonAnomalyFrequency}.
CERN-III measured \( a_{\mu} \) for positive and negative muons to \( 8.5 \times 10^{-9} \) (\SI{7.3}{ppm}). More details about the uncertainties are in Table \ref{tab:experiment_summary}.
The CERN-III measurement tested the  
6th order QED contribution $ C_6 \left( {\alpha}/{\pi} \right)^3 \sim 301.4 \times 10^{-9} $
in Equation \ref{eqQED} and Figure~\ref{amu22}. Also tested was the leading-order hadronic vacuum polarization contribution $a^{{\rm HVP-LO}}_\mu$, approximately \( 70 \times 10^{-9} \), labeled 
``HVP-LO'' in Figure~\ref{amu22}.

\begin{table}[h]
\centering
\begin{tabular}{lcccccc}
\toprule
\textbf{Experiment} & \textbf{Positrons counted} & \textbf{Stat. error} & \textbf{Syst. error $\omega_a$} & \textbf{Syst. error $\omega_p$} & \textbf{Total syst. error} & \textbf{Total error} \\
                    & ($\times 10^9$)            & [ppm]           & [ppm]                      & [ppm]                      & [ppm]                  & [ppm]            \\
\midrule
CERN-III & 0.134 (E $ > 0.7$ GeV)     & 7.0    & $<1$   & 1.5   & 2.0   & 7.3 \\
E821     & 8.55 (E $ > 1.8$ GeV)    & 0.46    & 0.21    & 0.17   & 0.28   & 0.54 \\
E989     & 143 (E $ > 1.7$ GeV)  & 0.098     & 0.053     & 0.057    & 0.081    & 0.127 \\
\bottomrule
\end{tabular}
\caption{Summary of the number of positrons analyzed and the associated statistical and systematic uncertainties for the CERN-III, E821, and E989 experiments. The uncertainties refer to the extracted value of the muon anomaly \(a_\mu\). Systematic uncertainties on \(\omega_a\) and \(\omega_p\) are weighted-averaged over the years of data-taking as obtained by the authors of this review. The \(\omega_a\) uncertainty includes beam dynamics corrections (such as electric field and pitch effects). Total systematic uncertainty also includes contributions from external parameters and fundamental constants ({\it e.g.} \SI{23}{ppb} for E989).
}
\label{tab:experiment_summary}
\end{table}

\subsection{The E821 experiment at Brookhaven}\label{sec:e821}

\begin{figure}[!htb]
    \centering
    \begin{subfigure}[b]{0.4\textwidth}
        \includegraphics[width=\textwidth]{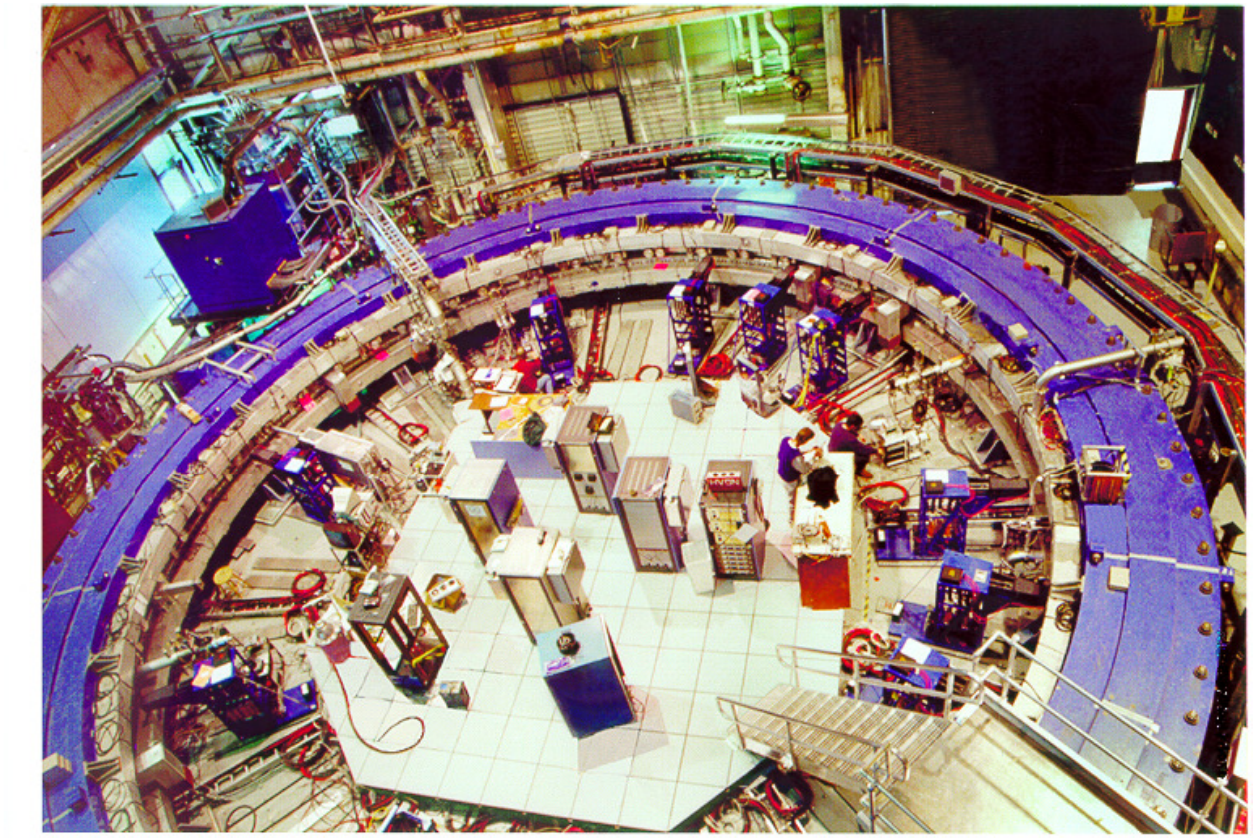}
        \caption{} 
    \end{subfigure}
    \hspace{0.05\textwidth} 
    \begin{subfigure}[b]{0.4\textwidth}
        \includegraphics[width=\textwidth]{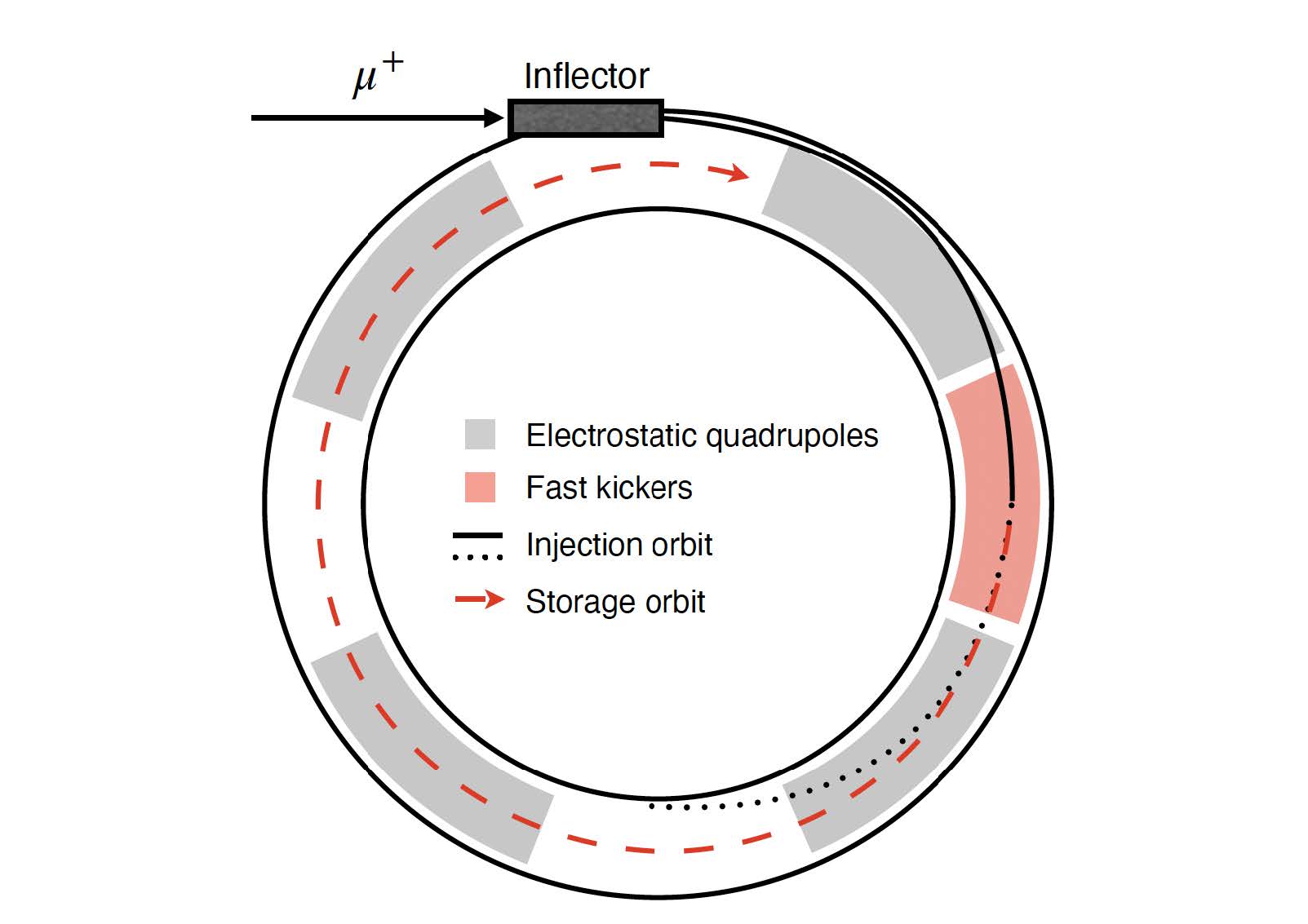}
        \caption{} 
    \end{subfigure}
    \caption{(a) Overview of the E821 storage ring at BNL (photo courtesy of R. Bowman). (b) Sketch of the muon beam injection system (from~\cite{Keshavarzi:2021eqa}).}
    \label{fig:bnl}
\end{figure}

Compared to what had been available at CERN, the Alternating Gradient Synchrotron (AGS) at the Brookhaven Laboratory produced 200 time the number of protons from which to make pions that decayed into muons. This provided the possibility to greatly reduce the statistical uncertainty that had limit the CERN experiments, if the magnetic field could be measured precisely enough.  Meanwhile, impressive theoretical progress had been made \cite{Kinoshita:1981vs,Kinoshita:1983xp,Kinoshita:1984it}.  Numerical SM field theory calculations were being extended into the 8th order,  $C_8 (\alpha/\pi)^4$ in Equation \ref{eq:SMTheory}. A large effort had gone into evaluating $a_\mu^{\rm{had}}$.  The $a_\mu^{\rm{HVP-LO}}$ contribution could not be entirely done from first principles, but it seemed possible to use measurements of cross sections at low-energy electron-positron colliders to deduce the SM prediction.  There was good reasons to test this evaluation of $a_\mu^{\rm{had}}$, and also to probe the weak interaction contribution, $a_\mu^{\rm{weak}}$ for the first time. A 20 times improved measurement would test these predictions, and look for potential new-physics such as would arise in supersymmetric additions of the SM~\cite{Martin:1997ns,Stockinger:2006zn}.

The $14$-m in diameter  storage ring built at Brookhaven for the E821 measurement (Figure~\ref{fig:bnl}a) improved upon the design that added a focusing electric field to the magnetic field of the storage ring for CERN-III \cite{CERN-Mainz-Daresbury:1978ccd}.  It also used the ``magic momentum'' to largely ameliorate the shifts of the observed anomaly frequency due to this electric field.  Instead of directly injecting pions, as in CERN-III, muons from pion decay in a $80$-m channel upstream were injected directly in the storage ring (Figure~\ref{fig:bnl}b) using a superconductive inflector magnet and three pulsed kickers.  To improve the homogeneity of the magnetic field to 1 ppm, three continuously wound superconducting coils were used instead of CERN-III's 40 segmented magnets. The $1.45\,$T field was extensively mapped with pulsed NMR probes \cite{Muong-2:2006rrc}.  A small trolley  carried $17$ probes along the muon beam path inside the storage ring, and was parked out of the muon beam inside the ring when muons were present.  In addition, $378$ fixed NMR probes continuously monitored the field homogeneity and stability.

\begin{figure}[h!]
\begin{center}
\subfloat[]{
\includegraphics[width=0.48\textwidth,angle=0]{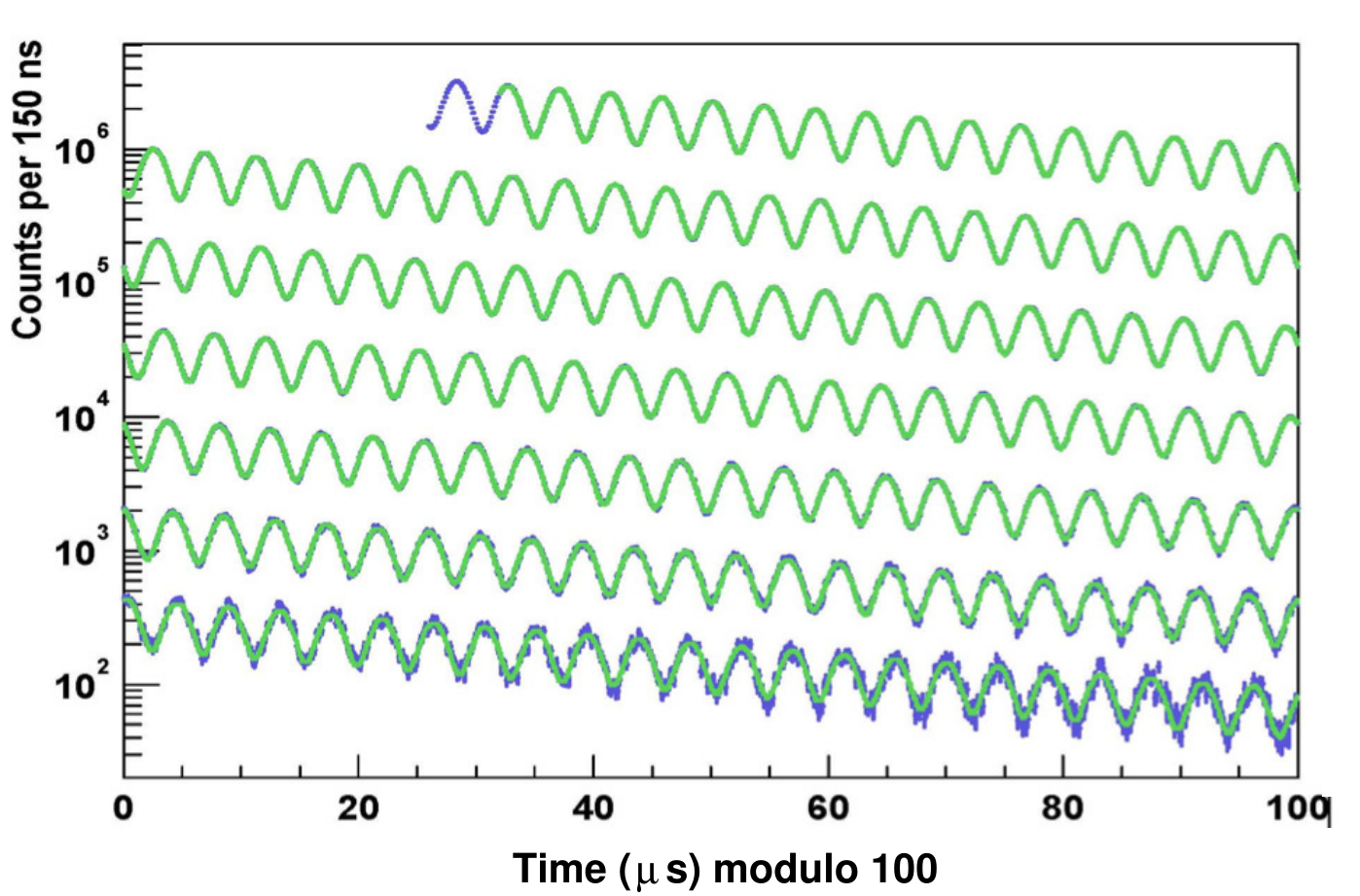}}
\subfloat[]{
 \includegraphics[width=0.32\textwidth,angle=0]{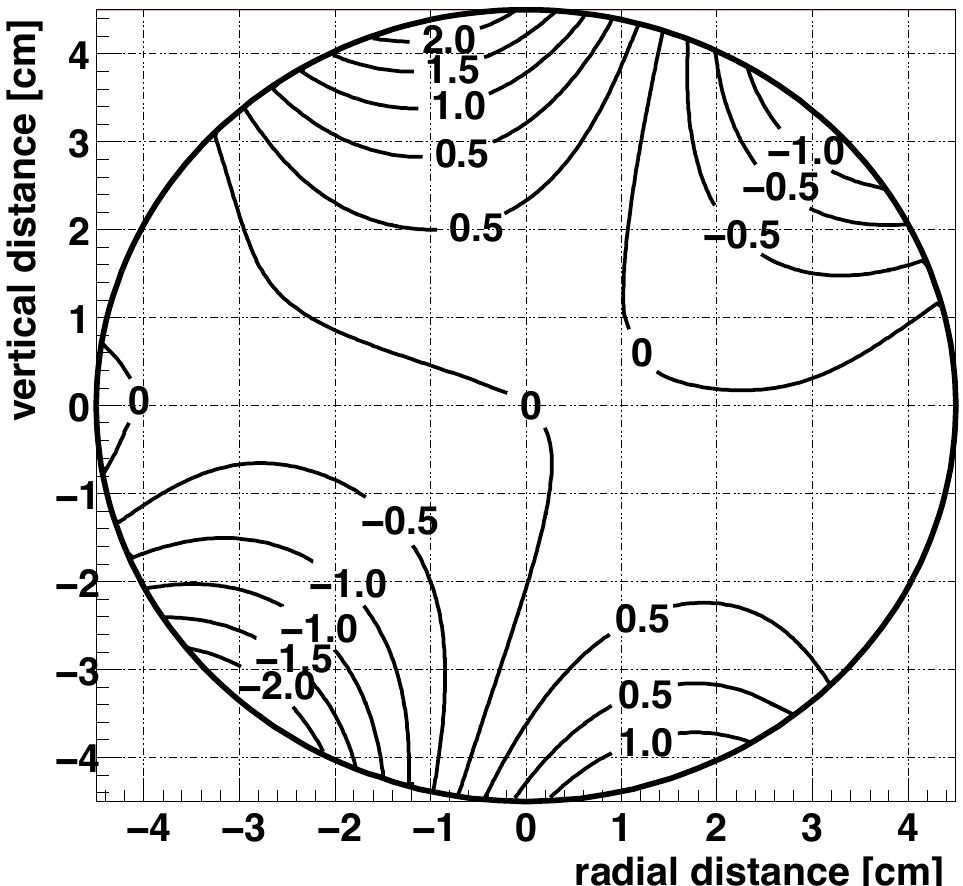}}
\caption{
(a) Positron arrival time spectrum from E821~\cite{Muong-2:2004fok}, corresponding to approximately \(3 \times 10^9\) positrons with energy \(> 1.8\,\mathrm{GeV}\) (from~\cite{Roberts:2018vsx}).  
(b) The magnetic field averaged over the full circumference of the storage ring from E821. The contour lines are in ppm. From~\cite{Muong-2:2006rrc}.
}
\label{fg:BNL-spectrum}
\end{center}
\end{figure}

 Decay positrons were detected using lead-scintillator calorimeters. The recorded positron count in Figure~\ref{fg:BNL-spectrum}a followed the characteristic oscillation pattern, showing a clear statistical improvement compared to CERN-III (Figure~\ref{fig:CERN-III}b). 
As the statistics increased, the fitting function had to account for a number of additional effects, including detector effects, muon losses, and the beam motion in the ring.
 The anomalous precession frequency $\omega_a$ was then extracted from a fit to the wiggles with average systematic uncertainty below \SI{0.3}{ppm}. 
 The field averaged over the full circumference of the storage ring, shown in Figure~\ref{fg:BNL-spectrum}b, was known with a systematic uncertainty below \SI{0.2}{ppm}~\cite{Muong-2:2006rrc,Carey:2009zzb}.
Because the positive and negative muon results agreed within their uncertainties, in accordance with CPT symmetry invariance, the two results were combined to give  an uncertainty of \( 63 \times 10^{-11} \) (\SI{0.54}{ppm}).  The statistical uncertainty was \SI{0.46}{ppm} and the systematic one was \SI{0.28}{ppm}.  Table~\ref{tab:experiment_summary} gives more details, and compares these to the CERN-III uncertainties.

\begin{figure}[h!]
\begin{center}
\subfloat[]{
 \includegraphics[width=0.9\textwidth,angle=0]{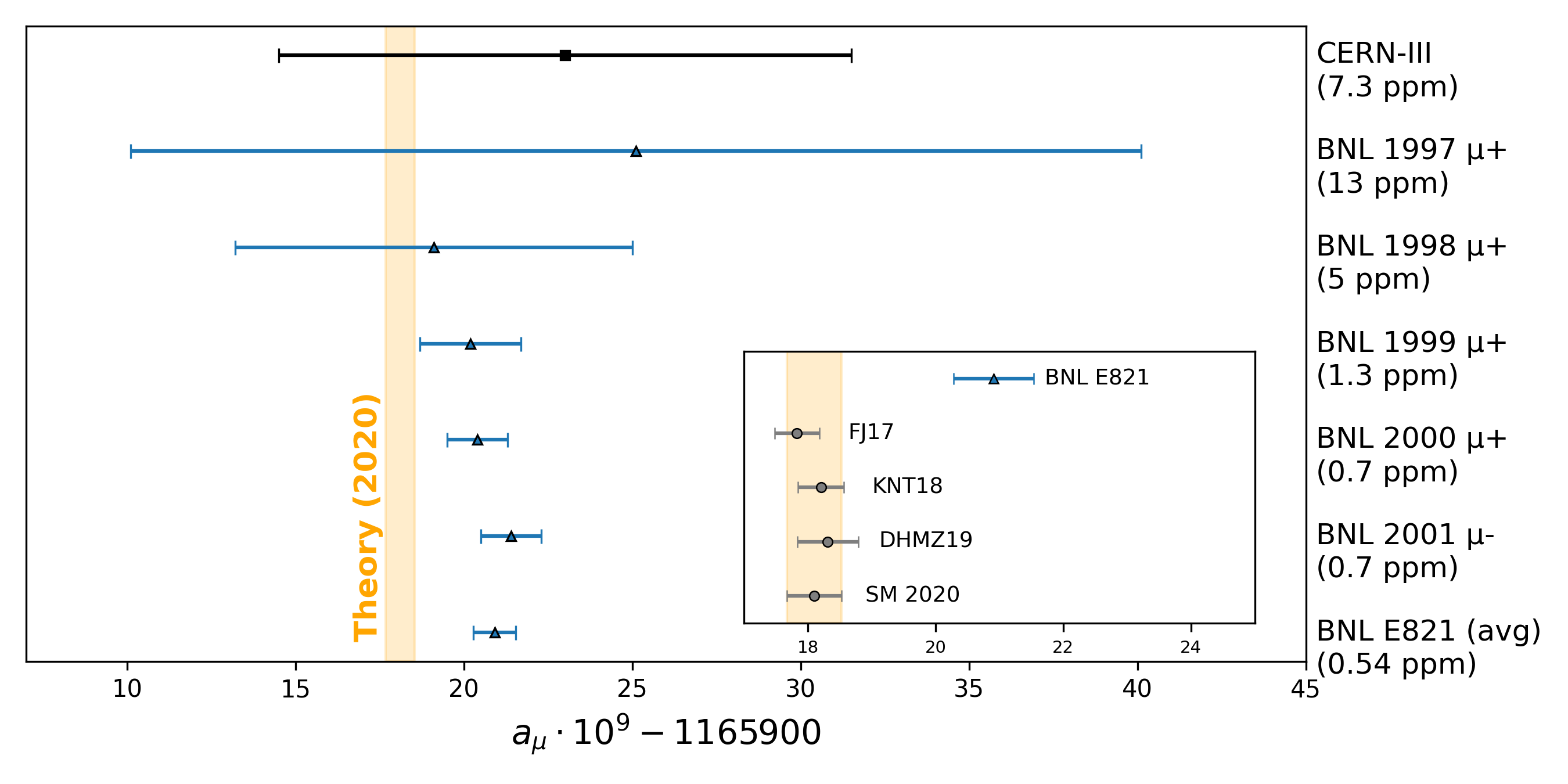}}
\caption{E821 result for \(a_{\mu}\) compared to theoretical predictions available at the time and to earlier CERN measurements. The points labeled CERN and BNL represent experimental measurements, while FJ17~\cite{Jegerlehner:2017gek}, KNT18~\cite{Keshavarzi:2018mgv}, DHMZ19~\cite{Davier:2019can}, and SM 2020~\cite{Aoyama:2020ynm} correspond to theoretical predictions. The orange band shows the 2020 Standard Model (SM) prediction, where the width of the band  represents the theoretical uncertainty of the prediction.
        Error bars indicate measurement uncertainties for experimental points.
}
\label{fg:BNL-TH}
\end{center}
\end{figure}

 Figure \ref{fig:muon_g_minus_2_comparison} shows the 14 times improved precision in the E821 final result \cite{Muong-2:2006rrc} compared to that from CERN, and the good agreement between the two measurements.  Figure~\ref{amu22} illustrates how the extent to which the measurements test the 8th order QED contribution $ C_8 \left( {\alpha}/{\pi} \right)^4 \sim 381 \times 10^{-11} $
in Equation \ref{eqQED}, the electroweak contributions  (EW1 and EW2 for a total contribution of $\sim 154 \times 10^{-11} $), and the hadronic contributions (HVP-(N)NLO and HLBL approximately $-87 \times 10^{-11}$ and $110 \times 10^{-11} $).  
Great excitement was generated by this measurement, and earlier Brookhaven measurements leading up to it that are shown with the CERN-III measurement in Figure~\ref{fg:BNL-TH}.  The reason was the evident, 3.7 standard deviation disagreement 
\begin{equation}
    \Delta a_{\mu} = (279\pm76)\times 10^{-11},
\end{equation}
with the best SM theory that was available as late as 2020 \cite{Aoyama:2020ynm}.  
This discrepancy stood for 20 years  as a potential indication of physics beyond the Standard Model.

\subsection{The E989 experiment at Fermilab}
\label{sec:e989}

\begin{figure}[!htb]
    \centering
    \includegraphics[width=.45\textwidth]{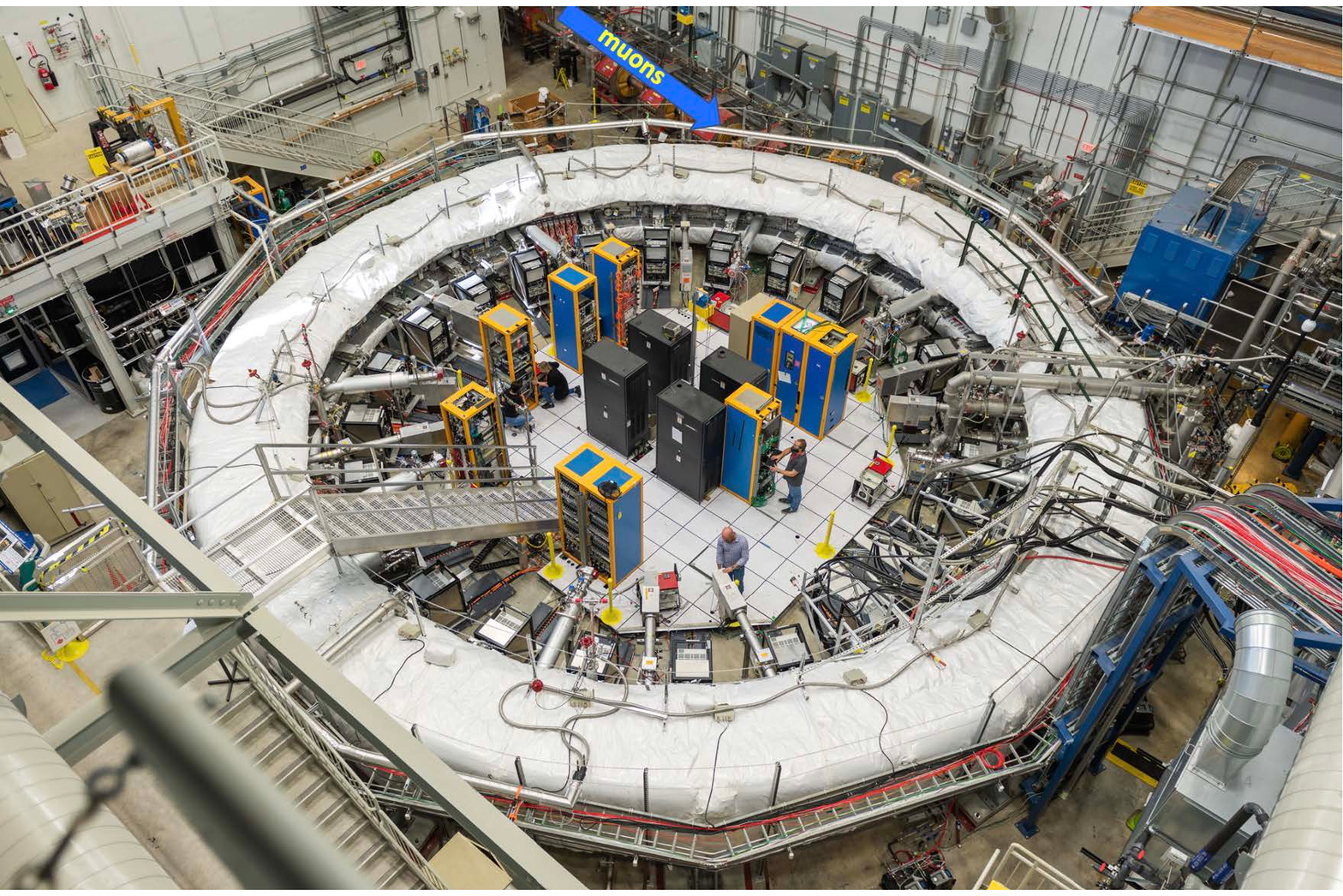}
    \caption{Aerial view of the experimental hall at Fermilab showing the muon injection and the storage ring covered with the thermal insulating blanket. Credit: Fermilab.}
    \label{g-2_ring}
\end{figure}

\noindent{\bf Epic journey of an intact storage ring}
\smallskip

The clear discrepancy between the Brookhaven measurements of the muon magnetic moment and the SM prediction that caused excitement starting in about 2000 was intriguing. Was this the first evidence of a breakdown in the SM, or was there some measurement or theory problem? It was extremely important to verify and even improve upon this result. Fermilab near Chicago had a muon source that could provide 20 times the muons that were available for the last Brookhaven measurement to reduce the statistical uncertainty by a factor of $\sqrt{20}=4.5$.  
However, the 15-meter wide storage ring that could be used to make a magnetic moment measurement was 1/3 the way across a continent ``as the crow flies".  A very expensive storage ring that could not be dismantled without destroying its superconducting magnet coils, and which was not designed to be moved or even stressed at all, cannot fly like a crow nor fit within any available aircraft.

\begin{figure}[!htb]
    \centering
    \begin{subfigure}[b]{0.45\textwidth}
        \includegraphics[width=\textwidth]{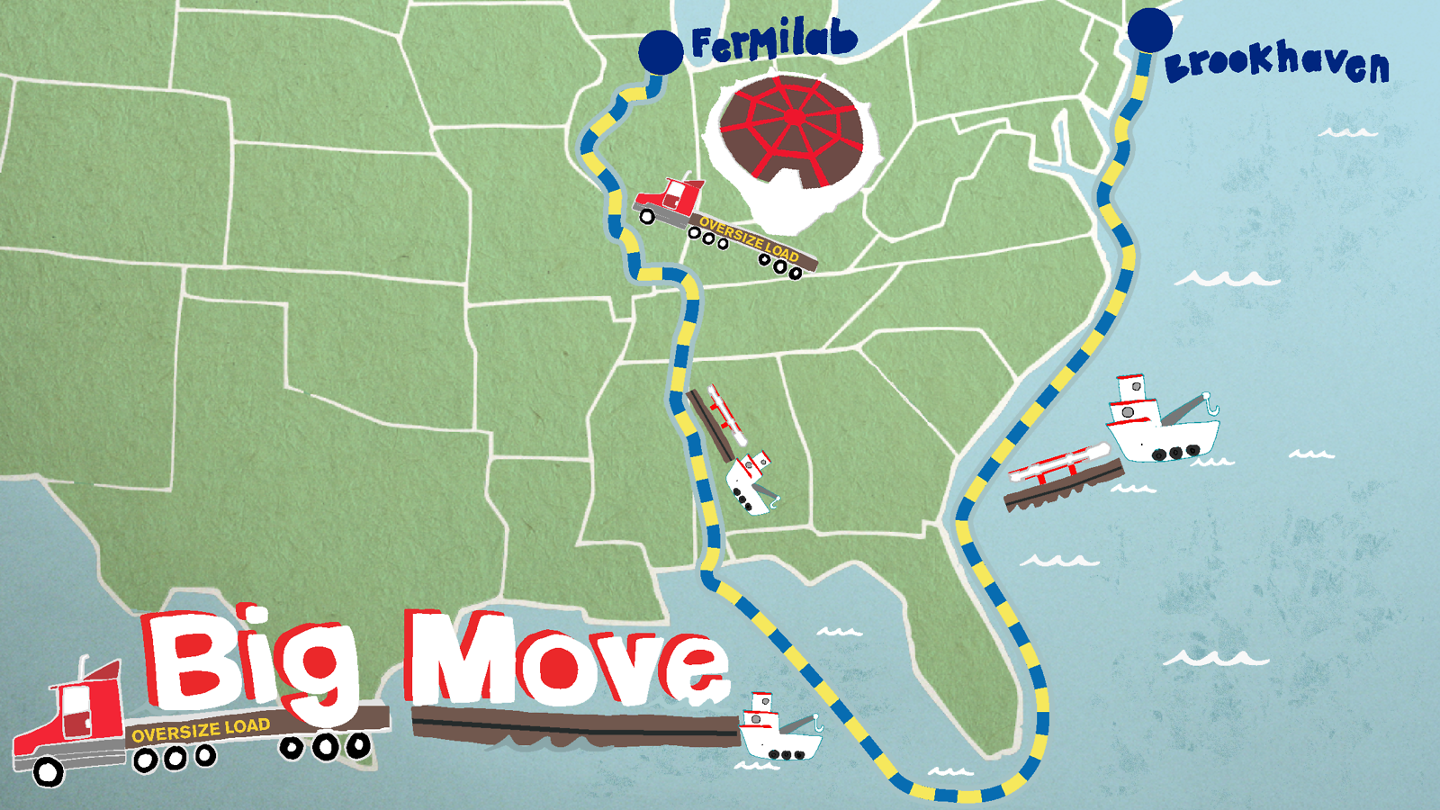}
        \caption{}
    \end{subfigure}
    \hspace{0.05\textwidth}
    \begin{subfigure}[b]{0.38\textwidth}
        \includegraphics[width=\textwidth]{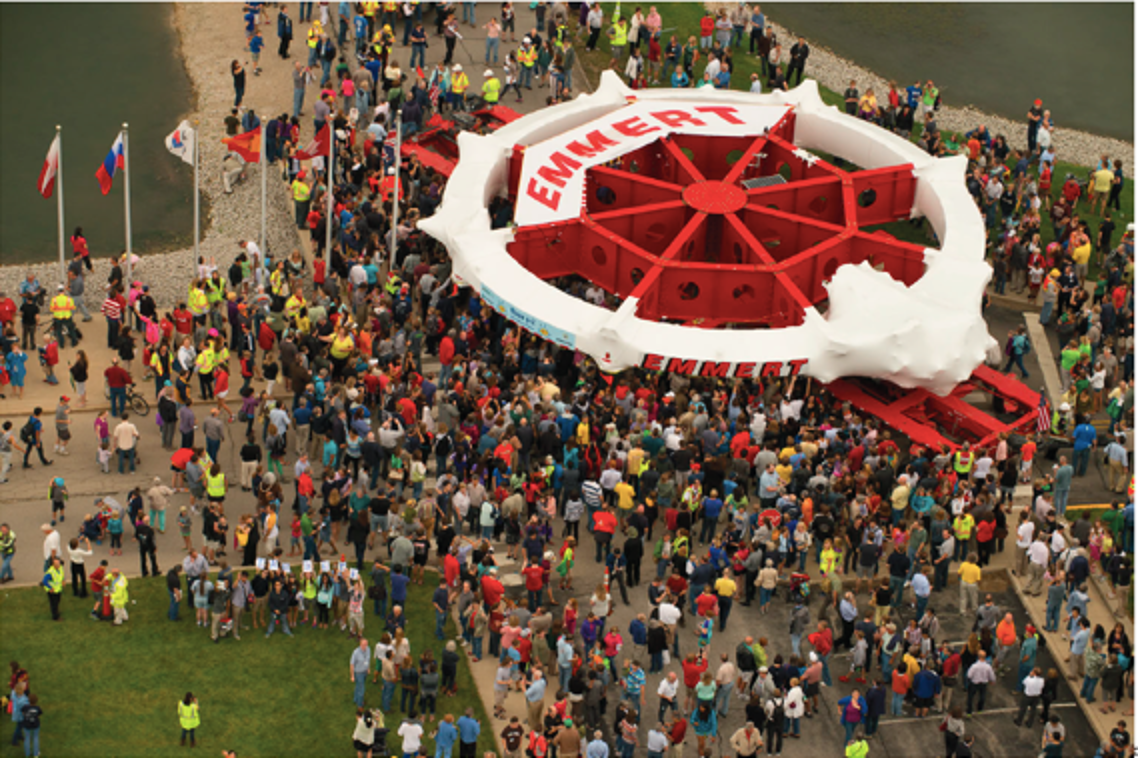}
        \caption{}
    \end{subfigure}
    \caption{(a) A ``5,000-kilometer journey of science''  map showing the route taken by the muon $g-2$ storage ring in 2013: from BNL in New York by road to the coast, then by barge along the Atlantic Ocean, around Florida, up the Mississippi River, and finally by truck again to Fermilab in Illinois. (b) Hundreds crowd around the Muon $g-2$ magnet during the celebration on Friday, July 26, 2013 commemorating the end of the Big Move. Credit: Fermilab.}
    \label{fig:bigm}
\end{figure}

The epic move of an intact storage ring  was closely monitored by many over the internet. This ``Big Move'' was more than a month-long logistical feat in which the delicate storage ring that was 15-m in diameter and weighed 17 tons was moved 3100 miles by land, sea and river (Figure~\ref{fig:bigm}a).  It was carefully lifted onto a custom-built truck, driven at walking pace to a port, then loaded onto a barge that navigated the Atlantic coast, the Gulf of Mexico, and the Mississippi River~\cite{FNALBigMove2019}. Remarkably, the ring arrived intact and in excellent condition at Fermilab after more than a month of travel (Figure~\ref{fig:bigm}b), where it was reassembled, powered, and cooled to a cryogenic temperature~\cite{FNAL2015} (Figure~\ref{g-2_ring}).

\bigskip
\noindent{\bf Measurement of the ratio of the anomalous precession frequency and the proton NMR frequency}
\smallskip

Fermilab delivered the muons needed for a 22 times larger data set for the muon magnetic moment measurement.  Compared to BNL, the muon injection rate was higher (\SI{11.4}{Hz} versus \SI{4.4}{Hz} at BNL) and a better beam quality provided a higher muon storage efficiency.  Muons were also available for 30 months~\cite{Muong-2:2024hpx} rather than 6 months as in BNL~\cite{Carey:2009zzb}.  The \SI{\sim 2500}{m} pion decay channel (compared to \SI{80}{m} at BNL) allows most pions to decay before reaching the storage ring.   The ``pion flash'' background and the resulting systematic uncertainties it caused was thereby  reduced by about a factor of 20 compared to at BNL.  The  E989 experiment at Fermilab \cite{Muong-2:2025xyk} reduced the statistical uncertainty by a factor of 4.7.   Table~\ref{tab:experiment_summary} gives more details on the uncertainties reached in E989 compared with the previous CERN and E821 experiments.

The detection of decay positrons was carried out 
using 24 highly segmented Cherenkov calorimeters  located around the inner circumference of the storage ring~\cite{Kaspar:2016ofv,Anastasi:2016luh,Muong-2:2019beb}. Each included a \(6 \times 9\) array of lead fluoride (PbF$_2$) crystals read out by silicon photomultipliers (SiPMs) with stabilized gain, offering excellent spatial granularity without perturbing the magnetic field.  
Compared to the monolithic detectors in E821, the ability to resolve overlapping events (pileup) was greatly improved. With greatly improved timing and data readout, a GPU-based real-time system processed these data, isolating relevant positron pulses for further analysis.   

\begin{figure}[h!]
\begin{center}
\subfloat[]{
\includegraphics[width=0.42\textwidth,angle=0]{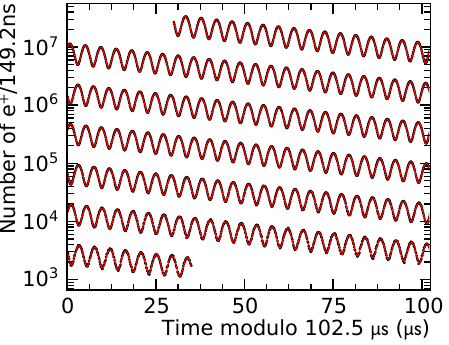}}
\subfloat[]{
 \includegraphics[width=0.4\textwidth,angle=0]{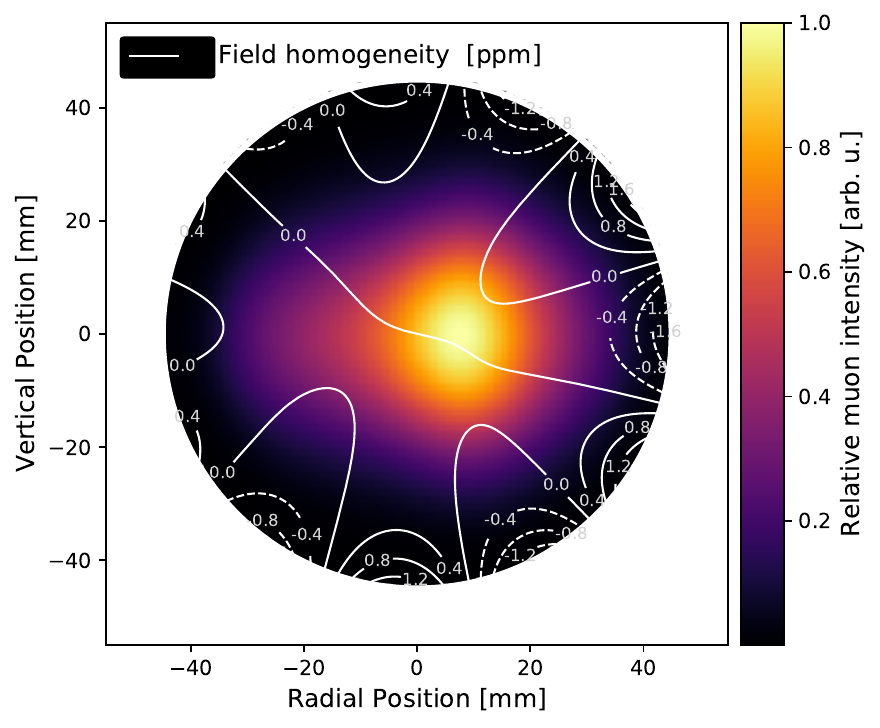}}
\caption{
(a) Example of a ``wiggle plot'' for a subset of E989 data (corresponding to about $32\times 10^9$ positrons with energy $> 1.7$ GeV). (b) Example of a magnetic field contours averaged over the full circumference of the storage ring, weighted by the muon orbit distribution. From~\cite{Muong-2:2023cdq,Muong-2:2024hpx}.
}
\label{fig:FNAL-spectrum}
\end{center}
\end{figure}

An example of a measured ``wiggle plot'' is in Figure \ref{fig:FNAL-spectrum}a.  The anomalous precession frequency $\omega_a$ is determined by analyzing the positrons detected in the calorimeters above a threshold energy, the optimal determined to be 1.7 GeV, during  700~$\mu$s after each injection \cite{Muong-2:2021vma,Muong-2:2024hpx}.
As for E821, the ideal anomaly oscillation as a function of time in Equation~\ref{eqw1} had to be corrected for detector and beam dynamics ~\cite{Muong-2:2021ojo,Muong-2:2021xzz,Muong-2:2021vma,Muong-2:2023cdq,Muong-2:2024hpx,Muong-2:2025xyk}. 
Examples included
\begin{itemize}

\item Muons injected with a radial displacement and a range of momenta undergo oscillations about their equilibrium paths at calculable frequencies.

\item Muon losses and systematic deviations arise from electrostatic quadrupoles and beam motion parallel to the magnetic field.

\item Pileup events (defined as two or more particles which impact in the calorimeter so close in time that
only one hit is recorded) and resulting variations in calorimeter gain that can bias the extracted $\omega_a$ due to their impact on positron energy distributions.
    
\end{itemize}
To ensure the accuracy and internal consistency of the $\omega_a$ analysis, independent teams employed distinct methods to fit the data after each applying corrections for systematic effects. Each analysis team was blind to the exact value of their result and to the values obtained by other teams.

The analysis made use of two straw-tracker detectors located inside the vacuum chamber which provided the decay positron trajectories from the muon vertex decay and positron momentum~\cite{King_2022}. Each tracker consisted of 32 planes of straw-tube detectors filled with an Ar-Ethane gas mixture. Tracker data refine the extraction of the anomalous precession frequency $\omega_a$ by providing accurate knowledge of the stored muon distribution and enabling improved corrections for electric field, pitch and other beam dynamics effects (see Section~\ref{expmet}). 
      
The magnetic field in the storage ring was carefully shimmed to reduce RMS azimuthal variations from 240 ppb (E821) to 70 ppb (E989)~\cite{Muong-2:2021ovs}. The field was frequently mapped using a movable trolley equipped with 17 Nuclear Magnetic Resonance (NMR) probes, which scans the ring approximately every few days, for more than 500 trolley runs in total. Between trolley scans, an array of 378 fixed NMR probes continuously monitors field stability. 
The  uncertainty in the angular proton NMR frequency $\omega_p$ was thereby reduced  from 170 ppb in E821 to about 55 ppb at E989 (Table~\ref{tab:experiment_summary}).

What was actually measured was the ratio of the anomaly frequency and the proton NMR frequency, $\omega_a/\omega_p$.  This ratio was  averaged to correct for the measured gradients in the magnetic field within the storage ring and the measured muon beam orbit distribution.  Corrections were also made for transient magnetic field perturbations that are synchronous with the muon injection \cite{Muong-2:2021ojo,Muong-2:2023cdq,Muong-2:2024hpx}.

\bigskip
\noindent{\bf The measured muon anomaly $a_\mu$ }
\label{expmet}
\smallskip

The muon anomaly is again determined using $a_{\mu}  =  \omega_a ({m_\mu}/ {eB})$ from 
equation~\ref{eq:MuonAnomalyFrequency}.  Because the Fermilab measurement determines $\omega_a/\omega_p$ rather than ${m_\mu}/ {eB}$, the anomaly is written in terms of this ratio and three ratios measured in other experiments, 
\begin{equation}
a_{\mu}  = 
 \frac{\omega_a}{\omega_p} \frac{g_e}{2}
\frac{m_\mu}{m_e} \frac{\mu_p}{\mu_e}.
\label{eq:MuonaDetermineation}
\end{equation}
The energy of the proton magnetic moment $\mu_p$ is  $\,\hbar \omega_p = 2 \mu_e B$. The measured dimensionless magnetic moment magnitude, $g_e/2$,   (from Section \ref{sec:Electron} and Equation \ref{eq:geOver2} \cite{NorthwesternMagneticMoment2023}) has  a 0.26 parts per trillion uncertainty -- small enough to not contribute uncertainty to $a_\mu$.  The ratio of the muon and electron masses, $m_\mu/m_e$,  determined from muonium hyperfine splitting~\cite{Liu:1999iz} is determined to 22 parts per billion~\cite{Mohr:2024kco}.  The ratio of the magnetic moments of the proton and electron, $\mu_\mu/{\mu_e},$ is determined from hydrogen spectroscopy to 4 parts per billion~\cite{Mohr:2024kco}.

\begin{figure}[htbp]
    \centering
    \includegraphics[width=0.9\textwidth]{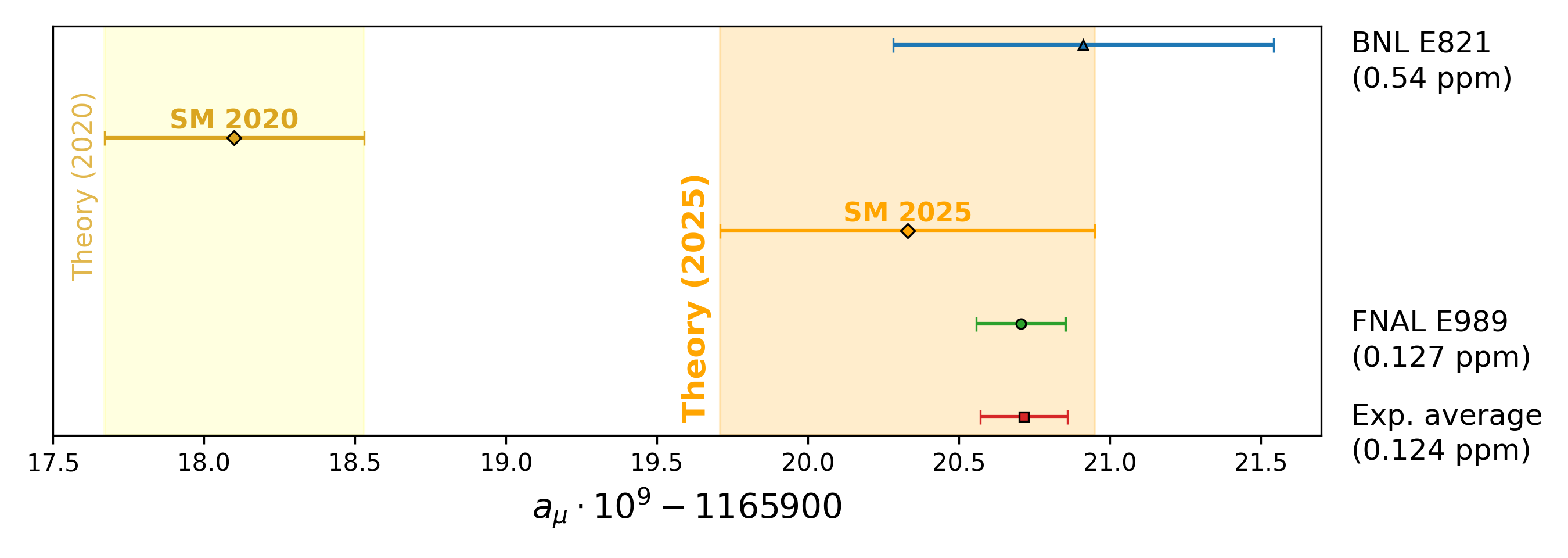}
    \caption{Comparison of the most recent measurements and theoretical predictions of the muon magnetic anomaly \( a_\mu \). 
        The horizontal axis shows the difference from a reference value ($1165900 \times 10^{-9}$), while each point represents a determination of $a_\mu$ with its associated uncertainty. The blue triangle corresponds to the BNL E821 experiment, while the green and red points represent the Fermilab E989 result and the combined experimental average, respectively. The orange and yellow diamonds correspond respectively to the Standard Model theory predictions 2025~\cite{Aliberti:2025beg}  and 2020~\cite{Aoyama:2020ynm}, with shaded bands indicating their theoretical uncertainties. As can be noted the SM 2025 prediction has moved significantly close to the experimental values.
    }
    \label{fig:muon_gm2_summary}
\end{figure}

\begin{figure}[htbp!]
\centering
\includegraphics[width=0.7\linewidth]{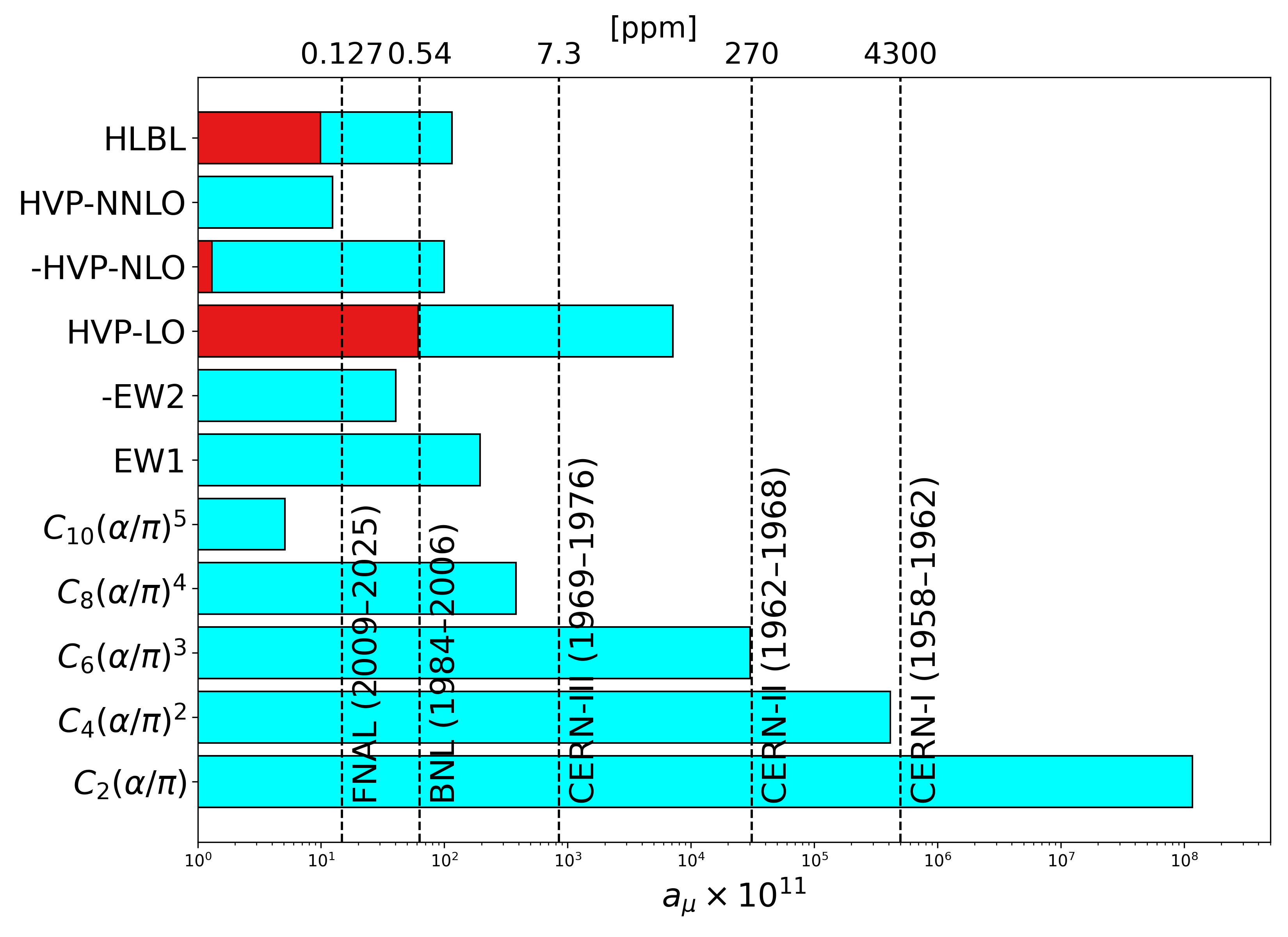}
\caption{%
Contributions to the the muon anomaly, $a_\mu$, expressed in units of $10^{11}$.
Each cyan bar represents the value of a Standard Model contribution—grouped as quantum electrodynamics (QED), electroweak (EW), and hadronic (HVP and HLBL)—with red overlays indicating the corresponding uncertainties. QED terms are labeled as $C_n (\alpha/\pi)^n$ for increasing loop order ($n = 1$ to $5$), see also Equation~\ref{eqQED}, while EW is divided into a leading order term (EW1) and higher-order corrections (EW2). The hadronic contributions are grouped by leading (LO), next-to-leading (NLO), and next-to-next-to-leading order (NNLO) corrections, including the hadronic light-by-light (HLBL) term. Dashed vertical lines indicate the uncertainties on the experimental measurements of $a_\mu$ from CERN (I–III), BNL E821, and the latest FNAL E989 experiment. The numbers shown above the dashed vertical lines indicate the relative experimental precision in parts per million.The logarithmic scale on 
the horizontal axis emphasizes the wide range of values across different contributions. Note that EW2 and HVP-NLO have negative values and therefore appear with negative sign. Theoretical values from~\cite{Aoyama:2020ynm,Aliberti:2025beg}.
}
\label{amu22}
\end{figure}

During six years of measurements, E989 at Fermilab used about 320 billion positrons with energy above 1 GeV to determine the muon anomaly 
\begin{equation}
a_\mu^{\text{E989}} = 116\,592\,0705 \,(114)_{stat} \,(95)_{syst} \,(148)_{tot}\times 10^{-12}.
\end{equation}
This \SI{0.127}{ppm} measurement\footnote{\SI{0.127}{ppm} on $a_\mu$ corresponds to a precision of \SI{0.148}{ppb} on $g_\mu$.} is the green point in Figure \ref{fig:muon_gm2_summary}.  It agrees well with the Brookhaven result that is the blue point in the figure.  The weighted average of these two measurements, the red point in the figure,
is not meaningfully different.  Figure~\ref{amu22} illustrates how deeply each of the SM contributions is tested. This measurement precision now approaches a precision approximately one-tenth of the electroweak contribution which enables a meaningful comparison with subtle higher-order SM contributions.
From CERN-I to E989, the precision on \(a_{\mu}\) improved by approximately 4.5 orders of magnitude, making it one of the most precise tests of the Standard Model of particle physics and a stringent test for possible physics that may be missing from the SM.

The Fermilab measurement remains in disagreement with the SM prediction compiled by the large Muon g-2 Theory Initiative~\cite{Aoyama:2020ynm} labeled SM 2020 in Figure \ref{fig:muon_gm2_summary}.  This prediction relied upon measured cross sections at electron-positron colliders operating in the energy region up to 10 GeV, to evaluate the hadronic contribution which could not be calculated from first principles. However, in a surprising recent development, a lattice gauge evaluation of the hadronic contribution~\cite{Borsanyi:2020mff}, supported by a new cross-section result~\cite{CMD-3:2023alj,CMD-3:2023rfe}, agree with the muon $g-2$ measurement.  
Although the theoretical uncertainty is currently four times larger than the experimental error, ongoing and planned theoretical efforts, including novel approaches for determining $a_\mu^{\rm HVP-LO}$~\cite{CarloniCalame:2015obs,MUonE:2016hru}, show promise for a reduction to the level of the experimental uncertainty in years~\cite{Aliberti:2025beg}.

.

\subsection{Proposed J-PARC Muon g-2/EDM experiment}

\begin{figure}[h]
\centering
\begin{subfigure}[b]{0.5\linewidth}
    \includegraphics[width=\textwidth]{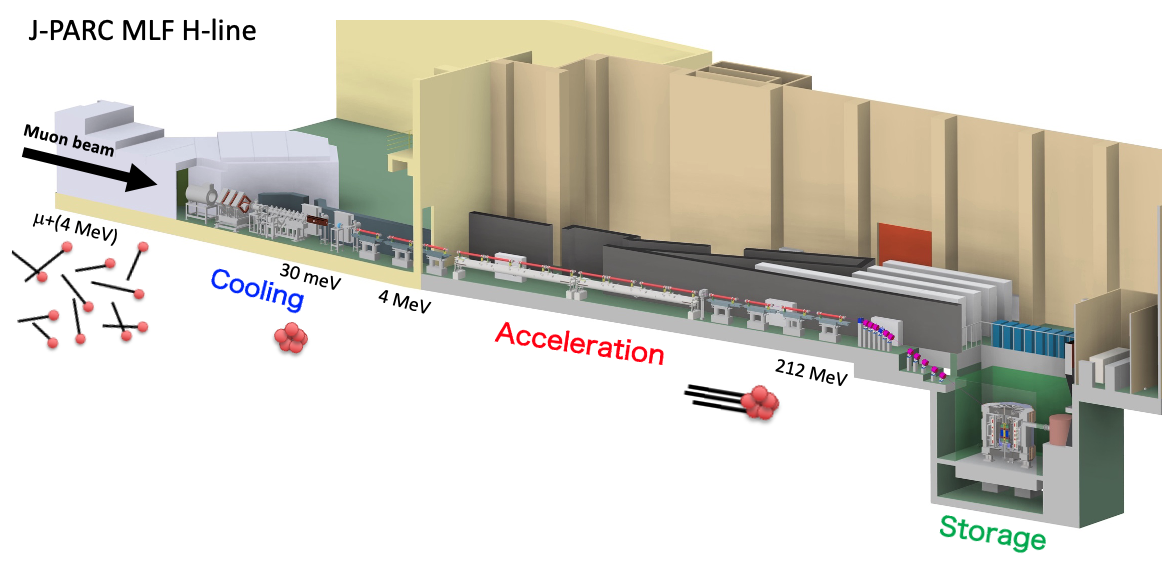}
    \caption{}
\end{subfigure}
\hspace{0.05\linewidth}
\begin{subfigure}[b]{0.35\linewidth}
    \includegraphics[width=\textwidth]{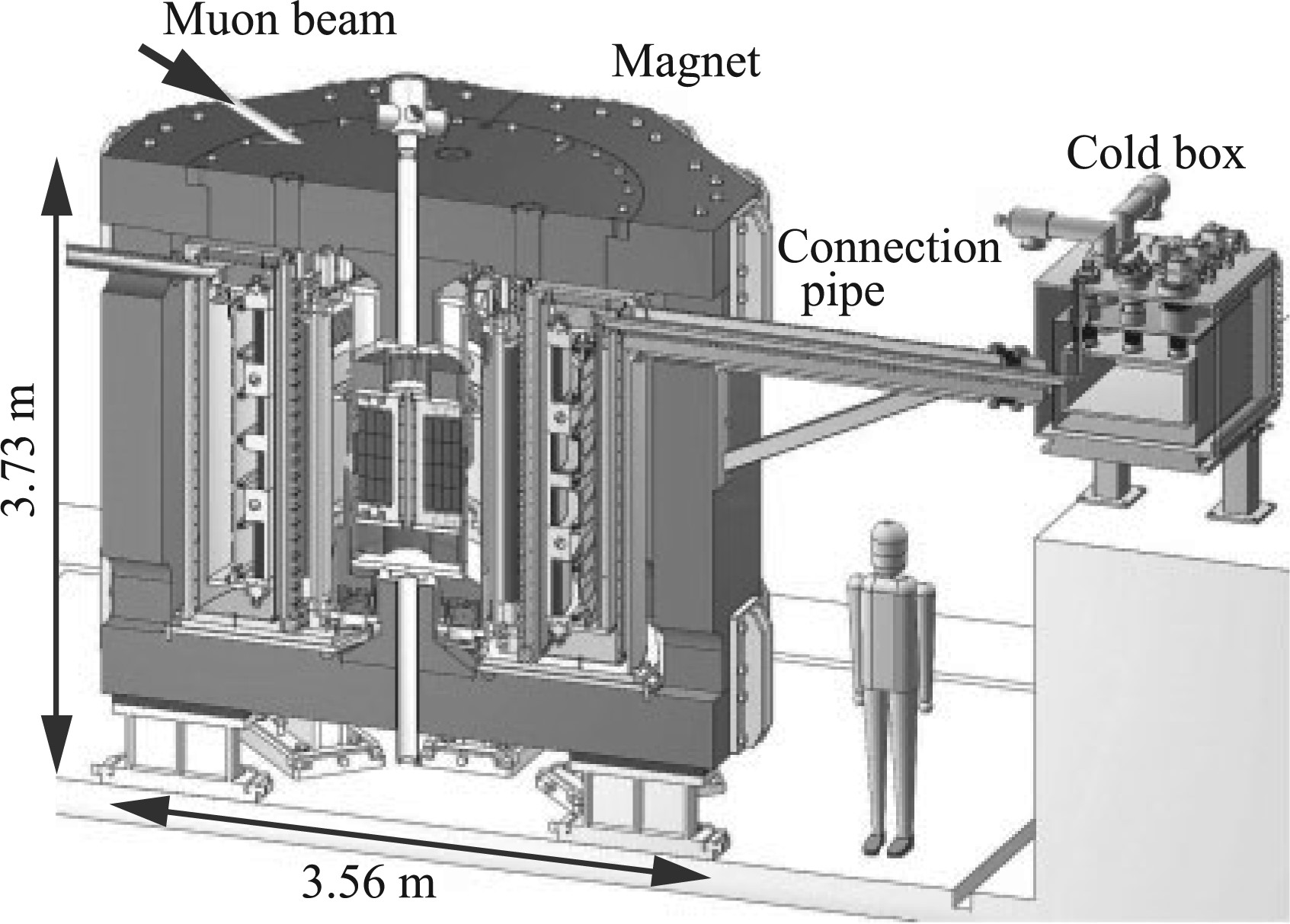}
    \caption{}
\end{subfigure}
\caption{(a) Schematic view of the accelerator complex for the muon $g-2$/EDM experiment at J-PARC (courtesy of T. Mibe). (b) Overview of the muon storage magnet (from~\cite{10.1093/ptep/ptz030}).}
\label{fig:jparc}
\end{figure}

The Muon g-2/EDM experiment at the Japan Proton Accelerator Research Complex (J-PARC) is scheduled to be commissioned around 2030.  It is designed to reach a systematic uncertainty comparable to that of the Fermilab (FNAL) experiment, while employing a very different  experimental approach~\cite{10.1093/ptep/ptz030,Abe:2018tmp, Iinuma:2016zfu,Otani:2016swo,Aritome:2024jiv,10.1093/ptep/ptac059,PhysRevLett.74.4811,BAKULE2008335,Zhang:2022ilj,Nakazawa:2022yae,Shimomura:2024puh,Kawamura:2018apy,10.1093/ptep/ptu116,10.1093/ptep/ptaa145,10.1093/ptep/ptt080,PhysRevAccelBeams.21.050101}.  (Another goal is a measurement of the muon electric dipole moment, EDM).  A 3-T magnetic field with \SI{1}{ppm} local magnetic field uniformity from a superconducting MRI-type magnet 
(Figure~\ref{fig:jparc}b) for the muon storage region with an orbit diameter of 66 cm, will be used with no electric field for focusing.  This allows for a simplified interpretation of the spin precession and no need to rely upon the ``magic momentum'' \cite{10.1093/ptep/ptz030}, and a different set of systematic uncertainties. $g-2$ oscillations in the decay of stored muons (\(\mu^+ \to e^+ \nu_e \bar{\nu}_\mu\)) are observed with 40 detector vanes of silicon strip sensors that achieves over 90\% efficiency in reconstructing the trajectories of decay positrons with energies between 200 and 275 MeV. The muon anomaly will then be determined from the measured anomaly frequency and the magnetic field determined by an NMR probe as has been discussed for earlier experiments.

A 300 MeV/c muon beam is produced using new methods.  A 3 GeV proton beam (Figure~\ref{fig:jparc}a) is used to generate a surface muon beam with a momentum of 27 MeV/c, which is directed onto a silica aerogel target\footnote{A laser-ablated silica aerogel target is employed to achieve an order-of-magnitude increase in muonium production.}. There, muonium atoms (\(e^-\mu^+\)) are formed and subsequently ionized using Lyman-\(\alpha\) laser pulses. This process produces thermal muons with a momentum of 2.3 keV/c and extremely low emittance.
These thermal muons are reaccelerated to a momentum of 300 MeV/c using a linear accelerator system composed of a radio-frequency quadrupole (RFQ), an interdigital H-type drift tube linac (IH-DTL), a disk-and-washer structure (DAW), and a disk-loaded traveling wave structure (DLS). The reaccelerated muons are then directed to (MRI)-type solenoid storage ring with the  high uniform magnetic field which contains the silicon strip sensors detector to measure positron tracks from decay of the stored muon beam.  A first demonstration of the acceleration of muons from thermal energy to 100 keV has been achieved~\cite{Aritome:2024jiv}.

\section{Tau magnetic moment}\label{sec:Tau}
Measurement of the magnetic anomaly of the $\tau$ lepton $a_\tau$ is particularly challenging because of its extremely short lifetime (\(2.9 \times 10^{-13}\) s), which prevents direct observation of spin precession in penning traps or storage rings, as done for electrons and muons. Consequently, indirect methods have been developed to estimate its value, primarily
through high-precision studies of tau pair production in high-energy collisions, comparing the observed cross sections with Standard Model predictions. Experiments such as those conducted at LEP and LHC (CERN) have analyzed the production rate and angular distribution of $\tau$ lepton pairs, as the magnetic moment influences these observables through virtual-photon interactions. Similarly, $\tau$-lepton decay studies, particularly those involving muons and neutrinos, provide another avenue for estimating possible deviations from the Standard Model prediction.  
Over time, these indirect approaches have led to increasingly precise limits on $a_\tau$. Until recently, the best limit came from earlier experiments at LEP~\cite{DELPHI:2003nah}. Newer studies at the LHC, including a recent CMS measurement, have improved the precision by a factor of five ($-0.0042< a_\tau < 0.0062$ at 95\% CL~\cite{CMS:2024qjo}), although the result is still a few times less precise than the Standard Model prediction, $a_\tau = 1.2 \times 10^{-3}$~\cite{Eidelman:2007sb}. Current and future efforts at the LHC, flavor factories, and possible future colliders may improve the experimental limits to a level sensitive enough to potentially reveal new physics (for a review of the activities see ~\cite{Crivellin:2021spu} and the references therein).

%% file: neutrino.tex
\section{Neutrino magnetic moment}

In the Standard Model extended to include non-zero neutrino masses, the neutrino magnetic moment arises at one-loop level and is directly proportional to the neutrino mass. For a Dirac neutrino, the expected magnetic moment is~\cite{Fujikawa1980}:
\begin{equation}
\mu_\nu^{SM} \approx 3 \times 10^{-19} \left( \frac{m_\nu}{\text{eV}} \right) \mu_B
\end{equation}
where \( \mu_B \) is the Bohr magneton. This prediction is many orders of magnitude below current experimental sensitivities. However, the search for a neutrino magnetic moment is a powerful probe of new physics beyond the Standard Model, including possible couplings to new light particles or electromagnetic interactions beyond minimal extensions.

Experimental limits on the neutrino magnetic moment derive from reactor and accelerator neutrino experiments, solar neutrino experiments and astrophysical observations (for a detailed review see ~\cite{Giunti:2014ixa,Giunti:2024gec,PDG2024Neutrino}).
The most stringent laboratory bounds come from the GEMMA experiment~\cite{Beda:2012zz} which reports:
\begin{equation}
\mu_\nu < 2.9 \times 10^{-11} \mu_B \quad \text{(90\% C.L.)}
\end{equation}
Astrophysical constraints, though more model-dependent, are typically stronger, with limits as low as $10^{-12} \mu_B$.
Future experiments with improved detector sensitivity, lower thresholds, and higher neutrino fluxes (e.g., next-generation reactor or coherent neutrino scattering detectors) aim to improve these bounds significantly.

%% file: Conclusion.tex

\section{Conclusions}

\label{sec:conclusions}

The electron and muon magnetic moments, the highest precision predictions of the Standard Model (SM), provide the highest precision tests of the SM.  Stable electrons, readily available, are suspended in a cm-size trap for a long as needed to cool them to the ground state of their motion.  Quantum jump spectroscopy and other quantum methods determine the electron magnetic moment more accurately than any other property of an elementary particle. Muons, produced only in high-energy collisions at large facilities, are studied as they orbit at nearly the speed of light, in a 14-m diameter storage ring,  during the several microseconds that they live.  The measurement of the electron and muon moments stimulated substantial theoretical advances.  The good agreement between the predictions and measurements at an unprecedented precision is a great triumph of the SM that establishes  important limits on possible physics beyond the SM. Measurements currently underway promise to provide more stringent SM tests along with deeper probes for BSM physics. These efforts ensure that this remains a vibrant and dynamic field of research.

\begin{ack}[Acknowledgments]%

Lepton magnetic moment measurements are the work of many students and colleagues over many years. 
We are grateful to those who carried out the measurements of the electron magnetic moment at Harvard and Northwestern Universities, and the muon $g\!-\!2$ experiments at CERN, Brookhaven, and Fermilab, including those who designed, built, and operated the large facilities required. We are also grateful to the theorists who devised clever methods to compute extremely intricate SM calculations, and to the funding agencies of many countries who supported the measurements and  the calculations. 
GG gratefully acknowledges the support of the U.S. National Science Foundation, the Templeton Foundation and the DOE SQMS Center.
GV gratefully acknowledges the support of the Leverhulme Trust (grant LIP-2021-014) and the Istituto Nazionale di Fisica Nucleare (Italy).

\end{ack}


\seealso{Chapter on lepton magnetic moment theory}

%% file: output.bbl
\begin{thebibliography*}{100}
\providecommand{\bibtype}[1]{}
\providecommand{\url}[1]{{\tt #1}}
\providecommand{\urlprefix}{URL }
\expandafter\ifx\csname urlstyle\endcsname\relax
  \providecommand{\doi}[1]{doi:\discretionary{}{}{}#1}\else
  \providecommand{\doi}{doi:\discretionary{}{}{}\begingroup \urlstyle{rm}\Url}\fi
\providecommand{\bibinfo}[2]{#2}
\providecommand{\eprint}[2][]{\url{#2}}
\makeatletter\def\@biblabel#1{\bibinfo{label}{[#1]}}\makeatother

\bibtype{Article}%
\bibitem{Gabrielse2013}
\bibinfo{author}{Gerald Gabrielse}, \bibinfo{title}{The Standard Model's Greatest Triumph}, \bibinfo{journal}{Physics Today} \bibinfo{volume}{66} (\bibinfo{number}{12}) (\bibinfo{year}{2013}) \bibinfo{pages}{64--65}, \bibinfo{doi}{\doi{10.1063/PT.3.2223}}, \bibinfo{note}{december issue}, \bibinfo{url}{\urlprefix\url{https://doi.org/10.1063/PT.3.2223}}.

\bibtype{Article}%
\bibitem{DiracTheoryOriginal}
\bibinfo{author}{P.~A.~M. {Dirac}}, \bibinfo{title}{{The Quantum Theory of the Electron. Part II}}, \bibinfo{journal}{Proceedings of the Royal Society of London Series A} \bibinfo{volume}{118} (\bibinfo{number}{779}) (\bibinfo{year}{1928}) \bibinfo{pages}{351--361}, \bibinfo{doi}{\doi{10.1098/rspa.1928.0056}}.

\bibtype{Article}%
\bibitem{Schwinger}
\bibinfo{author}{Julian Schwinger}, \bibinfo{title}{On Quantum-Electrodynamics and the Magnetic Moment of the Electron}, \bibinfo{journal}{Phys. Rev.} \bibinfo{volume}{73} (\bibinfo{year}{1948}) \bibinfo{pages}{416--417}, \bibinfo{doi}{\doi{10.1103/PhysRev.73.416}}, \bibinfo{url}{\urlprefix\url{https://link.aps.org/doi/10.1103/PhysRev.73.416}}.

\bibtype{Article}%
\bibitem{Petermann}
\bibinfo{author}{A. Petermann}, \bibinfo{title}{{Fourth Order Magnetic Moment of the Electron}}, \bibinfo{journal}{Helv. Phys. Acta} \bibinfo{volume}{30} (\bibinfo{year}{1957}) \bibinfo{pages}{407--408}.

\bibtype{Article}%
\bibitem{Sommerfield:1958}
\bibinfo{author}{C.~M. Sommerfield}, \bibinfo{title}{{The Magnetic Moment of the Electron}}, \bibinfo{journal}{{Ann.\ Phys.\ (N.Y.)}} \bibinfo{volume}{5} (\bibinfo{year}{1958}) \bibinfo{pages}{26--57}.

\bibtype{Article}%
\bibitem{Kinoshita:1995}
\bibinfo{author}{Toichiro Kinoshita}, \bibinfo{title}{New Value of the ${\ensuremath{\alpha}}^{3}$ Electron Anomalous Magnetic Moment}, \bibinfo{journal}{Phys. Rev. Lett.} \bibinfo{volume}{75} (\bibinfo{year}{1995}) \bibinfo{pages}{4728--4731}, \bibinfo{doi}{\doi{10.1103/PhysRevLett.75.4728}}, \bibinfo{url}{\urlprefix\url{https://link.aps.org/doi/10.1103/PhysRevLett.75.4728}}.

\bibtype{Article}%
\bibitem{LaportaRemiddi:1996}
\bibinfo{author}{S. Laporta}, \bibinfo{author}{E. Remiddi}, \bibinfo{title}{{Analytical Value of the Electron ($g-2$) at Order $a^3$ in {QED}}}, \bibinfo{journal}{Phys. Lett. B} \bibinfo{volume}{379} (\bibinfo{year}{1996}) \bibinfo{pages}{283--291}.

\bibtype{Article}%
\bibitem{QED_C8_Lapo}
\bibinfo{author}{Stefano Laporta}, \bibinfo{title}{High-precision calculation of the 4-loop contribution to the electron g-2 in QED}, \bibinfo{journal}{Physics Letters B} \bibinfo{volume}{772} (\bibinfo{year}{2017}) \bibinfo{pages}{232 -- 238}, ISSN \bibinfo{issn}{0370-2693}, \bibinfo{doi}{\doi{https://doi.org/10.1016/j.physletb.2017.06.056}}, \bibinfo{url}{\urlprefix\url{http://www.sciencedirect.com/science/article/pii/S0370269317305324}}.

\bibtype{Article}%
\bibitem{KURZ2014_HeavyLeptons}
\bibinfo{author}{Alexander Kurz}, \bibinfo{author}{Tao Liu}, \bibinfo{author}{Peter Marquard}, \bibinfo{author}{Matthias Steinhauser}, \bibinfo{title}{Anomalous magnetic moment with heavy virtual leptons}, \bibinfo{journal}{Nuclear Physics B} \bibinfo{volume}{879} (\bibinfo{year}{2014}) \bibinfo{pages}{1--18}, ISSN \bibinfo{issn}{0550-3213}, \bibinfo{doi}{\doi{https://doi.org/10.1016/j.nuclphysb.2013.11.018}}, \bibinfo{url}{\urlprefix\url{https://www.sciencedirect.com/science/article/pii/S0550321313005828}}.

\bibtype{Article}%
\bibitem{TenthOrderQED2012}
\bibinfo{author}{T. Aoyama}, \bibinfo{author}{M. Hayakawa}, \bibinfo{author}{T. Kinoshita}, \bibinfo{author}{M. Nio}, \bibinfo{title}{Tenth-Order QED Contribution to the Electron g - 2 and an Improved Value of the Fine Structure Constant}, \bibinfo{journal}{Phys. Rev. Lett.} \bibinfo{volume}{109} (\bibinfo{year}{2012}) \bibinfo{pages}{111807}.

\bibtype{Article}%
\bibitem{Aoyama:2012wk}
\bibinfo{author}{Tatsumi Aoyama}, \bibinfo{author}{Masashi Hayakawa}, \bibinfo{author}{Toichiro Kinoshita}, \bibinfo{author}{Makiko Nio}, \bibinfo{title}{{Complete Tenth-Order QED Contribution to the Muon g-2}}, \bibinfo{journal}{Phys. Rev. Lett.} \bibinfo{volume}{109} (\bibinfo{year}{2012}) \bibinfo{pages}{111808}, \bibinfo{doi}{\doi{10.1103/PhysRevLett.109.111808}}, \eprint{1205.5370}.

\bibtype{Article}%
\bibitem{Laporta:1994md}
\bibinfo{author}{S. Laporta}, \bibinfo{title}{{Analytical and numerical contributions of some tenth order graphs containing vacuum polarization insertions to the muon (g-2) in QED}}, \bibinfo{journal}{Phys. Lett. B} \bibinfo{volume}{328} (\bibinfo{year}{1994}) \bibinfo{pages}{522--527}, \bibinfo{doi}{\doi{10.1016/0370-2693(94)91513-X}}, \eprint{hep-ph/9404204}.

\bibtype{Article}%
\bibitem{Baikov:2013ula}
\bibinfo{author}{P.~A. Baikov}, \bibinfo{author}{A. Maier}, \bibinfo{author}{P. Marquard}, \bibinfo{title}{{The QED vacuum polarization function at four loops and the anomalous magnetic moment at five loops}}, \bibinfo{journal}{Nucl. Phys. B} \bibinfo{volume}{877} (\bibinfo{year}{2013}) \bibinfo{pages}{647--661}, \bibinfo{doi}{\doi{10.1016/j.nuclphysb.2013.10.020}}, \eprint{1307.6105}.

\bibtype{Article}%
\bibitem{QED_C10_nio}
\bibinfo{author}{Tatsumi Aoyama}, \bibinfo{author}{Toichiro Kinoshita}, \bibinfo{author}{Makiko Nio}, \bibinfo{title}{Revised and improved value of the QED tenth-order electron anomalous magnetic moment}, \bibinfo{journal}{Phys. Rev. D} \bibinfo{volume}{97} (\bibinfo{year}{2018}) \bibinfo{pages}{036001}, \bibinfo{doi}{\doi{10.1103/PhysRevD.97.036001}}, \bibinfo{url}{\urlprefix\url{https://link.aps.org/doi/10.1103/PhysRevD.97.036001}}.

\bibtype{Article}%
\bibitem{Volkov:2019phy}
\bibinfo{author}{Sergey Volkov}, \bibinfo{title}{{Calculating the five-loop QED contribution to the electron anomalous magnetic moment: Graphs without lepton loops}}, \bibinfo{journal}{Phys. Rev. D} \bibinfo{volume}{100} (\bibinfo{number}{9}) (\bibinfo{year}{2019}) \bibinfo{pages}{096004}, \bibinfo{doi}{\doi{10.1103/PhysRevD.100.096004}}, \eprint{1909.08015}.

\bibtype{Article}%
\bibitem{Volkov:2024yzc}
\bibinfo{author}{Sergey Volkov}, \bibinfo{title}{{Calculation of the total 10th order QED contribution to the electron magnetic moment}}, \bibinfo{journal}{Phys. Rev. D} \bibinfo{volume}{110} (\bibinfo{number}{3}) (\bibinfo{year}{2024}) \bibinfo{pages}{036001}, \bibinfo{doi}{\doi{10.1103/PhysRevD.110.036001}}, \eprint{2404.00649}.

\bibtype{Article}%
\bibitem{Aoyama:2024aly}
\bibinfo{author}{Tatsumi Aoyama}, \bibinfo{author}{Masashi Hayakawa}, \bibinfo{author}{Akira Hirayama}, \bibinfo{author}{Makiko Nio}, \bibinfo{title}{{Verification of the tenth-order QED contribution to the anomalous magnetic moment of the electron from diagrams without fermion loops}}, \bibinfo{journal}{Phys. Rev. D} \bibinfo{volume}{111} (\bibinfo{number}{3}) (\bibinfo{year}{2025}) \bibinfo{pages}{L031902}, \bibinfo{doi}{\doi{10.1103/PhysRevD.111.L031902}}, \eprint{2412.06473}.

\bibtype{Article}%
\bibitem{Jegerlehner:2009ry}
\bibinfo{author}{Fred Jegerlehner}, \bibinfo{author}{Andreas Nyffeler}, \bibinfo{title}{{The Muon g-2}}, \bibinfo{journal}{Phys. Rept.} \bibinfo{volume}{477} (\bibinfo{year}{2009}) \bibinfo{pages}{1--110}, \bibinfo{doi}{\doi{10.1016/j.physrep.2009.04.003}}, \eprint{0902.3360}.

\bibtype{Book}%
\bibitem{Jegerlehner:2017gek}
\bibinfo{author}{Friedrich Jegerlehner}, \bibinfo{title}{{The Anomalous Magnetic Moment of the Muon}}, \bibinfo{comment}{vol.} \bibinfo{volume}{274}, \bibinfo{publisher}{Springer}, \bibinfo{address}{Cham} \bibinfo{year}{2017}, \bibinfo{doi}{\doi{10.1007/978-3-319-63577-4}}.

\bibtype{Article}%
\bibitem{Aoyama:2020ynm}
\bibinfo{author}{T. Aoyama}, et al., \bibinfo{title}{{The anomalous magnetic moment of the muon in the Standard Model}}, \bibinfo{journal}{Phys. Rept.} \bibinfo{volume}{887} (\bibinfo{year}{2020}) \bibinfo{pages}{1--166}, \bibinfo{doi}{\doi{10.1016/j.physrep.2020.07.006}}, \eprint{2006.04822}.

\bibtype{Article}%
\bibitem{Aliberti:2025beg}
\bibinfo{author}{R. Aliberti}, et al., \bibinfo{title}{{The anomalous magnetic moment of the muon in the Standard Model: an update}}  (\bibinfo{year}{2025}), \eprint{2505.21476}.

\bibtype{Article}%
\bibitem{WeakCondtribution1}
\bibinfo{author}{Kazuo Fujikawa}, \bibinfo{author}{Benjamin~W. Lee}, \bibinfo{author}{A.~I. Sanda}, \bibinfo{title}{Generalized Renormalizable Gauge Formulation of Spontaneously Broken Gauge Theories}, \bibinfo{journal}{Phys. Rev. D} \bibinfo{volume}{6} (\bibinfo{year}{1972}) \bibinfo{pages}{2923--2943}, \bibinfo{doi}{\doi{10.1103/PhysRevD.6.2923}}, \bibinfo{url}{\urlprefix\url{https://link.aps.org/doi/10.1103/PhysRevD.6.2923}}.

\bibtype{Article}%
\bibitem{WeakCondtribution2}
\bibinfo{author}{Andrzej Czarnecki}, \bibinfo{author}{Bernd Krause}, \bibinfo{author}{William~J. Marciano}, \bibinfo{title}{Electroweak Corrections to the Muon Anomalous Magnetic Moment}, \bibinfo{journal}{Phys. Rev. Lett.} \bibinfo{volume}{76} (\bibinfo{year}{1996}) \bibinfo{pages}{3267--3270}, \bibinfo{doi}{\doi{10.1103/PhysRevLett.76.3267}}, \bibinfo{url}{\urlprefix\url{https://link.aps.org/doi/10.1103/PhysRevLett.76.3267}}.

\bibtype{Article}%
\bibitem{WeakCondtribution3}
\bibinfo{author}{Marc Knecht}, \bibinfo{author}{Michel Perrottet}, \bibinfo{author}{Eduardo de Rafael}, \bibinfo{author}{Santiago Peris}, \bibinfo{title}{Electroweak Hadronic Contributions to the muon$<$i$>$g$<$/i$>$-2}, \bibinfo{journal}{Journal of High Energy Physics} \bibinfo{volume}{11} (\bibinfo{year}{2002}) \bibinfo{pages}{003--003}, \bibinfo{doi}{\doi{10.1088/1126-6708/2002/11/003}}, \bibinfo{url}{\urlprefix\url{https://doi.org/10.1088/1126-6708/2002/11/003}}.

\bibtype{Article}%
\bibitem{WeakCondtribution4}
\bibinfo{author}{Andrzej Czarnecki}, \bibinfo{author}{William~J. Marciano}, \bibinfo{author}{Arkady Vainshtein}, \bibinfo{title}{Refinements in electroweak contributions to the muon anomalous magnetic moment}, \bibinfo{journal}{Phys. Rev. D} \bibinfo{volume}{67} (\bibinfo{year}{2003}) \bibinfo{pages}{073006}, \bibinfo{doi}{\doi{10.1103/PhysRevD.67.073006}}, \bibinfo{url}{\urlprefix\url{https://link.aps.org/doi/10.1103/PhysRevD.67.073006}}.

\bibtype{Article}%
\bibitem{Gnendiger:2013pva}
\bibinfo{author}{C. Gnendiger}, \bibinfo{author}{D. St{\"o}ckinger}, \bibinfo{author}{H. St{\"o}ckinger-Kim}, \bibinfo{title}{{The electroweak contributions to $(g-2)_\mu$ after the Higgs boson mass measurement}}, \bibinfo{journal}{Phys. Rev. D} \bibinfo{volume}{88} (\bibinfo{year}{2013}) \bibinfo{pages}{053005}, \bibinfo{doi}{\doi{10.1103/PhysRevD.88.053005}}, \eprint{1306.5546}.

\bibtype{Article}%
\bibitem{MullerAlpha2018}
\bibinfo{author}{Richard~H. Parker}, \bibinfo{author}{Chenghui Yu}, \bibinfo{author}{Weicheng Zhong}, \bibinfo{author}{Brian Estey}, \bibinfo{author}{Holger M{\"u}ller}, \bibinfo{title}{Measurement of the fine-structure constant as a test of the Standard Model}, \bibinfo{journal}{Science} \bibinfo{volume}{360} (\bibinfo{number}{6385}) (\bibinfo{year}{2018}) \bibinfo{pages}{191--195}, ISSN \bibinfo{issn}{0036-8075}, \bibinfo{doi}{\doi{10.1126/science.aap7706}}, \eprint{https://science.sciencemag.org/content/360/6385/191.full.pdf}, \bibinfo{url}{\urlprefix\url{https://science.sciencemag.org/content/360/6385/191}}.

\bibtype{Article}%
\bibitem{RbAlpha2020}
\bibinfo{author}{L{\'e}o Morel}, \bibinfo{author}{Zhibin Yao}, \bibinfo{author}{Pierre Clad{\'e}}, \bibinfo{author}{Sa{\"\i}da Guellati-Kh{\'e}lifa}, \bibinfo{title}{Determination of the fine-structure constant with an accuracy of 81 parts per trillion}, \bibinfo{journal}{Nature} \bibinfo{volume}{588} (\bibinfo{number}{7836}) (\bibinfo{year}{2020}) \bibinfo{pages}{61--65}, \bibinfo{doi}{\doi{10.1038/s41586-020-2964-7}}, \bibinfo{url}{\urlprefix\url{https://doi.org/10.1038/s41586-020-2964-7}}.

\bibtype{Article}%
\bibitem{SternGerlach}
\bibinfo{author}{W. Gerlach}, \bibinfo{author}{O. Stery}, \bibinfo{title}{The experimental proof of directional quantization in the magnetic field} \bibinfo{volume}{9} \bibinfo{pages}{349}.

\bibtype{Article}%
\bibitem{DehmeltMagneticMoment}
\bibinfo{author}{{R. S. Van Dyck, Jr.}}, \bibinfo{author}{P.~B. Schwinberg}, \bibinfo{author}{H.~G. Dehmelt}, \bibinfo{title}{{New High-Precision Comparison of Electron and Positron $g$ Factors}}, \bibinfo{journal}{Phys. Rev. Lett.} \bibinfo{volume}{59} (\bibinfo{year}{1987}) \bibinfo{pages}{26--29}.

\bibtype{Article}%
\bibitem{QuantumCyclotron}
\bibinfo{author}{S. Peil}, \bibinfo{author}{G. Gabrielse}, \bibinfo{title}{{Observing the Quantum Limit of an Electron Cyclotron: {QND} Measurements of Quantum Jumps Between Fock States}}, \bibinfo{journal}{Phys. Rev. Lett.} \bibinfo{volume}{83} (\bibinfo{year}{1999}) \bibinfo{pages}{1287--1290}.

\bibtype{Article}%
\bibitem{NorthwesternMagneticMoment2023}
\bibinfo{author}{X. Fan}, \bibinfo{author}{T.~G. Myers}, \bibinfo{author}{B.~A.~D. Sukra}, \bibinfo{author}{G. Gabrielse}, \bibinfo{title}{Measurement of the Electron Magnetic Moment}, \bibinfo{journal}{Phys. Rev. Lett.} \bibinfo{volume}{130} (\bibinfo{year}{2023}) \bibinfo{pages}{071801}, \bibinfo{doi}{\doi{10.1103/PhysRevLett.130.071801}}.

\bibtype{Article}%
\bibitem{HistoryOfElectrong}
\bibinfo{author}{B. Lautrup}, \bibinfo{author}{H. Zinkernagel}, \bibinfo{title}{{g-2 and the Trust in Experimental Results}}, \bibinfo{journal}{Stud. Hist. Phil. Mod. Phys.} \bibinfo{volume}{30} (\bibinfo{year}{1999}) \bibinfo{pages}{85 -- 110}.

\bibtype{Article}%
\bibitem{GraffMeasuringCyclotronAndSpinResonancesForElectrons1968}
\bibinfo{author}{G. Graeff}, \bibinfo{author}{F.~G. Major}, \bibinfo{author}{R.~W.~T Roeder}, \bibinfo{author}{G. Werth}, \bibinfo{title}{Method for Measuring the Cyclotron and Spin Resonance of Free Electrons}, \bibinfo{journal}{Phys. Rev. Lett.} \bibinfo{volume}{21} (\bibinfo{year}{1968}) \bibinfo{pages}{340}.

\bibtype{Article}%
\bibitem{DehmeltWalls1968}
\bibinfo{author}{H. Dehmelt}, \bibinfo{author}{F. Walls}, \bibinfo{title}{"Bolometric" Technique for the rf Spectroscopy of Stored Ions}, \bibinfo{journal}{Phys. Rev. Lett.} \bibinfo{volume}{21} (\bibinfo{year}{1968}) \bibinfo{pages}{127}.

\bibtype{Article}%
\bibitem{GraffAnomalyResonanceForElectrons1969}
\bibinfo{author}{G. Graeff}, \bibinfo{author}{E. Klempt}, \bibinfo{author}{G. Werth}, \bibinfo{title}{Method for measuring the anomalous magnetic moment of free electrons}, \bibinfo{journal}{Z. Phys.} \bibinfo{volume}{222} (\bibinfo{year}{1969}) \bibinfo{pages}{201}.

\bibtype{Article}%
\bibitem{WallsSteinAnomalyForElectrons1973}
\bibinfo{author}{F. Walls}, \bibinfo{author}{T. Stein}, \bibinfo{title}{Observation of the $g-2$ Resonance of a Stored Electron Gas Using a Bolometric Technique}, \bibinfo{journal}{Phys. Rev. Lett.} \bibinfo{volume}{31} (\bibinfo{year}{1973}) \bibinfo{pages}{975}.

\bibtype{Article}%
\bibitem{FirstSingleElectron1973}
\bibinfo{author}{D. Wineland}, \bibinfo{author}{P. Ekstrom}, \bibinfo{author}{H. Dehmelt}, \bibinfo{title}{Monoelectron Oscillator}, \bibinfo{journal}{Phys. Rev. Lett.} \bibinfo{volume}{31} (\bibinfo{year}{1973}) \bibinfo{pages}{1279}.

\bibtype{Article}%
\bibitem{OrthogonalCompensate}
\bibinfo{author}{G. Gabrielse}, \bibinfo{title}{{Relaxation Calculation of the Electrostatic Properties of Compensated Penning Traps with Hyperbolic Electrodes}}, \bibinfo{journal}{Phys. Rev. A} \bibinfo{volume}{27} (\bibinfo{year}{1983}) \bibinfo{pages}{2277--2290}.

\bibtype{Article}%
\bibitem{Gabrielse84h}
\bibinfo{author}{G. Gabrielse}, \bibinfo{title}{{Detection, Damping and Translating the Center of the Axial Oscillation of a Charged Particle in a Penning Trap with Hyperbolic Electrodes}}, \bibinfo{journal}{Phys. Rev. A} \bibinfo{volume}{29} (\bibinfo{year}{1984}) \bibinfo{pages}{462--469}.

\bibtype{Article}%
\bibitem{DarkPhotonFromQuantumCyclotron}
\bibinfo{author}{Xing Fan}, \bibinfo{author}{Gerald Gabrielse}, \bibinfo{author}{Peter~W. Graham}, \bibinfo{author}{Roni Harnik}, \bibinfo{author}{Thomas~G. Myers}, \bibinfo{author}{Harikrishnan Ramani}, \bibinfo{author}{Benedict A.~D. Sukra}, \bibinfo{author}{Samuel S.~Y. Wong}, \bibinfo{author}{Yawen Xiao}, \bibinfo{title}{{One-Electron Quantum Cyclotron as a Milli-eV Dark-Photon Detector}}, \bibinfo{journal}{Phys. Rev. Lett.} \bibinfo{volume}{129} (\bibinfo{number}{26}) (\bibinfo{year}{2022}) \bibinfo{pages}{261801}, \bibinfo{doi}{\doi{10.1103/PhysRevLett.129.261801}}, \eprint{2208.06519}.

\bibtype{Article}%
\bibitem{SelfShieldingSolenoid}
\bibinfo{author}{G. Gabrielse}, \bibinfo{author}{J. Tan}, \bibinfo{title}{Self-Shielding Superconducting Solenoid Systems}, \bibinfo{journal}{J. Appl. Phys.} \bibinfo{volume}{63} (\bibinfo{year}{1988}) \bibinfo{pages}{5143--5148}.

\bibtype{Article}%
\bibitem{Atoms2019TowardImprovedMeasurement}
\bibinfo{author}{G. Gabrielse}, \bibinfo{author}{S. Fayer}, \bibinfo{author}{T. Myers}, \bibinfo{author}{X. Fan}, \bibinfo{title}{Towards an Improved Test of the Standard Model's Most Precise Prediction}, \bibinfo{journal}{Atoms} \bibinfo{volume}{7} (\bibinfo{number}{2}) (\bibinfo{year}{2019}) \bibinfo{pages}{45}, ISSN \bibinfo{issn}{2218-2004}, \bibinfo{doi}{\doi{10.3390/atoms7020045}}, \bibinfo{url}{\urlprefix\url{http://dx.doi.org/10.3390/atoms7020045}}.

\bibtype{Article}%
\bibitem{Helium3NMR2019}
\bibinfo{author}{X. Fan}, \bibinfo{author}{S.~E. Fayer}, \bibinfo{author}{G. Gabrielse}, \bibinfo{title}{Gaseous 3He nuclear magnetic resonance probe for cryogenic environments}, \bibinfo{journal}{Review of Scientific Instruments} \bibinfo{volume}{90} (\bibinfo{number}{8}) (\bibinfo{year}{2019}) \bibinfo{pages}{083107}, \bibinfo{doi}{\doi{10.1063/1.5099379}}, \eprint{https://doi.org/10.1063/1.5099379}, \bibinfo{url}{\urlprefix\url{https://doi.org/10.1063/1.5099379}}.

\bibtype{Article}%
\bibitem{CylindricalPenningTrap}
\bibinfo{author}{G. Gabrielse}, \bibinfo{author}{F.~Colin MacKintosh}, \bibinfo{title}{{Cylindrical Penning Traps with Orthogonalized Anharmonicity Compensation}}, \bibinfo{journal}{Intl. J. of Mass Spec. and Ion Proc.} \bibinfo{volume}{57} (\bibinfo{year}{1984}) \bibinfo{pages}{1--17}.

\bibtype{Article}%
\bibitem{CylindricalPenningTrapDemonstrated}
\bibinfo{author}{J. Tan}, \bibinfo{author}{G. Gabrielse}, \bibinfo{title}{{One Electron in an Orthogonalized Cylindrical Penning Trap}}, \bibinfo{journal}{Appl. Phys. Lett.} \bibinfo{volume}{55} (\bibinfo{year}{1989}) \bibinfo{pages}{2144--2146}.

\bibtype{Article}%
\bibitem{RenormalizedModesPRL}
\bibinfo{author}{L.~S. Brown}, \bibinfo{author}{G. Gabrielse}, \bibinfo{author}{K. Helmerson}, \bibinfo{author}{J. Tan}, \bibinfo{title}{{Cyclotron Motion in a Microwave Cavity: Possible shifts of the Measured Electron $g$ Factor}}, \bibinfo{journal}{Phys. Rev. Lett.} \bibinfo{volume}{55} (\bibinfo{year}{1985}) \bibinfo{pages}{44--47}.

\bibtype{Article}%
\bibitem{RenormalizedModesPRA}
\bibinfo{author}{L.~S. Brown}, \bibinfo{author}{G. Gabrielse}, \bibinfo{author}{K. Helmerson}, \bibinfo{author}{J. Tan}, \bibinfo{title}{Cyclotron Motion in a Microwave Cavity: Lifetime and Frequency Shifts}, \bibinfo{journal}{Phys. Rev. A} \bibinfo{volume}{32} (\bibinfo{year}{1985}) \bibinfo{pages}{3204--3218}.

\bibtype{Article}%
\bibitem{HarvardMagneticMoment2008}
\bibinfo{author}{D. Hanneke}, \bibinfo{author}{S. Fogwell}, \bibinfo{author}{G. Gabrielse}, \bibinfo{title}{{New Measurement of the Electron Magnetic Moment and the Fine Structure Constant}}, \bibinfo{journal}{Phys. Rev. Lett.} \bibinfo{volume}{100} (\bibinfo{year}{2008}) \bibinfo{pages}{120801}.

\bibtype{Article}%
\bibitem{HarvardMagneticMoment2011}
\bibinfo{author}{D. Hanneke}, \bibinfo{author}{S. Fogwell~Hoogerheide}, \bibinfo{author}{G. Gabrielse}, \bibinfo{title}{Cavity control of a single-electron quantum cyclotron: Measuring the electron magnetic moment}, \bibinfo{journal}{Phys. Rev. A} \bibinfo{volume}{83} (\bibinfo{year}{2011}) \bibinfo{pages}{052122}, \bibinfo{doi}{\doi{10.1103/PhysRevA.83.052122}}, \bibinfo{url}{\urlprefix\url{https://link.aps.org/doi/10.1103/PhysRevA.83.052122}}.

\bibtype{Inbook}%
\bibitem{VanDyckMagnetronCoolingLimit}
\bibinfo{author}{R.~S. {Van Dyck, Jr.}}, \bibinfo{author}{P.~B. Schwinberg}, \bibinfo{author}{H.~G. Dehmelt}, \bibinfo{title}{New Frontiers in High Energy Physics}, \bibinfo{publisher}{{Plenum}}, \bibinfo{address}{New York} \bibinfo{year}{1978} p. \bibinfo{pages}{159}.

\bibtype{Article}%
\bibitem{Review}
\bibinfo{author}{L.~S. Brown}, \bibinfo{author}{G. Gabrielse}, \bibinfo{title}{{Geonium Theory: Physics of a Single Electron or Ion in a Penning Trap}}, \bibinfo{journal}{Rev. Mod. Phys.} \bibinfo{volume}{58} (\bibinfo{year}{1986}) \bibinfo{pages}{233--311}.

\bibtype{Article}%
\bibitem{InvarianceTheorem}
\bibinfo{author}{L.~S. Brown}, \bibinfo{author}{G. Gabrielse}, \bibinfo{title}{{Precision Spectroscopy of a Charged Particle in an Imperfect Penning Trap}}, \bibinfo{journal}{Phys. Rev. A} \bibinfo{volume}{25} (\bibinfo{year}{1982}) \bibinfo{pages}{2423--2425}.

\bibtype{Article}%
\bibitem{InvarianceTheoremApplications2}
\bibinfo{author}{G. Gabrielse}, \bibinfo{title}{Whys is Sideband Mass Spectroscopy Possible with Ions in a Penning Trap?} \bibinfo{volume}{102} \bibinfo{pages}{172501}.

\bibtype{Article}%
\bibitem{TrueCyclotronFrequency}
\bibinfo{author}{G. Gabrielse}, \bibinfo{title}{True Cyclotron Frequency for Particles and Ions in a Penning Trap}, \bibinfo{journal}{Int. J. Mass Spectrom.} \bibinfo{volume}{279} (\bibinfo{year}{2009}) \bibinfo{pages}{107}.

\bibtype{Article}%
\bibitem{Gabrielse85e_RelativisticMassBistableHysteresis}
\bibinfo{author}{G. Gabrielse}, \bibinfo{author}{H. Dehmelt}, \bibinfo{author}{W. Kells}, \bibinfo{title}{``Observation of a Relativistic Bistable Hysteresis in the Cyclotron Motion of a Single Electron''}, \bibinfo{journal}{Phys. Rev. Lett.} \bibinfo{volume}{54} (\bibinfo{year}{1985}) \bibinfo{pages}{537--540}.

\bibtype{Article}%
\bibitem{HarvardMagneticMoment2006}
\bibinfo{author}{B. Odom}, \bibinfo{author}{D. Hanneke}, \bibinfo{author}{B. D'Urso}, \bibinfo{author}{G. Gabrielse}, \bibinfo{title}{{New Measurement of the Electron Magnetic Moment Using a One-Electron Quantum Cyclotron}}, \bibinfo{journal}{Phys. Rev. Lett.} \bibinfo{volume}{97} (\bibinfo{year}{2006}) \bibinfo{pages}{030801}.

\bibtype{Article}%
\bibitem{Day:2025waz}
\bibinfo{author}{Hannah Day}, \bibinfo{author}{Roni Harnik}, \bibinfo{author}{Yonatan Kahn}, \bibinfo{author}{Shashin Pavaskar}, \bibinfo{author}{Kevin Zhou}, \bibinfo{title}{{Quantum Calculations of the Cavity Shift in Electron Magnetic Moment Measurements}}  (\bibinfo{year}{2025}), \eprint{2511.07514}.

\bibtype{Article}%
\bibitem{GreatistTriumph}
\bibinfo{author}{G. Gabrielse}, \bibinfo{title}{The Standard Model's Grestest Triumph} \bibinfo{volume}{64} \bibinfo{pages}{64 --}.

\bibtype{Misc}%
\bibitem{DysonLetter}
\bibinfo{author}{F. Dyson}, \bibinfo{title}{Letter to {G. Gabrielse}}, \bibinfo{howpublished}{{quoted in Physics Today, August 2006, pp. 15--17}}.

\bibtype{Article}%
\bibitem{Fan2020BackActionPRL}
\bibinfo{author}{X. Fan}, \bibinfo{author}{G. Gabrielse}, \bibinfo{title}{Circumventing Detector Backaction on a Quantum Cyclotron}, \bibinfo{journal}{Phys. Rev. Lett.} \bibinfo{volume}{126} (\bibinfo{year}{2021}) \bibinfo{pages}{070402}, \bibinfo{doi}{\doi{10.1103/PhysRevLett.126.070402}}, \bibinfo{url}{\urlprefix\url{https://link.aps.org/doi/10.1103/PhysRevLett.126.070402}}.

\bibtype{Article}%
\bibitem{Lee:1956qn}
\bibinfo{author}{T.~D. Lee}, \bibinfo{author}{Chen-Ning Yang}, \bibinfo{title}{{Question of Parity Conservation in Weak Interactions}}, \bibinfo{journal}{Phys. Rev.} \bibinfo{volume}{104} (\bibinfo{year}{1956}) \bibinfo{pages}{254--258}, \bibinfo{doi}{\doi{10.1103/PhysRev.104.254}}.

\bibtype{Phdthesis}%
\bibitem{Fienberg:2019nwt}
\bibinfo{author}{Aaron~T. Fienberg}, \bibinfo{title}{{Measuring the Precession Frequency in the E989 Muon g-2 Experiment}}, \bibinfo{comment}{Ph.D. thesis}, \bibinfo{school}{U. Washington, Seattle (main)} \bibinfo{year}{2019}.

\bibtype{Article}%
\bibitem{Garwin:1957hc}
\bibinfo{author}{R.~L. Garwin}, \bibinfo{author}{L.~M. Lederman}, \bibinfo{author}{M. Weinrich}, \bibinfo{title}{{Observations of the Failure of Conservation of Parity and Charge Conjugation in Meson Decays: The Magnetic Moment of the Free Muon}}, \bibinfo{journal}{Phys. Rev.} \bibinfo{volume}{105} (\bibinfo{year}{1957}) \bibinfo{pages}{1415--1417}, \bibinfo{doi}{\doi{10.1103/PhysRev.105.1415}}.

\bibtype{Article}%
\bibitem{Cassels:1957}
\bibinfo{author}{J~M Cassels}, \bibinfo{author}{T~W O'Keeffe}, \bibinfo{author}{M Rigby}, \bibinfo{author}{A~M Wetherell}, \bibinfo{author}{J~R Wormald}, \bibinfo{title}{Experiments with a Polarized Muon Beam}, \bibinfo{journal}{Proceedings of the Physical Society. Section A} \bibinfo{volume}{70} (\bibinfo{number}{7}) (\bibinfo{year}{1957}) \bibinfo{pages}{543--546}, \bibinfo{doi}{\doi{10.1088/0370-1298/70/7/412}}, \bibinfo{url}{\urlprefix\url{https://doi.org/10.1088/0370-1298/70/7/412}}.

\bibtype{Article}%
\bibitem{Garwin:1960zz}
\bibinfo{author}{R.~L. Garwin}, \bibinfo{author}{D.~P. Hutchinson}, \bibinfo{author}{S. Penman}, \bibinfo{author}{G. Shapiro}, \bibinfo{title}{{Accurate Determination of the mu+ Magnetic Moment}}, \bibinfo{journal}{Phys. Rev.} \bibinfo{volume}{118} (\bibinfo{year}{1960}) \bibinfo{pages}{271--283}, \bibinfo{doi}{\doi{10.1103/PhysRev.118.271}}.

\bibtype{Article}%
\bibitem{Louisell}
\bibinfo{author}{W.~H. Louisell}, \bibinfo{author}{R.~W. Pidd}, \bibinfo{author}{H.~R. Crane}, \bibinfo{journal}{Phys. Rev.} \bibinfo{volume}{91} (\bibinfo{year}{1953}) \bibinfo{pages}{475}.

\bibtype{Article}%
\bibitem{Schupp:1961zz}
\bibinfo{author}{A.~A. Schupp}, \bibinfo{author}{R.~W. Pidd}, \bibinfo{author}{H.~R. Crane}, \bibinfo{title}{{Measurement of the g Factor of Free, High-Energy Electrons}}, \bibinfo{journal}{Phys. Rev.} \bibinfo{volume}{121} (\bibinfo{year}{1961}) \bibinfo{pages}{1--17}, \bibinfo{doi}{\doi{10.1103/PhysRev.121.1}}.

\bibtype{Article}%
\bibitem{Thomas:1926dy}
\bibinfo{author}{L.~H. Thomas}, \bibinfo{title}{{The motion of a spinning electron}}, \bibinfo{journal}{Nature} \bibinfo{volume}{117} (\bibinfo{year}{1926}) \bibinfo{pages}{514}, \bibinfo{doi}{\doi{10.1038/117514a0}}.

\bibtype{Article}%
\bibitem{Thomas:1927yu}
\bibinfo{author}{L.~H. Thomas}, \bibinfo{title}{{The Kinematics of an electron with an axis}}, \bibinfo{journal}{Phil. Mag. Ser. 7} \bibinfo{volume}{3} (\bibinfo{year}{1927}) \bibinfo{pages}{1--21}, \bibinfo{doi}{\doi{10.1080/14786440108564170}}.

\bibtype{Article}%
\bibitem{Farley:1979yb}
\bibinfo{author}{F.~J.~M. Farley}, \bibinfo{author}{E. Picasso}, \bibinfo{title}{{The Muon ($g-2$) Experiments at {CERN}}}, \bibinfo{journal}{Ann. Rev. Nucl. Part. Sci.} \bibinfo{volume}{29} (\bibinfo{year}{1979}) \bibinfo{pages}{243--282}, \bibinfo{doi}{\doi{10.1146/annurev.ns.29.120179.001331}}.

\bibtype{Article}%
\bibitem{Farley:2004hp}
\bibinfo{author}{F.~J.~M. Farley}, \bibinfo{author}{Y.~K. Semertzidis}, \bibinfo{title}{{The 47 years of muon g-2}}, \bibinfo{journal}{Prog. Part. Nucl. Phys.} \bibinfo{volume}{52} (\bibinfo{year}{2004}) \bibinfo{pages}{1--83}, \bibinfo{doi}{\doi{10.1016/j.ppnp.2003.09.004}}.

\bibtype{Article}%
\bibitem{Charpak:1961mz}
\bibinfo{author}{Georges Charpak}, \bibinfo{author}{F.~J.~M. Farley}, \bibinfo{author}{R.~L. Garwin}, \bibinfo{author}{T. Muller}, \bibinfo{author}{J.~C. Sens}, \bibinfo{author}{V.~L. Telegdi}, \bibinfo{author}{A. Zichichi}, \bibinfo{title}{{Measurement of the anomalous magnetic moment of the muon}}, \bibinfo{journal}{Phys. Rev. Lett.} \bibinfo{volume}{6} (\bibinfo{year}{1961}) \bibinfo{pages}{128--132}, \bibinfo{doi}{\doi{10.1103/PhysRevLett.6.128}}.

\bibtype{Article}%
\bibitem{Charpak:1962zz}
\bibinfo{author}{G. Charpak}, \bibinfo{author}{F.~J.~M. Farley}, \bibinfo{author}{R.~L. Garwin}, \bibinfo{title}{{A New Measurement of the Anomalous Magnetic Moment of the Muon}}, \bibinfo{journal}{Phys. Lett.} \bibinfo{volume}{1} (\bibinfo{year}{1962}) \bibinfo{pages}{16}, \bibinfo{doi}{\doi{10.1016/0031-9163(62)90263-9}}.

\bibtype{Article}%
\bibitem{Bailey:1968rxd}
\bibinfo{author}{J. Bailey}, \bibinfo{author}{W. Bartl}, \bibinfo{author}{G. Von~Bochmann}, \bibinfo{author}{R.~C.~A. Brown}, \bibinfo{author}{F.~J.~M. Farley}, \bibinfo{author}{H. Joestlein}, \bibinfo{author}{E. Picasso}, \bibinfo{author}{R.~W. Williams}, \bibinfo{title}{{Precision measurement of the anomalous magnetic moment of the muon}}, \bibinfo{journal}{Phys. Lett. B} \bibinfo{volume}{28} (\bibinfo{year}{1968}) \bibinfo{pages}{287--290}, \bibinfo{doi}{\doi{10.1016/0370-2693(68)90261-X}}.

\bibtype{Article}%
\bibitem{Bailey:1972eu}
\bibinfo{author}{J. Bailey}, \bibinfo{author}{W. Bartl}, \bibinfo{author}{G. von Bochmann}, \bibinfo{author}{R.~C.~A. Brown}, \bibinfo{author}{F.~J.~M. Farley}, \bibinfo{author}{M. Giesch}, \bibinfo{author}{H. Jostlein}, \bibinfo{author}{S. van~der Meer}, \bibinfo{author}{E. Picasso}, \bibinfo{author}{R.~W. Williams}, \bibinfo{title}{{Precise Measurement of the Anomalous Magnetic Moment of the Muon}}, \bibinfo{journal}{Nuovo Cim. A} \bibinfo{volume}{9} (\bibinfo{year}{1972}) \bibinfo{pages}{369--432}, \bibinfo{doi}{\doi{10.1007/BF02785248}}.

\bibtype{Article}%
\bibitem{Farley66}
\bibinfo{author}{F.~J.~M. Farley}, \bibinfo{author}{J. Bailey}, \bibinfo{author}{R.~C.~A. Brown}, \bibinfo{author}{M Giesch}, \bibinfo{author}{H. J\'ostlein}, \bibinfo{author}{S. van~der Meer}, \bibinfo{author}{E. Picasso}, \bibinfo{author}{M. Tannenbaum}, \bibinfo{title}{The anomalous magnetic moment of the negative muon}, \bibinfo{journal}{Nuovo Cimento} \bibinfo{volume}{45} (\bibinfo{year}{1966}) \bibinfo{pages}{281--286}.

\bibtype{Techreport}%
\bibitem{cernIII}
\bibinfo{author}{J. Bailey}, et al., \bibinfo{title}{Proposal For A Measurement Of The Anomalous Magnetic Moment Of The MUon At The Level Of 10-20 PPM}, \bibinfo{type}{\bibinfo{comment}{tech. rep.}}, \bibinfo{note}{muon g-2 proposal to the CERN PS, CERN-PH-I-COM-69-20 (19 May 1969)}, \bibinfo{url}{\urlprefix\url{https://cds.cern.ch/record/732381/files/cm-p00052390.pdf}}.

\bibtype{Article}%
\bibitem{CERNMuonStorageRing:1975gyf}
\bibinfo{author}{J. Bailey}, et al. (\bibinfo{collaboration}{CERN Muon Storage Ring}), \bibinfo{title}{{New Measurement of (G-2) of the Muon}}, \bibinfo{journal}{Phys. Lett. B} \bibinfo{volume}{55} (\bibinfo{year}{1975}) \bibinfo{pages}{420--424}, \bibinfo{doi}{\doi{10.1016/0370-2693(75)90374-3}}.

\bibtype{Article}%
\bibitem{CERNMuonStorageRing:1977bbe}
\bibinfo{author}{J. Bailey}, et al. (\bibinfo{collaboration}{CERN Muon Storage Ring}), \bibinfo{title}{{The Anomalous Magnetic Moment of Positive and Negative Muons}}, \bibinfo{journal}{Phys. Lett. B} \bibinfo{volume}{67} (\bibinfo{year}{1977}) \bibinfo{pages}{225}, \bibinfo{doi}{\doi{10.1016/0370-2693(77)90199-X}}.

\bibtype{Article}%
\bibitem{CERN-Mainz-Daresbury:1978ccd}
\bibinfo{author}{J. Bailey}, et al. (\bibinfo{collaboration}{CERN-Mainz-Daresbury}), \bibinfo{title}{{Final Report on the CERN Muon Storage Ring Including the Anomalous Magnetic Moment and the Electric Dipole Moment of the Muon, and a Direct Test of Relativistic Time Dilation}}, \bibinfo{journal}{Nucl. Phys. B} \bibinfo{volume}{150} (\bibinfo{year}{1979}) \bibinfo{pages}{1--75}, \bibinfo{doi}{\doi{10.1016/0550-3213(79)90292-X}}.

\bibtype{Article}%
\bibitem{Keshavarzi:2021eqa}
\bibinfo{author}{Alex Keshavarzi}, \bibinfo{author}{Kim~Siang Khaw}, \bibinfo{author}{Tamaki Yoshioka}, \bibinfo{title}{{Muon g\ensuremath{-}2: A review}}, \bibinfo{journal}{Nucl. Phys. B} \bibinfo{volume}{975} (\bibinfo{year}{2022}) \bibinfo{pages}{115675}, \bibinfo{doi}{\doi{10.1016/j.nuclphysb.2022.115675}}, \eprint{2106.06723}.

\bibtype{Article}%
\bibitem{Kinoshita:1981vs}
\bibinfo{author}{T. Kinoshita}, \bibinfo{author}{W.~B. Lindquist}, \bibinfo{title}{{Eighth Order Anomalous Magnetic Moment of the electron}}, \bibinfo{journal}{Phys. Rev. Lett.} \bibinfo{volume}{47} (\bibinfo{year}{1981}) \bibinfo{pages}{1573}, \bibinfo{doi}{\doi{10.1103/PhysRevLett.47.1573}}.

\bibtype{Article}%
\bibitem{Kinoshita:1983xp}
\bibinfo{author}{T. Kinoshita}, \bibinfo{author}{B. Nizic}, \bibinfo{author}{Y. Okamoto}, \bibinfo{title}{{Improved Theory of the Muon Anomalous Magnetic Moment}}, \bibinfo{journal}{Phys. Rev. Lett.} \bibinfo{volume}{52} (\bibinfo{year}{1984}) \bibinfo{pages}{717}, \bibinfo{doi}{\doi{10.1103/PhysRevLett.52.717}}.

\bibtype{Article}%
\bibitem{Kinoshita:1984it}
\bibinfo{author}{T. Kinoshita}, \bibinfo{author}{B. Nizic}, \bibinfo{author}{Y. Okamoto}, \bibinfo{title}{{Hadronic Contributions to the Anomalous Magnetic Moment of the Muon}}, \bibinfo{journal}{Phys. Rev. D} \bibinfo{volume}{31} (\bibinfo{year}{1985}) \bibinfo{pages}{2108}, \bibinfo{doi}{\doi{10.1103/PhysRevD.31.2108}}.

\bibtype{Article}%
\bibitem{Martin:1997ns}
\bibinfo{author}{Stephen~P. Martin}, \bibinfo{title}{{A Supersymmetry primer}}, \bibinfo{journal}{Adv. Ser. Direct. High Energy Phys.} \bibinfo{volume}{18} (\bibinfo{year}{1998}) \bibinfo{pages}{1--98}, \bibinfo{doi}{\doi{10.1142/9789812839657_0001}}, \eprint{hep-ph/9709356}.

\bibtype{Article}%
\bibitem{Stockinger:2006zn}
\bibinfo{author}{Dominik Stockinger}, \bibinfo{title}{{The Muon Magnetic Moment and Supersymmetry}}, \bibinfo{journal}{J. Phys. G} \bibinfo{volume}{34} (\bibinfo{year}{2007}) \bibinfo{pages}{R45--R92}, \bibinfo{doi}{\doi{10.1088/0954-3899/34/2/R01}}, \eprint{hep-ph/0609168}.

\bibtype{Article}%
\bibitem{Muong-2:2006rrc}
\bibinfo{author}{G.~W. Bennett}, et al. (\bibinfo{collaboration}{Muon g-2}), \bibinfo{title}{{Final Report of the Muon E821 Anomalous Magnetic Moment Measurement at BNL}}, \bibinfo{journal}{Phys. Rev. D} \bibinfo{volume}{73} (\bibinfo{year}{2006}) \bibinfo{pages}{072003}, \bibinfo{doi}{\doi{10.1103/PhysRevD.73.072003}}, \eprint{hep-ex/0602035}.

\bibtype{Article}%
\bibitem{Muong-2:2004fok}
\bibinfo{author}{G.~W. Bennett}, et al. (\bibinfo{collaboration}{Muon g-2}), \bibinfo{title}{{Measurement of the negative muon anomalous magnetic moment to 0.7 ppm}}, \bibinfo{journal}{Phys. Rev. Lett.} \bibinfo{volume}{92} (\bibinfo{year}{2004}) \bibinfo{pages}{161802}, \bibinfo{doi}{\doi{10.1103/PhysRevLett.92.161802}}, \eprint{hep-ex/0401008}.

\bibtype{Article}%
\bibitem{Roberts:2018vsx}
\bibinfo{author}{B.~Lee Roberts}, \bibinfo{title}{{The History of the Muon $(g-2)$ Experiments}}, \bibinfo{journal}{SciPost Phys. Proc.} \bibinfo{volume}{1} (\bibinfo{year}{2019}) \bibinfo{pages}{032}, \bibinfo{doi}{\doi{10.21468/SciPostPhysProc.1.032}}, \eprint{1811.06974}.

\bibtype{Techreport}%
\bibitem{Carey:2009zzb}
\bibinfo{author}{R.~M. Carey}, et al., \bibinfo{title}{{The New (g-2) Experiment: A proposal to measure the muon anomalous magnetic moment to +-0.14 ppm precision}}, \bibinfo{type}{\bibinfo{comment}{tech. rep.}} \bibinfo{year}{2009}, \bibinfo{doi}{\doi{10.2172/952029}}, \bibinfo{url}{\urlprefix\url{https://lss.fnal.gov/archive/test-proposal/0000/fermilab-proposal-0989.pdf}}.

\bibtype{Article}%
\bibitem{Keshavarzi:2018mgv}
\bibinfo{author}{Alexander Keshavarzi}, \bibinfo{author}{Daisuke Nomura}, \bibinfo{author}{Thomas Teubner}, \bibinfo{title}{{Muon $g-2$ and $\alpha(M_Z^2)$: a new data-based analysis}}, \bibinfo{journal}{Phys. Rev. D} \bibinfo{volume}{97} (\bibinfo{number}{11}) (\bibinfo{year}{2018}) \bibinfo{pages}{114025}, \bibinfo{doi}{\doi{10.1103/PhysRevD.97.114025}}, \eprint{1802.02995}.

\bibtype{Article}%
\bibitem{Davier:2019can}
\bibinfo{author}{M. Davier}, \bibinfo{author}{A. Hoecker}, \bibinfo{author}{B. Malaescu}, \bibinfo{author}{Z. Zhang}, \bibinfo{title}{{A new evaluation of the hadronic vacuum polarisation contributions to the muon anomalous magnetic moment and to $\mathbf{\boldsymbol\alpha(m_Z^2)}$}}, \bibinfo{journal}{Eur. Phys. J. C} \bibinfo{volume}{80} (\bibinfo{number}{3}) (\bibinfo{year}{2020}) \bibinfo{pages}{241}, \bibinfo{doi}{\doi{10.1140/epjc/s10052-020-7792-2}}, \bibinfo{note}{[Erratum: Eur.Phys.J.C 80, 410 (2020)]}, \eprint{1908.00921}.

\bibtype{Misc}%
\bibitem{FNALBigMove2019}
\bibinfo{author}{{Fermilab Muon g‑2 Collaboration}}, \bibinfo{title}{The Big Move}, \bibinfo{howpublished}{\url{https://muon-g-2.fnal.gov/bigmove/}} \bibinfo{year}{2019}, \bibinfo{note}{last modified July 1, 2019. Retrieved July 3, 2025}.

\bibtype{Misc}%
\bibitem{FNAL2015}
\bibinfo{author}{{Fermi National Accelerator Laboratory}}, \bibinfo{title}{Muon g-2 Superconducting Magnet Commissioning Preparation} \bibinfo{year}{2015}, \bibinfo{note}{accessed: 2025-07-03}, \bibinfo{url}{\urlprefix\url{https://www.youtube.com/watch?v=zPngQ_9yugM}}.

\bibtype{Article}%
\bibitem{Muong-2:2024hpx}
\bibinfo{author}{D.~P. Aguillard}, et al. (\bibinfo{collaboration}{Muon g-2}), \bibinfo{title}{{Detailed report on the measurement of the positive muon anomalous magnetic moment to 0.20~ppm}}, \bibinfo{journal}{Phys. Rev. D} \bibinfo{volume}{110} (\bibinfo{number}{3}) (\bibinfo{year}{2024}) \bibinfo{pages}{032009}, \bibinfo{doi}{\doi{10.1103/PhysRevD.110.032009}}, \eprint{2402.15410}.

\bibtype{Article}%
\bibitem{Muong-2:2025xyk}
\bibinfo{author}{D.~P. Aguillard}, et al. (\bibinfo{collaboration}{Muon g-2}), \bibinfo{title}{{Measurement of the Positive Muon Anomalous Magnetic Moment to 127 ppb}}  (\bibinfo{year}{2025}), \eprint{2506.03069}.

\bibtype{Article}%
\bibitem{Kaspar:2016ofv}
\bibinfo{author}{J. Kaspar}, et al., \bibinfo{title}{{Design and performance of SiPM-based readout of PbF$_2$ crystals for high-rate, precision timing applications}}, \bibinfo{journal}{JINST} \bibinfo{volume}{12} (\bibinfo{number}{01}) (\bibinfo{year}{2017}) \bibinfo{pages}{P01009}, \bibinfo{doi}{\doi{10.1088/1748-0221/12/01/P01009}}, \eprint{1611.03180}.

\bibtype{Article}%
\bibitem{Anastasi:2016luh}
\bibinfo{author}{A. Anastasi}, et al., \bibinfo{title}{{Electron beam test of key elements of the laser-based calibration system for the muon $g$ $-$ $2$ experiment}}, \bibinfo{journal}{Nucl. Instrum. Meth. A} \bibinfo{volume}{842} (\bibinfo{year}{2017}) \bibinfo{pages}{86--91}, \bibinfo{doi}{\doi{10.1016/j.nima.2016.10.047}}, \eprint{1610.03210}.

\bibtype{Article}%
\bibitem{Muong-2:2019beb}
\bibinfo{author}{K.~S. Khaw}, et al. (\bibinfo{collaboration}{Muon g-2}), \bibinfo{title}{{Performance of the Muon $g-2$ calorimeter and readout systems measured with test beam data}}, \bibinfo{journal}{Nucl. Instrum. Meth. A} \bibinfo{volume}{945} (\bibinfo{year}{2019}) \bibinfo{pages}{162558}, \bibinfo{doi}{\doi{10.1016/j.nima.2019.162558}}, \eprint{1905.04407}.

\bibtype{Article}%
\bibitem{Muong-2:2021vma}
\bibinfo{author}{T. Albahri}, et al. (\bibinfo{collaboration}{Muon g-2}), \bibinfo{title}{{Measurement of the anomalous precession frequency of the muon in the Fermilab Muon $g-2$ Experiment}}, \bibinfo{journal}{Phys. Rev. D} \bibinfo{volume}{103} (\bibinfo{number}{7}) (\bibinfo{year}{2021}) \bibinfo{pages}{072002}, \bibinfo{doi}{\doi{10.1103/PhysRevD.103.072002}}, \eprint{2104.03247}.

\bibtype{Article}%
\bibitem{Muong-2:2021ojo}
\bibinfo{author}{B. Abi}, et al. (\bibinfo{collaboration}{Muon g-2}), \bibinfo{title}{{Measurement of the Positive Muon Anomalous Magnetic Moment to 0.46 ppm}}, \bibinfo{journal}{Phys. Rev. Lett.} \bibinfo{volume}{126} (\bibinfo{number}{14}) (\bibinfo{year}{2021}) \bibinfo{pages}{141801}, \bibinfo{doi}{\doi{10.1103/PhysRevLett.126.141801}}, \eprint{2104.03281}.

\bibtype{Article}%
\bibitem{Muong-2:2021xzz}
\bibinfo{author}{T. Albahri}, et al. (\bibinfo{collaboration}{Muon g-2}), \bibinfo{title}{{Beam dynamics corrections to the Run-1 measurement of the muon anomalous magnetic moment at Fermilab}}, \bibinfo{journal}{Phys. Rev. Accel. Beams} \bibinfo{volume}{24} (\bibinfo{number}{4}) (\bibinfo{year}{2021}) \bibinfo{pages}{044002}, \bibinfo{doi}{\doi{10.1103/PhysRevAccelBeams.24.044002}}, \eprint{2104.03240}.

\bibtype{Article}%
\bibitem{Muong-2:2023cdq}
\bibinfo{author}{D.~P. Aguillard}, et al. (\bibinfo{collaboration}{Muon g-2}), \bibinfo{title}{{Measurement of the Positive Muon Anomalous Magnetic Moment to 0.20~ppm}}, \bibinfo{journal}{Phys. Rev. Lett.} \bibinfo{volume}{131} (\bibinfo{number}{16}) (\bibinfo{year}{2023}) \bibinfo{pages}{161802}, \bibinfo{doi}{\doi{10.1103/PhysRevLett.131.161802}}, \eprint{2308.06230}.

\bibtype{Article}%
\bibitem{King_2022}
\bibinfo{author}{B.T. King}, et al., \bibinfo{title}{{The straw tracking detector for the Fermilab Muon g-2 Experiment}}, \bibinfo{journal}{J. Instrum.} \bibinfo{volume}{17} (\bibinfo{number}{02}) (\bibinfo{year}{2022}) \bibinfo{pages}{P02035}, \bibinfo{doi}{\doi{10.1088/1748-0221/17/02/P02035}}, \bibinfo{url}{\urlprefix\url{https://dx.doi.org/10.1088/1748-0221/17/02/P02035}}.

\bibtype{Article}%
\bibitem{Muong-2:2021ovs}
\bibinfo{author}{T. Albahri}, et al. (\bibinfo{collaboration}{Muon g-2}), \bibinfo{title}{{Magnetic-field measurement and analysis for the Muon $g-2$ Experiment at Fermilab}}, \bibinfo{journal}{Phys. Rev. A} \bibinfo{volume}{103} (\bibinfo{number}{4}) (\bibinfo{year}{2021}) \bibinfo{pages}{042208}, \bibinfo{doi}{\doi{10.1103/PhysRevA.103.042208}}, \eprint{2104.03201}.

\bibtype{Article}%
\bibitem{Liu:1999iz}
\bibinfo{author}{Weiwen Liu}, et al., \bibinfo{title}{{High precision measurements of the ground state hyperfine structure interval of muonium and of the muon magnetic moment}}, \bibinfo{journal}{Phys. Rev. Lett.} \bibinfo{volume}{82} (\bibinfo{year}{1999}) \bibinfo{pages}{711--714}, \bibinfo{doi}{\doi{10.1103/PhysRevLett.82.711}}.

\bibtype{Article}%
\bibitem{Mohr:2024kco}
\bibinfo{author}{Peter~J. Mohr}, \bibinfo{author}{David~B. Newell}, \bibinfo{author}{Barry~N. Taylor}, \bibinfo{author}{Eite Tiesinga}, \bibinfo{title}{{CODATA recommended values of the fundamental physical constants: 2022*}}, \bibinfo{journal}{Rev. Mod. Phys.} \bibinfo{volume}{97} (\bibinfo{number}{2}) (\bibinfo{year}{2025}) \bibinfo{pages}{025002}, \bibinfo{doi}{\doi{10.1103/RevModPhys.97.025002}}, \eprint{2409.03787}.

\bibtype{Article}%
\bibitem{Borsanyi:2020mff}
\bibinfo{author}{Sz. Borsanyi}, et al., \bibinfo{title}{{Leading hadronic contribution to the muon magnetic moment from lattice QCD}}, \bibinfo{journal}{Nature} \bibinfo{volume}{593} (\bibinfo{number}{7857}) (\bibinfo{year}{2021}) \bibinfo{pages}{51--55}, \bibinfo{doi}{\doi{10.1038/s41586-021-03418-1}}, \eprint{2002.12347}.

\bibtype{Article}%
\bibitem{CMD-3:2023alj}
\bibinfo{author}{F.~V. Ignatov}, et al. (\bibinfo{collaboration}{CMD-3}), \bibinfo{title}{{Measurement of the e+e-{\textrightarrow}{\ensuremath{\pi}}+{\ensuremath{\pi}}- cross section from threshold to 1.2~GeV with the CMD-3 detector}}, \bibinfo{journal}{Phys. Rev. D} \bibinfo{volume}{109} (\bibinfo{number}{11}) (\bibinfo{year}{2024}) \bibinfo{pages}{112002}, \bibinfo{doi}{\doi{10.1103/PhysRevD.109.112002}}, \eprint{2302.08834}.

\bibtype{Article}%
\bibitem{CMD-3:2023rfe}
\bibinfo{author}{F.~V. Ignatov}, et al. (\bibinfo{collaboration}{CMD-3}), \bibinfo{title}{{Measurement of the Pion Form Factor with CMD-3 Detector and its Implication to the Hadronic Contribution to Muon (g-2)}}, \bibinfo{journal}{Phys. Rev. Lett.} \bibinfo{volume}{132} (\bibinfo{number}{23}) (\bibinfo{year}{2024}) \bibinfo{pages}{231903}, \bibinfo{doi}{\doi{10.1103/PhysRevLett.132.231903}}, \eprint{2309.12910}.

\bibtype{Article}%
\bibitem{CarloniCalame:2015obs}
\bibinfo{author}{C.~M. Carloni~Calame}, \bibinfo{author}{M. Passera}, \bibinfo{author}{L. Trentadue}, \bibinfo{author}{G. Venanzoni}, \bibinfo{title}{{A new approach to evaluate the leading hadronic corrections to the muon $g$-2}}, \bibinfo{journal}{Phys. Lett. B} \bibinfo{volume}{746} (\bibinfo{year}{2015}) \bibinfo{pages}{325--329}, \bibinfo{doi}{\doi{10.1016/j.physletb.2015.05.020}}, \eprint{1504.02228}.

\bibtype{Article}%
\bibitem{MUonE:2016hru}
\bibinfo{author}{G. Abbiendi}, et al. (\bibinfo{collaboration}{MUonE}), \bibinfo{title}{{Measuring the leading hadronic contribution to the muon g-2 via $\mu e$ scattering}}, \bibinfo{journal}{Eur. Phys. J. C} \bibinfo{volume}{77} (\bibinfo{number}{3}) (\bibinfo{year}{2017}) \bibinfo{pages}{139}, \bibinfo{doi}{\doi{10.1140/epjc/s10052-017-4633-z}}, \eprint{1609.08987}.

\bibtype{Article}%
\bibitem{10.1093/ptep/ptz030}
\bibinfo{author}{M Abe}, et al., \bibinfo{title}{{A new approach for measuring the muon anomalous magnetic moment and electric dipole moment}}, \bibinfo{journal}{Prog. Theor. Exp. Phys.} \bibinfo{volume}{2019} (\bibinfo{year}{2019}) \bibinfo{pages}{053C02}, \bibinfo{doi}{\doi{10.1093/ptep/ptz030}}.

\bibtype{Article}%
\bibitem{Abe:2018tmp}
\bibinfo{author}{M. Abe}, \bibinfo{author}{Y. Murata}, \bibinfo{author}{H. Iinuma}, \bibinfo{author}{T. Ogitsu}, \bibinfo{author}{N. Saito}, \bibinfo{author}{K. Sasaki}, \bibinfo{author}{T. Mibe}, \bibinfo{author}{H. Nakayama}, \bibinfo{title}{{Magnetic design and method of a superconducting magnet for muon g\ensuremath{-}2 /EDM precise measurements in a cylindrical volume with homogeneous magnetic field}}, \bibinfo{journal}{Nucl. Instrum. Meth. A} \bibinfo{volume}{890} (\bibinfo{year}{2018}) \bibinfo{pages}{51--63}, \bibinfo{doi}{\doi{10.1016/j.nima.2018.01.026}}.

\bibtype{Article}%
\bibitem{Iinuma:2016zfu}
\bibinfo{author}{Hiromi Iinuma}, \bibinfo{author}{Hisayoshi Nakayama}, \bibinfo{author}{Katsunobu Oide}, \bibinfo{author}{Ken-ichi Sasaki}, \bibinfo{author}{Naohito Saito}, \bibinfo{author}{Tsutomu Mibe}, \bibinfo{author}{Mitsushi Abe}, \bibinfo{title}{{Three-dimensional spiral injection scheme for the g-2/EDM experiment at J-PARC}}, \bibinfo{journal}{Nucl. Instrum. Meth. A} \bibinfo{volume}{832} (\bibinfo{year}{2016}) \bibinfo{pages}{51--62}, \bibinfo{doi}{\doi{10.1016/j.nima.2016.05.126}}.

\bibtype{Article}%
\bibitem{Otani:2016swo}
\bibinfo{author}{M. Otani}, \bibinfo{author}{T. Mibe}, \bibinfo{author}{M. Yoshida}, \bibinfo{author}{K. Hasegawa}, \bibinfo{author}{Y. Kondo}, \bibinfo{author}{N. Hayashizaki}, \bibinfo{author}{Y. Iwashita}, \bibinfo{author}{Y. Iwata}, \bibinfo{author}{R. Kitamura}, \bibinfo{author}{N. Saito}, \bibinfo{title}{{Interdigital H-mode drift-tube linac design with alternative phase focusing for muon linac}}, \bibinfo{journal}{Phys. Rev. Accel. Beams} \bibinfo{volume}{19} (\bibinfo{number}{4}) (\bibinfo{year}{2016}) \bibinfo{pages}{040101}, \bibinfo{doi}{\doi{10.1103/PhysRevAccelBeams.19.040101}}.

\bibtype{Article}%
\bibitem{Aritome:2024jiv}
\bibinfo{author}{S. Aritome}, et al., \bibinfo{title}{{Acceleration of positive muons by a radio-frequency cavity}}  (\bibinfo{year}{2024}), \eprint{2410.11367}.

\bibtype{Article}%
\bibitem{10.1093/ptep/ptac059}
\bibinfo{author}{Yu Hamada}, \bibinfo{author}{Ryuichiro Kitano}, \bibinfo{author}{Ryutaro Matsudo}, \bibinfo{author}{Hiromasa Takaura}, \bibinfo{author}{Mitsuhiro Yoshida}, \bibinfo{title}{{$\mu$TRISTAN}}, \bibinfo{journal}{Prog. Theor. Exp. Phys.} \bibinfo{volume}{2022} (\bibinfo{year}{2022}) \bibinfo{pages}{053B02}, \bibinfo{doi}{\doi{10.1093/ptep/ptac059}}.

\bibtype{Article}%
\bibitem{PhysRevLett.74.4811}
\bibinfo{author}{K. Nagamine}, \bibinfo{author}{Y. Miyake}, \bibinfo{author}{K. Shimomura}, \bibinfo{author}{P. Birrer}, \bibinfo{author}{J.~P. Marangos}, \bibinfo{author}{M. Iwasaki}, \bibinfo{author}{P. Strasser}, \bibinfo{author}{T. Kuga}, \bibinfo{title}{{Ultraslow Positive-Muon Generation by Laser Ionization of Thermal Muonium from Hot Tungsten at Primary Proton Beam}}, \bibinfo{journal}{Phys. Rev. Lett.} \bibinfo{volume}{74} (\bibinfo{year}{1995}) \bibinfo{pages}{4811--4814}, \bibinfo{doi}{\doi{10.1103/PhysRevLett.74.4811}}.

\bibtype{Article}%
\bibitem{BAKULE2008335}
\bibinfo{author}{Pavel Bakule}, \bibinfo{author}{Yasuyuki Matsuda}, \bibinfo{author}{Yasuhiro Miyake}, \bibinfo{author}{Kanetada Nagamine}, \bibinfo{author}{Masahiko Iwasaki}, \bibinfo{author}{Yutaka Ikedo}, \bibinfo{author}{Koichiro Shimomura}, \bibinfo{author}{Patrick Strasser}, \bibinfo{author}{Shunshuke Makimura}, \bibinfo{title}{Pulsed source of ultra low energy positive muons for near-surface $\mu\mathrm{SR}$ studies}, \bibinfo{journal}{Nucl. Instrum. Methods Phys. Res. B} \bibinfo{volume}{266} (\bibinfo{year}{2008}) \bibinfo{pages}{335--346}, \bibinfo{doi}{\doi{https://doi.org/10.1016/j.nimb.2007.11.009}}.

\bibtype{Article}%
\bibitem{Zhang:2022ilj}
\bibinfo{author}{C. Zhang}, et al., \bibinfo{title}{{Modeling the diffusion of muonium in silica aerogel and its application to a novel design of multi-layer target for thermal muon generation}}, \bibinfo{journal}{Nucl. Instrum. Meth. A} \bibinfo{volume}{1042} (\bibinfo{year}{2022}) \bibinfo{pages}{167443}, \bibinfo{doi}{\doi{10.1016/j.nima.2022.167443}}.

\bibtype{Article}%
\bibitem{Nakazawa:2022yae}
\bibinfo{author}{Y. Nakazawa}, et al., \bibinfo{title}{{High-power test of an interdigital H-mode drift tube linac for the J-PARC muon g\ensuremath{-}2 and electric dipole moment experiment}}, \bibinfo{journal}{Phys. Rev. Accel. Beams} \bibinfo{volume}{25} (\bibinfo{number}{11}) (\bibinfo{year}{2022}) \bibinfo{pages}{110101}, \bibinfo{doi}{\doi{10.1103/PhysRevAccelBeams.25.110101}}.

\bibtype{Article}%
\bibitem{Shimomura:2024puh}
\bibinfo{author}{Koichiro Shimomura}, et al., \bibinfo{title}{{Pulsed muon facility of J-PARC MUSE}}, \bibinfo{journal}{Hyperfine Interact.} \bibinfo{volume}{245} (\bibinfo{number}{1}) (\bibinfo{year}{2024}) \bibinfo{pages}{31}, \bibinfo{doi}{\doi{10.1007/s10751-024-01863-8}}.

\bibtype{Article}%
\bibitem{Kawamura:2018apy}
\bibinfo{author}{Naritoshi Kawamura}, et al., \bibinfo{title}{{New concept for a large-acceptance general-purpose muon beamline}}, \bibinfo{journal}{PTEP} \bibinfo{volume}{2018} (\bibinfo{number}{11}) (\bibinfo{year}{2018}) \bibinfo{pages}{113G01}, \bibinfo{doi}{\doi{10.1093/ptep/pty116}}.

\bibtype{Article}%
\bibitem{10.1093/ptep/ptu116}
\bibinfo{author}{G.~A. Beer}, et al., \bibinfo{title}{{Enhancement of muonium emission rate from silica aerogel with a laser-ablated surface}}, \bibinfo{journal}{Prog. Theor. Exp. Phys.} \bibinfo{volume}{2014} (\bibinfo{year}{2014}) \bibinfo{pages}{091C01}, \bibinfo{doi}{\doi{10.1093/ptep/ptu116}}.

\bibtype{Article}%
\bibitem{10.1093/ptep/ptaa145}
\bibinfo{author}{J Beare}, et al., \bibinfo{title}{{Study of muonium emission from laser-ablated silica aerogel}}, \bibinfo{journal}{Prog. Theor. Exp. Phys.} \bibinfo{volume}{2020} (\bibinfo{year}{2020}) \bibinfo{pages}{123C01}, \bibinfo{doi}{\doi{10.1093/ptep/ptaa145}}.

\bibtype{Article}%
\bibitem{10.1093/ptep/ptt080}
\bibinfo{author}{P. Bakule}, et al., \bibinfo{title}{{Measurement of muonium emission from silica aerogel}}, \bibinfo{journal}{Prog. Theor. Exp. Phys.} \bibinfo{volume}{2013} (\bibinfo{year}{2013}) \bibinfo{pages}{103C01}, \bibinfo{doi}{\doi{10.1093/ptep/ptt080}}.

\bibtype{Article}%
\bibitem{PhysRevAccelBeams.21.050101}
\bibinfo{author}{S. Bae}, et al., \bibinfo{title}{First muon acceleration using a radio-frequency accelerator}, \bibinfo{journal}{Phys. Rev. Accel. Beams} \bibinfo{volume}{21} (\bibinfo{year}{2018}) \bibinfo{pages}{050101}, \bibinfo{doi}{\doi{10.1103/PhysRevAccelBeams.21.050101}}.

\bibtype{Article}%
\bibitem{DELPHI:2003nah}
\bibinfo{author}{J. Abdallah}, et al. (\bibinfo{collaboration}{DELPHI}), \bibinfo{title}{{Study of tau-pair production in photon-photon collisions at LEP and limits on the anomalous electromagnetic moments of the tau lepton}}, \bibinfo{journal}{Eur. Phys. J. C} \bibinfo{volume}{35} (\bibinfo{year}{2004}) \bibinfo{pages}{159--170}, \bibinfo{doi}{\doi{10.1140/epjc/s2004-01852-y}}, \eprint{hep-ex/0406010}.

\bibtype{Article}%
\bibitem{CMS:2024qjo}
\bibinfo{author}{Aram Hayrapetyan}, et al. (\bibinfo{collaboration}{CMS}), \bibinfo{title}{{Observation of $\gamma\gamma\to\tau\tau$ in proton-proton collisions and limits on the anomalous electromagnetic moments of the $\tau$ lepton}}, \bibinfo{journal}{Rept. Prog. Phys.} \bibinfo{volume}{87} (\bibinfo{number}{10}) (\bibinfo{year}{2024}) \bibinfo{pages}{107801}, \bibinfo{doi}{\doi{10.1088/1361-6633/ad6fcb}}, \eprint{2406.03975}.

\bibtype{Article}%
\bibitem{Eidelman:2007sb}
\bibinfo{author}{S. Eidelman}, \bibinfo{author}{M. Passera}, \bibinfo{title}{{Theory of the tau lepton anomalous magnetic moment}}, \bibinfo{journal}{Mod. Phys. Lett. A} \bibinfo{volume}{22} (\bibinfo{year}{2007}) \bibinfo{pages}{159--179}, \bibinfo{doi}{\doi{10.1142/S0217732307022694}}, \eprint{hep-ph/0701260}.

\bibtype{Article}%
\bibitem{Crivellin:2021spu}
\bibinfo{author}{Andreas Crivellin}, \bibinfo{author}{Martin Hoferichter}, \bibinfo{author}{J.~Michael Roney}, \bibinfo{title}{{Toward testing the magnetic moment of the tau at one part per million}}, \bibinfo{journal}{Phys. Rev. D} \bibinfo{volume}{106} (\bibinfo{number}{9}) (\bibinfo{year}{2022}) \bibinfo{pages}{093007}, \bibinfo{doi}{\doi{10.1103/PhysRevD.106.093007}}, \eprint{2111.10378}.

\bibtype{Article}%
\bibitem{Fujikawa1980}
\bibinfo{author}{Kazuo Fujikawa}, \bibinfo{author}{Robert~E. Shrock}, \bibinfo{title}{The Magnetic Moment of a Massive Neutrino and Neutrino-Spin Rotation}, \bibinfo{journal}{Physical Review Letters} \bibinfo{volume}{45} (\bibinfo{year}{1980}) \bibinfo{pages}{963--966}, \bibinfo{doi}{\doi{10.1103/PhysRevLett.45.963}}.

\bibtype{Article}%
\bibitem{Giunti:2014ixa}
\bibinfo{author}{Carlo Giunti}, \bibinfo{author}{Alexander Studenikin}, \bibinfo{title}{{Neutrino electromagnetic interactions: a window to new physics}}, \bibinfo{journal}{Rev. Mod. Phys.} \bibinfo{volume}{87} (\bibinfo{year}{2015}) \bibinfo{pages}{531}, \bibinfo{doi}{\doi{10.1103/RevModPhys.87.531}}, \eprint{1403.6344}.

\bibtype{Article}%
\bibitem{Giunti:2024gec}
\bibinfo{author}{Carlo Giunti}, \bibinfo{author}{Konstantin Kouzakov}, \bibinfo{author}{Yu-Feng Li}, \bibinfo{author}{Alexander Studenikin}, \bibinfo{title}{{Neutrino Electromagnetic Properties}}  (\bibinfo{year}{2024}), \bibinfo{doi}{\doi{10.1146/annurev-nucl-102122-023242}}, \eprint{2411.03122}.

\bibtype{Misc}%
\bibitem{PDG2024Neutrino}
\bibinfo{author}{{Particle Data Group}}, \bibinfo{title}{Neutrino Properties} \bibinfo{year}{2024}, \bibinfo{url}{\urlprefix\url{https://pdg.lbl.gov/2024/web/viewer.html?file=../listings/rpp2024-list-neutrino-prop.pdf}}.

\bibtype{Article}%
\bibitem{Beda:2012zz}
\bibinfo{author}{A.~G. Beda}, \bibinfo{author}{V.~B. Brudanin}, \bibinfo{author}{V.~G. Egorov}, \bibinfo{author}{D.~V. Medvedev}, \bibinfo{author}{V.~S. Pogosov}, \bibinfo{author}{M.~V. Shirchenko}, \bibinfo{author}{A.~S. Starostin}, \bibinfo{title}{{The results of search for the neutrino magnetic moment in GEMMA experiment}}, \bibinfo{journal}{Adv. High Energy Phys.} \bibinfo{volume}{2012} (\bibinfo{year}{2012}) \bibinfo{pages}{350150}, \bibinfo{doi}{\doi{10.1155/2012/350150}}.

\end{thebibliography*}